\documentclass[twocolumn,english,prd,showpacs,superscriptaddress,nofootinbib]{revtex4}
\pdfoutput=1

\usepackage[latin9]{inputenc}
\usepackage{xcolor}
\usepackage{babel}
\usepackage{amsmath}
\usepackage{amssymb}
\usepackage{esint}
\usepackage[unicode=true,
 bookmarks=false,
 breaklinks=false,pdfborder={0 0 1},backref=false,colorlinks=true]
 {hyperref}
\hypersetup{
 pdfauthor={Hamid Afshar, Daniel Grumiller, Wout Merbis, Alfredo Perez, David Tempo and Ricardo Troncoso},
 linkcolor=blue,urlcolor=magenta,filecolor=green,citecolor=red,pdfstartview=FitV,bookmarksopen=true}
\usepackage{breakurl}

\makeatletter
\@ifundefined{textcolor}{}
{%
 \definecolor{BLACK}{gray}{0}
 \definecolor{WHITE}{gray}{1}
 \definecolor{RED}{rgb}{1,0,0}
 \definecolor{GREEN}{rgb}{0,1,0}
 \definecolor{BLUE}{rgb}{0,0,1}
 \definecolor{CYAN}{cmyk}{1,0,0,0}
 \definecolor{MAGENTA}{cmyk}{0,1,0,0}
 \definecolor{YELLOW}{cmyk}{0,0,1,0}
}

\usepackage{babel}
\usepackage{babel}
\usepackage{babel}
\usepackage{babel}
\usepackage{babel}

\usepackage{verbatim}\usepackage{graphicx}\usepackage{color}

\DeclareFontFamily{OT1}{rsfs}{}
\DeclareFontShape{OT1}{rsfs}{m}{n}{ <-7> rsfs5 <7-10> rsfs7 <10->rsfs10}{} 
\DeclareMathAlphabet{\mycal}{OT1}{rsfs}{m}{n}

\newcommand{\unity}{1\hspace{-0.243em}\text{l}}

\newcommand{\be}[1]{ \begin{equation}\label{#1} }
\newcommand{\ee}{\end{equation}}
\newcommand{\bea}[1]{\begin{eqnarray}\label{#1} }
\newcommand{\eea}{\end{eqnarray}}

\newcommand{\eq}[2]{\begin{equation} #1 \label{#2} \end{equation}}

\newcommand{\ga}{\gamma}
\newcommand{\de}{\delta}

\newcommand{\Om}{\Omega}

\DeclareMathOperator{\extdm}{d}

\newcommand{\extd}{\extdm \!}

\newcommand{\cM}{{\cal M}}
\newcommand{\cN}{{\cal L}} 

\newcommand{\beq}{\begin{equation}}
\newcommand{\eeq}{\end{equation}}
\newcommand{\bi}{\begin{itemize}}
\newcommand{\ei}{\end{itemize}}
\newcommand{\bt}{\begin{tabular}}
\newcommand{\et}{\end{tabular}}
\newcommand{\bc}{\begin{center}}
\newcommand{\ec}{\end{center}}

\def\one{{\hbox{ 1\kern-.8mm l}}}
\newcommand{\Dslash}{\not{\hbox{\kern-4pt $D$}}}
\newcommand{\pdslash}{\not{\hbox{\kern-2pt $\partial$}}}

\newcommand{\ba}{\begin{array}}
\newcommand{\ea}{\end{array}}

\def\bbox{{\,\lower0.9pt\vbox{\hrule \hbox{\vrule height 0.2 cm
\hskip 0.2 cm \vrule height 0.2 cm}\hrule}\,}}
\newcommand{\dsl}{\pa \kern-0.5em /}

\newcommand{\vp}{\varphi}

\makeatother

\begin{document}
\global\long\def\mytitle{Soft hairy horizons in three spacetime dimensions}

\title{\mytitle}

\author{Hamid Afshar}
\email{afshar@ipm.ir}
\affiliation{School of Physics, Institute for Research in Fundamental Sciences
(IPM), P.O.Box 19395-5531, Tehran, Iran}

\author{Daniel Grumiller}
\email{grumil@hep.itp.tuwien.ac.at}
\affiliation{Institute for Theoretical Physics, TU Wien, Wiedner Hauptstrasse
8--10/136, A-1040 Vienna, Austria}
\affiliation{Centro de Estudios Cient\'ificos (CECs), Av. Arturo Prat 514, Valdivia,
Chile}

\author{Wout Merbis}
\email{merbis@hep.itp.tuwien.ac.at}
\affiliation{Institute for Theoretical Physics, TU Wien, Wiedner Hauptstrasse
8--10/136, A-1040 Vienna, Austria}

\author{Alfredo Perez}
\email{aperez@cecs.cl}
\affiliation{Centro de Estudios Cient\'ificos (CECs), Av. Arturo Prat 514, Valdivia,
Chile}

\author{David Tempo}
\email{tempo@cecs.cl}
\affiliation{Centro de Estudios Cient\'ificos (CECs), Av. Arturo Prat 514, Valdivia,
Chile}

\author{Ricardo Troncoso}
\email{troncoso@cecs.cl}
\affiliation{Centro de Estudios Cient\'ificos (CECs), Av. Arturo Prat 514, Valdivia,
Chile}

\date{\today}

\preprint{CECS-PHY-16/05, TUW--16--23}
\begin{abstract}
We discuss some aspects of soft hairy black holes and a new kind of ``soft hairy cosmologies'', including a detailed derivation of the metric formulation, results on flat space, and novel observations concerning the entropy. Remarkably, like in the case with negative cosmological constant, we find that the asymptotic symmetries for locally flat spacetimes with a horizon are governed by infinite copies of the Heisenberg algebra that generate soft hair descendants. It is also shown that the generators of the three-dimensional Bondi--Metzner--Sachs algebra arise from composite operators of the affine $\hat{u}(1)$ currents through a twisted Sugawara-like construction. We then discuss entropy macroscopically, thermodynamically and microscopically and discover that a microscopic formula derived recently for boundary conditions associated to the Korteweg--de~Vries hierarchy fits perfectly our results for entropy and ground state energy. We conclude with a comparison to related approaches.
\end{abstract}


\maketitle
\tableofcontents{}


\section{Introduction}

Black holes and cosmological spacetimes exhibit generically non-extremal horizons. In the near horizon limit such spacetimes universally are approximated by Rindler space \cite{Rindler:1966zz}. Whenever one is interested in asking conditional questions, like ``given a black hole, what are the scattering amplitudes in a given channel?'' or ``given a cosmological horizon, what are the allowed states that remain in the physical Hilbert space and how could they be related through symmetries?'' or ``given a black hole or cosmological horizon, can we microscopically account for the Bekenstein--Hawking entropy?'', it is crucial to impose boundary conditions that make sure that the condition in the question is met \cite{Carlip:1999cy}. In other words, we are searching for a consistent set of boundary conditions that guarantees the existence of a (regular, non-extremal) horizon. This motivates the invention of suitable near horizon boundary conditions.

Recently, near-horizon-inspired boundary conditions were proposed for three-dimensional spacetimes with negative cosmological constant \cite{Afshar:2016wfy}, which allowed to discuss novel aspects of soft hair (in the sense of Hawking, Perry and Strominger \cite{Hawking:2016msc, Hawking:2016sgy}), black hole entropy and black hole complementarity.

In the present work we give more details on the metric formulation, and generalize the discussion to cosmological spacetimes in the absence of a cosmological constant.

One of the main results of our flat space analysis is that the asymptotic symmetry algebra turns out to be precisely the same as in Anti-de Sitter (AdS) space \cite{Afshar:2016wfy}, namely infinite copies of the Heisenberg algebra supplemented by two Casimirs, one of which is the Hamiltonian. This supports the interpretation that our symmetry algebra can be naturally considered from a near horizon perspective, since the near horizon physics is expected to be insensitive to the presence or absence of a cosmological constant, as opposed to the usual expectation in the asymptotic region. For this reason hereafter we shall refer to this algebra sometimes as ``near horizon symmetry algebra'' (NHSA). However, we stress that our boundary conditions and all results based upon them can be interpreted also from an asymptotic observers perspective, which sometimes is more useful than the near horizon perspective.

Remarkably, the Bekenstein--Hawking entropy 
\begin{equation}
S_{\textrm{\tiny BH}}=\frac{A}{4G} = 2\pi\big(J_0^+ + J_0^-\big)\;,\label{eq:r1}
\end{equation}
once expressed in terms of our global charges, also acquires a unique expression that depends only on the zero modes $J_0^\pm$ and turns out to be insensitive to the value of the cosmological constant.

This work is organized as follows. In section \ref{se:2} we recapitulate key results of \cite{Afshar:2016wfy}, recast in Gaussian normal coordinates. In section \ref{MetricFormulationAdS} we address in more detail aspects of the metric formulation, in particular the asymptotic Killing vectors, the Regge--Teitelboim charges and the boundary conditions on metric fluctuations near the horizon and in the asymptotic region. In section \ref{se:3} we formulate suitable boundary conditions in flat space and derive the asymptotic symmetry algebra, which turns out to be isomorphic to the one in AdS; we generalize the soft hairy discussion to flat space cosmologies. In section \ref{se:algebras} we focus on algebraic aspects and various Sugawara-like constructions based on our near horizon symmetry algebra, as well as algebraic generalizations to higher spins and higher dimensions. In section \ref{se:6} we discuss the entropy from various perspectives; first, we recover the macroscopic Bekenstein--Hawking result in the Chern--Simons formulation; second, we address microstate counting and observe that a special case of a microscopic entropy formula fits perfectly our results, which was originally derived for boundary conditions on AdS$_3$ where the boundary gravitons obey the equations of some representative of the Korteweg--de~Vries (KdV) hierarchy. In section \ref{se:42} we compare with various related approaches, in particular with the one in \cite{Donnay:2015abr}. In section \ref{se:7} we conclude with a brief summary and an outlook to future research directions. 

Before starting we mention some of our conventions. We work in $2+1$ dimensional spacetimes of signature $(-,+,+)$ and use the following sign convention for the Ricci tensor $R_{\mu\nu}=\partial_{\alpha}\Gamma^{\alpha}{}_{\mu\nu}+\dots$

\section{Summary of Anti-de~Sitter results}\label{se:2}

In this section we summarize, and partly make more explicit, the results of \cite{Afshar:2016wfy}. We start by presenting a special class of metrics that solves the three-dimensional Einstein equations with negative cosmological constant, 
\begin{equation}
R_{\mu\nu}-\frac{1}{2}g_{\mu\nu}R+\Lambda g_{\mu\nu}=0\label{eq:EEqs}
\end{equation}
and that approaches Rindler spacetime near the horizon in section \ref{se:2.1}, focusing first on Gaussian normal coordinates and then on Eddington--Finkelstein coordinates. In section \ref{se:2.2} we recapitulate the Chern--Simons formulation, in which the boundary conditions appear naturally in diagonal gauge, see section \ref{se:2.4}. Section \ref{se:2.3} displays the canonical charges and their asymptotic symmetry algebra. In section \ref{se:soft} we recall the definition of soft hair descendants.

\subsection{Soft hairy black hole metric}\label{se:2.1}

The near horizon metric in a co-rotating frame acquires the form 
\begin{equation}
\extd s^{2}=-a^{2}r^{2}\extd t^{2}+\extd r^{2}+\gamma^{2}\extd\varphi^{2}+\cdots\,\label{eq:r2}
\end{equation}
where $a$ is the Rindler acceleration, and $r=0$ stands for the location of the (Rindler) horizon. The angular coordinate is assumed to be periodic, $\vp\sim\vp+2\pi$, so that the horizon area is given by
\eq{
A=\oint\extd\varphi\,\gamma\,. 
}{eq:area}
The ellipsis in \eqref{eq:r2} corresponds to higher order terms in the radial coordinate $r$.

In \cite{Afshar:2016wfy} we continued the analysis in Eddington--Finkelstein coordinates. We summarize (and generalize) these results in appendix \ref{app:A}. Here we proceed instead in Gaussian normal coordinates, following the higher spin discussion \cite{Grumiller:2016kcp}.

Parametrizing the cosmological constant in terms of the AdS radius, $\Lambda=-\ell^{-2}$, the full metric that approaches \eqref{eq:r2} in the near horizon limit for a generic rotating frame and solves the Einstein equations \eqref{eq:EEqs} in Gaussian normal coordinates is given by 
\begin{align}
\extd s^{2}= & \extd r^{2}-\big((a^{2}\ell^{2}-\Omega^{2})\cosh^{2}\left(r/\ell\right)-a^{2}\ell^{2}\big)\,\extd t^{2}\nonumber \\
 & +2\big(\gamma\Omega\cosh^{2}\left(r/\ell\right)+a\omega\ell^{2}\sinh^{2}\left(r/\ell\right)\big)\,\extd t\extd\varphi\nonumber \\
 & +\big(\gamma^{2}\cosh^{2}\left(r/\ell\right)-\omega^{2}\ell^{2}\sinh^{2}\left(r/\ell\right)\big)\,\extd\varphi^{2} \,.\label{eq:tba}
\end{align}
The line-element \eqref{eq:tba} depends on four functions $a,\Omega,\gamma,\omega$ of time and the angular coordinate, subject to the on-shell conditions
\begin{equation}
\dot{\gamma}=\Omega^{\prime}\text{\quad\quad}\dot{\omega}=-a^{\prime}\label{eq:r6}
\end{equation}
where prime denotes $\partial_{\varphi}$ and dot denotes $\partial_{t}$.

This patch of coordinates covers the region outside the event horizon, $r\geq0$. Since the generic solution is not spherically symmetric, its geometry describes a sort of ``black flower'' (see e.g.~\cite{Barnich:2015dvt}). As explained in section \ref{se:2.3} and further elaborated in section \ref{MetricFormulationAdS}, the state-dependent functions $\omega$ and $\gamma$ cannot be gauged away because they correspond to the global charges. The quantities $a$ and $\Omega$ are interpreted as chemical potentials, meaning that they are arbitrary but fixed functions of $\varphi$ and $t$. For consistency we assume that the function determining the surface area is positive everywhere, i.e., $\gamma>0$, with no loss of generality.

In order to make explicit contact with the expression for the metric in \cite{Afshar:2016wfy}, for simplicity 
we choose $a$ to be constant and adopt a co-rotating frame ($\Omega=0$). Then, in (ingoing) Eddington--Finkelstein coordinates with advanced time $v$,
\begin{equation}
v=t-\frac{1}{2}a^{-1}\text{log}\Big(\frac{f\left(\rho\right)}{\rho}\Big)\qquad \rho=a\ell^{2}\Big(\text{cosh}\big(\frac{r}{\ell}\big)-1\Big)
\label{eq:ChangeofCoordAdS}
\end{equation}
the line element \eqref{eq:tba} reduces to 
\begin{multline}
\extd s^{2}=-2a\rho f\left(\rho\right)\,\extd v^{2}+2\extd v\extd\rho+4\omega\rho f\left(\rho\right)\,\extd v\extd\varphi\\
-2\frac{\omega}{a}\,\extd\varphi\extd\rho + \Big[\gamma^{2}+\frac{2\rho}{a\ell^{2}}(\gamma^{2}-\ell^{2}\omega^{2})f\left(\rho\right)\Big]\,\extd\varphi^{2} 
\label{eq:Null-AdS-metric}
\end{multline}
where $f\left(\rho\right)=1+\rho/(2a\ell^{2})$. The functions $\omega$ and $\gamma$ now depend on $\vp$, only. Note that in the case of constant values of $\omega$ and $\gamma$ the solution is spherically symmetric and reduces to the BTZ black hole \cite{Banados:1992wn,Banados:1992gq}.

To facilitate comparison with literature we translate now into standard BTZ variables. The BTZ black hole with outer and inner horizon radii $r_{\pm}^{\textrm{\tiny BTZ}}$ (in the Schwarzschild-type of coordinates introduced in \cite{Banados:1992wn}) is recovered by identifying 
\begin{equation}
r_{+}^{\text{BTZ}}=\gamma\text{\quad\quad}r_{-}^{\text{BTZ}}=\left|\omega\right|\ell\label{eq:BTZ}
\end{equation}
with constant $\gamma>0$ and constant $\omega$. The sign of $\omega$ determines the direction of the rotation, and the following BPS-like inequality holds
\begin{equation}
\gamma>\left|\omega\right|\ell\,.\label{eq:BPS}
\end{equation}
This can be seen either directly from the line-element \eqref{eq:tba} in the limit of large $r$ or from the usual BTZ-inequality $r_{+}\geq r_{-}$ together with \eqref{eq:BTZ}. 

We continue in the metric formulation in section \ref{MetricFormulationAdS}. In the remainder of this section we switch to the Chern--Simons formulation, which is technically more convenient.

\subsection{Chern--Simons formulation}\label{se:2.2}

In the Chern--Simons formulation the bulk action for Einstein
gravity reads \cite{Achucarro:1987vz,Witten:1988hc} 
\begin{equation}
I_{\text{CS}}=\frac{k}{4\pi}\int\text{\,}\left\langle {\cal A}\wedge\extd{\cal A}+\tfrac{2}{3}\text{\,}{\cal A}\wedge{\cal A}\wedge{\cal A}\right\rangle \,. \label{eq:r3}
\end{equation}

In presence of a negative cosmological constant the Chern--Simons level is given by $k=\ell/(4G)$. The gauge field ${\cal A}$ splits into two independent $sl(2,\mathbb{R})$ connections $A^{\pm}$. The generators then fulfill
\begin{equation}
[L_{n},L_{m}]=(n-m)L_{n+m}
\end{equation}
with $n,m=0,\pm1$, and the bilinear form $\left\langle \,,\right\rangle $
corresponds to the standard one for each copy of $sl(2,\mathbb{R})$, given by
\begin{equation}
\langle L_{1} ,\,L_{-1}\rangle=-1\;,\; \langle L_{\pm1},\,L_{0}\rangle=0\;,\; \langle L_{0},\,L_{0}\rangle=\tfrac{1}{2}\,
\end{equation}
once one includes a relative minus sign in one of the sectors, which most conveniently is put into the action. Explicitly, in the AdS case the action \eqref{eq:r3} then splits into left ($+$) and right ($-$) chiral parts 
\begin{multline}
I_{\text{CS}}=\frac{\ell}{16\pi G}\,\int\text{\,}\left\langle A^{+}\wedge\extd A^{+}+\tfrac{2}{3}\,A^{+}\wedge A^{+}\wedge A^{+}\right\rangle \\
-\frac{\ell}{16\pi G}\,\int\text{\,}\left\langle A^{-}\wedge\extd A^{-}+\tfrac{2}{3}\,A^{-}\wedge A^{-}\wedge A^{-}\right\rangle \label{eq:r200}
\end{multline}
and the spacetime metric is recovered from the gauge fields $A^{\pm}$ according to 
\begin{equation}
g_{\mu\nu}=\tfrac{\ell^{2}}{2}\left\langle \left(A_{\mu}^{+}-A_{\mu}^{-}\right)\left(A_{\nu}^{+}-A_{\nu}^{-}\right)\right\rangle \,.\label{eq:r8}
\end{equation}

In our conventions, $t,r,\gamma,\ell,G$ have length dimension one, $\vp,\omega,\Omega,k,A^{\pm},L_{n}$ are dimensionless, while the length dimension of the Rindler acceleration $a$ is minus one.

\subsection{Boundary conditions}\label{se:2.4}

The crucial role of boundary conditions in field theories and particularly in gravitational theories is well appreciated by now. In three spacetime dimensions the asymptotic AdS boundary conditions by Brown and Henneaux \cite{Brown:1986nw} provided an important precursor of AdS$_{3}$/CFT$_{2}$. Since then, these boundary conditions were modified (see e.g.~\cite{Compere:2013bya,Perez:2016vqo,Troessaert:2013fma,Afshar:2016wfy,Grumiller:2016pqb}) and generalized (see e.g.~\cite{Henneaux:2002wm,Henneaux:2004zi,Grumiller:2008es,Henneaux:2009pw,Oliva:2009ip,Barnich:2013sxa,Bunster:2014cna,Perez:2015jxn}) in numerous ways.

Let us now describe the boundary conditions proposed in \cite{Afshar:2016wfy}. The gauge fields compatible with these boundary conditions can be written as 
\begin{equation}
A^{\pm}=b_{\pm}^{-1}\big(\extd+\frak{a}^{\pm}\big)b_{\pm}\label{eq:r7}
\end{equation}
where the gauge group elements $b_{\pm}$ depend on the radial coordinate. In \cite{Afshar:2016wfy}, the remaining analysis was carried out in Eddington--Finkelstein coordinates. Hereafter, we prefer to continue with the development in Gaussian normal coordinates, since they provide certain advantages once dealing with the metric formulation (see section \ref{MetricFormulationAdS}).

A suitable choice of $b_{\pm}$ is then given by 
\begin{equation}
b_{\pm}=\exp\left(\pm\frac{r}{2\ell}\left(L_{1}-L_{-1}\right)\right)\label{r7GNC}
\end{equation}
so that the auxiliary connections $\frak{a}^{\pm}$ can be expressed through 
\begin{equation}
\frak{a}^{\pm}=L_{0}\left(\pm{\cal J}^{\pm}\text{\,}\extd\varphi+\zeta^{\pm}\text{\,}\extd t\right)\label{eq:r21}
\end{equation}
which depend only on time and the angular coordinate. Here we have
used the following definitions: 
\begin{equation}
{\cal J}^{\pm}=\gamma\ell^{-1}\pm\omega\text{\quad\quad}\zeta^{\pm}=-a\pm\Omega\ell^{-1}\,.\label{lalapetz}
\end{equation}
The field equations then imply the vanishing of the field strength
\begin{equation}
{\cal F}=\extd{\cal A}+{\cal A}\wedge{\cal A}=0\label{gaugeflat}
\end{equation}
which exactly hold provided 
\begin{equation}
\dot{{\cal J}}^{\pm}=\pm\zeta^{\pm\prime}\label{whatever}
\end{equation}
in agreement with \eqref{eq:r6}.

As explained in \cite{Henneaux:2013dra,Bunster:2014mua}, the components of the gauge fields along time turn out to be Lagrange multipliers, so that $\zeta^{\pm}$ can be naturally interpreted as chemical potentials and assumed to be fixed at the boundary. The remaining functions ${\cal J}^{\pm}$ then correspond to the dynamical fields.

Having formulated our boundary conditions in terms of the diagonal Chern--Simons connections \eqref{eq:r21}, it becomes very simple to check the regularity of the fields along a contractible cycle in the Euclidean continuation. Since the Lagrange multipliers are switched on, the range of the coordinates can be fixed so that the torus possesses a trivial modular parameter, i.e., $0\leq\varphi<2\pi$, $0\leq\tau<\beta=T^{-1}$, where $T$ stands for the Hawking temperature. Noteworthy, requiring the holonomy along the Euclidean time cycle to be trivial does not impose any restriction on the state-dependent functions ${\cal J}^{\pm}$. This regularity condition instead tells us that the (now complex) chemical potentials are constrained to be
\begin{equation}
\zeta^\pm\equiv-2\pi/\beta_\pm=-a\pm i\Omega\ell^{-1}=-2\pi/\beta\label{eq:cp}
\end{equation}
where $2/T=\beta_++\beta_-$ and $2\Omega/T=\beta_+-\beta_-$.

From a geometrical point of view this means that all the solutions satisfying the boundary conditions have a regular horizon, regardless of the value of ${\cal J}^{\pm}$, as long as $a/\left(2\pi\right)$ is identified with the Unruh temperature and $\Omega=0$. 

\subsection{Canonical charges and their algebra}\label{se:2.3}

The canonical charges associated with the theory defined by our boundary conditions \eqref{eq:r7}-\eqref{eq:r21} turn out to be finite and conserved in time, given by \cite{Afshar:2016wfy} 
\begin{equation}
{\cal Q}^{\pm}[\eta^{\pm}]=\mp\frac{k}{4\pi}\oint\extd\varphi\text{\,}\eta{}^{\pm}\text{\,}{\cal J}^{\pm}\,.\label{eq:Qmn}
\end{equation}

The algebra of the global charges captures all boundary condition preserving transformations 
\begin{equation}
\delta_{\epsilon^{\pm}}\frak{a}^{\pm}=\extd\epsilon^{\pm}+\left[\frak{a}^{\pm},\,\epsilon^{\pm}\right]\label{eq:bcpt}
\end{equation}
modulo trivial gauge transformations, where 
\begin{equation}
\epsilon^{\pm}=\eta^{\pm}L_{0}\label{eq:epsilon}
\end{equation}
and 
\begin{equation}
\delta{\cal J}_{\pm}=\pm\eta_{\pm}^{\prime}\label{eq:deltaJmn}
\end{equation}
with $\dot{\eta}=0$.

Expanding in Fourier modes 
\begin{equation}
J_{n}^{\pm}=\frac{k}{4\pi}\int\extd\varphi\,e^{\pm in\varphi}{\cal J}^{\pm}\label{eq:fourier}
\end{equation}
we found that their commutators are given by\footnote{%
Poisson brackets are replaced by commutators according to $i\left\{ ,\right\} \rightarrow\left[,\right]$.}
\begin{equation}
\left[J_{n}^{\pm},\,J_{m}^{\pm}\right]=\tfrac{1}{2}kn\,\delta_{n+m,\text{\,}0}\text{\quad\quad}\left[J_{n}^{+},\,J_{m}^{-}\right]=0\label{eq:NHSA}
\end{equation}
which consist of two $\hat{u}(1)$ current algebras with the same levels ($k/2$).

Linearly combining the generators as $P_{0}=J_{0}^{+}+J_{0}^{-}$, $P_{n}=\tfrac{i}{kn}\,(J_{-n}^{+}+J_{n}^{-})$ if $n\neq0$, $X_{n}=J_{n}^{+}-J_{-n}^{-}$, the algebra \eqref{eq:NHSA} reads 
\begin{align}
\left[X_{n},\,X_{m}\right] & =\left[P_{n},\,P_{m}\right]=\left[X_{0},\,P_{n}\right]=\left[P_{0},\,X_{n}\right]=0\label{eq:heisenberg1}\\
\left[X_{n},\,P_{m}\right] & =i\delta_{n,m}\quad\textrm{if}\;n\neq0\label{eq:heisenberg}
\end{align}
which corresponds to the commutation relations for Casimir--Darboux coordinates, where $X_{0}$, $P_{0}$ stand for the Casimirs and the remaining $X_{n},\,P_{n}$ form canonical pairs. Note that \eqref{eq:heisenberg} is the Heisenberg algebra.

It is worth highlighting that the global charges \eqref{eq:Qmn} are manifestly independent of the radial coordinate, and therefore the analysis holds for an arbitrary fixed value of it, regardless whether the boundary is chosen to be near the horizon or at infinity.

\subsection{Soft hair descendants}\label{se:soft}

Some consequences of these results were discussed in \cite{Afshar:2016wfy}. A striking one is the existence of ``soft hair'' excitations
\eq{
|\psi_s(\{n_i^\pm\})\rangle \propto \prod_{n_i^\pm>0} J_{-n_i^+}^+ J_{-n_i^-}^- |\psi\rangle
}{eq:soft}
of some arbitrary state $|\psi\rangle$, which can e.g.~be a black hole state. Due to the facts that the Hamiltonian is proportional to $P_0 = J_0^+ + J_0^-$ and that $P_0$ is a Casimir operator, all soft hair descendants $|\psi_s(\{n_i^\pm\})\rangle$ have the same energy as the original state $|\psi\rangle$, for any set of positive integers $\{n_i^\pm\}$. In accordance with the nomenclature introduced by Hawking, Perry and Strominger \cite{Hawking:2016msc} we call the zero energy excitations generated by raising operators $J_{-n_i^\pm}^\pm$ ``soft hair''.

We shall generalize this discussion to the case of locally flat spacetimes in section \ref{se:3}. Before doing this we recover the key results above in the metric formulation, spelled out in section \ref{MetricFormulationAdS}.

\section{Metric formulation}\label{MetricFormulationAdS}

We show now that the results reviewed in the previous section can also be directly obtained in the metric formulation. This is hardly surprising, as the Chern--Simons formulation \eqref{eq:r3} is classically equivalent to the metric formulation. However, it is still useful to perform this exercise in order to get some physical intuition into the meaning of our boundary conditions from a purely metric perspective.

Indeed, since we have already shown that the analysis does not depend on the precise choice of the fixed value of the radial coordinate where the boundary is located, for simplicity in section \ref{se:III.1} we first proceed within the phase space defined through the family of metrics defined in Eq.~\eqref{eq:tba}; where $a$ and $\Omega$ (or equivalently $\zeta^{\pm}$) are assumed to be arbitrary fixed functions of $t$, $\varphi$, without functional variation. Nonetheless, the results are also explicitly carried out 
for relaxed boundary conditions that allow sub-subleading fluctuations either near the horizon or in the asymptotic region in sections \ref{se:NH} and \ref{se:AS}, respectively.

\subsection{Asymptotic Killing vectors and charges}\label{se:III.1}

The asymptotic Killing vectors can be seen to correspond to the diffeomorphisms that maintain the form of the metric \eqref{eq:tba} within the same family. Hence, we look for diffeomorphisms spanned by $\xi=\xi^{\mu}\partial_{\mu}$, whose action on the spacetime metric coincides with its functional variation, i.e., 
\begin{equation}
\delta_{\xi}g_{\mu\nu}={\cal L}_{\xi}g_{\mu\nu}\;.\label{eq:deltagmn}
\end{equation}
Taking into account the compatibility of these symmetries with the time evolution of the dynamical fields, given by Eq.~\eqref{whatever}, the relation in \eqref{eq:deltagmn} then implies that the components of $\xi$ are given by 
\begin{subequations}
\label{eq:xi-mu}
\begin{align}
\xi^{t} & =\frac{\eta^{+}\mathcal{J}^{-}+\eta^{-}\mathcal{J}^{+}}{\zeta^{+}\mathcal{J}^{-}+\zeta^{-}\mathcal{J}^{+}} \\
\xi^{\varphi} & =\frac{\eta^{+}\zeta^{-}-\eta^{-}\zeta^{+}}{\zeta^{+}\mathcal{J}^{-}+\zeta^{-}\mathcal{J}^{+}}\label{eq:MinisuperAKV-AdS} \\
\xi^{r} & =0
\end{align}
\end{subequations}
where $\eta^{\pm}$ stand for arbitrary functions of $\varphi$, while the transformation law of the dynamical fields ${\cal J}^{\pm}$ is found to precisely agree with \eqref{eq:deltaJmn}. 

Note that the result in \eqref{eq:xi-mu} naturally agrees with the one previously derived in terms of gauge fields in section \ref{se:2.3}. Indeed, diffeomorphisms acting on gauge-flat connections are equivalent to gauge transformations spanned by Lie-algebra-valued parameters fulfilling (see e.g.~\cite{Witten:1988hc})
\begin{equation}
\epsilon^{\pm}=\frak{a}_{\mu}^{\pm}\xi^{\mu}\,,\label{eq:AKVs4}
\end{equation}
which is certainly so for $\epsilon^{\pm}$ and $\frak{a}_{\mu}^{\pm}$ given by \eqref{eq:epsilon} and \eqref{eq:r21}, respectively, provided \eqref{eq:xi-mu} holds.

Following the Regge--Teitelboim approach \cite{Regge:1974zd}, the variation of the canonical generators associated to the symmetries spanned by $\xi$, are given by the following surface integrals 
\begin{align}
\delta Q\left[\xi\right] & =\int \extd S_{l}\Big[G^{ijkl}\left(\varepsilon^{\perp}\nabla_{k}\delta g_{ij}-\nabla_{k}\varepsilon^{\perp}\delta g_{ij}\right)\nonumber \\
 & +2\varepsilon^{j}\delta\left(g_{jk}\pi^{kl}\right)-\varepsilon^{l}\pi^{jk}\delta g_{jk}\Big]\,,\label{eq:deltaQ-H}
\end{align}
where $G^{ijkl}=\frac{1}{32\pi G}\,g^{1/2}\left(g^{ik}g^{jl}+g^{il}g^{jk}-2g^{ij}g^{kl}\right)$, and 
\[
\varepsilon^{\perp}=N^{\perp}\xi^{t}\qquad \varepsilon^{i}=\xi^{i}+N^{i}\xi^{t}\,.
\]
These surface integrals can be directly evaluated in terms of the metric in \eqref{eq:tba}, and the symmetries spanned by \eqref{eq:xi-mu}, so that they reduce to
\begin{equation}
\delta Q\left[\eta^{+},\eta^{-}\right]=\frac{\ell}{16\pi G}\int\extd\varphi\;\left(\eta^{+}\,\delta{\cal J}^{+}+\eta^{-}\,\delta{\cal J}^{-}\right)\,,
\end{equation}
which readily integrate as 
\begin{equation}
Q\left[\eta^{+},\eta^{-}\right]={\cal Q}^{+}\left[\eta^{+}\right]-{\cal Q}^{-}\left[\eta^{-}\right]\,,\label{eq:Q-Hamiltonian}
\end{equation}
with ${\cal Q}^{\pm}\left[\eta^{\pm}\right]$ given by \eqref{eq:Qmn}. It is then clear that their Poisson bracket algebra is given by two copies of the affine $\hat{u}\left(1\right)$ currents, coinciding with \eqref{eq:NHSA}.

Note that, as expected, the global charges obtained in the metric formalism do not depend on the radial coordinate, which implies that the same results have to be recovered from an asymptotic analysis performed for a wider class of metrics, either in the near horizon region or close to the asymptotic boundary.

\subsection{Near horizon behaviour}\label{se:NH}

One of the advantages of dealing with Gaussian normal coordinates is that the analysis of the spacetime structure, despite of being performed in the horizon neighborhood, can be suitably carried out following the canonical approach, as it is the case in \cite{Regge:1974zd} for the asymptotic region. The leading terms of the metric around the horizon, located at $r=0$, are then specified according to 
\begin{subequations}
\label{eq:gmn-Asymp}
\begin{align}
g_{tt} & =\Omega^{2}+\left(\Omega^{2}-\ell^{2}a^{2}\right)\frac{r^{2}}{\ell^{2}}
+{\cal O}\left(r^{3}\right) \\
g_{\varphi\varphi} & =\gamma^{2}+\left(\gamma^{2}-\ell^{2}\omega^{2}\right)\frac{r^{2}}{\ell^{2}}
+{\cal O}\left(r^{3}\right) \\
g_{t\varphi} & =\gamma\Omega+\left(\gamma\Omega+a\omega\ell^{2}\right)\frac{r^{2}}{\ell^{2}}
+{\cal O}\left(r^{3}\right) \\
g_{rr} & =1+{\cal O}\left(r^2\right)\\
g_{rt} & ={\cal O}\left(r^2\right) \\
g_{r\varphi} & ={\cal O}\left(r^2\right)\,. 
\end{align}
\end{subequations}

The near horizon symmetries are then found through solving Eq.~\eqref{eq:deltagmn} up to the subleading orders in \eqref{eq:gmn-Asymp}, which implies that the asymptotic Killing vectors are given by \eqref{eq:xi-mu} modulo subleading corrections of ${\cal O}\left(r^{3}\right)$, i.e.,
\begin{subequations}
\label{eq:xi_mu-Asympt}
\begin{align}
\xi^{t} & =\frac{\eta^{+}\mathcal{J}^{-}+\eta^{-}\mathcal{J}^{+}}{\zeta^{+}\mathcal{J}^{-}+\zeta^{-}\mathcal{J}^{+}}+{\cal O}\left(r^{3}\right) \\
\xi^{\varphi} & =\frac{\eta^{+}\zeta^{-}-\eta^{-}\zeta^{+}}{\zeta^{+}\mathcal{J}^{-}+\zeta^{-}\mathcal{J}^{+}}+{\cal O}\left(r^{3}\right)\\
\xi^{r} & ={\cal O}\left(r^3\right)
\end{align}
\end{subequations}
provided that the fields ${\cal J}^{\pm}$ transform as in \eqref{eq:deltaJmn}.

The global charges associated to these near horizon symmetries can then be readily obtained from \eqref{eq:deltaQ-H}, and they are found to agree with \eqref{eq:Q-Hamiltonian}. Hence, the canonical realization of the NHSA is described by two $\hat{u}\left(1\right)$ currents, given by \eqref{eq:NHSA}.

\subsection{Asymptotic behaviour}\label{se:AS}

Analogously, when the radial coordinate approaches infinity ($r\rightarrow\infty$) it is convenient to make the change $\frac{r}{\ell}\rightarrow\text{log}\left(\frac{r}{\ell}\right)$, so that the asymptotic behaviour of the spacetime metric reads 
\begin{subequations}
\label{eq:gmn-Asymptic}
\begin{align}
g_{tt} & =\left(\Omega^{2}-\ell^{2}a^{2}\right)\frac{r^{2}}{4\ell^{2}}+\tfrac{1}{2}\left(\Omega^{2}+\ell^{2}a^{2}\right)+{\cal O}\big(\tfrac1r\big) \\
g_{\varphi\varphi} & =\left(\gamma^{2}-\ell^{2}\omega^{2}\right)\frac{r^{2}}{4\ell^{2}}+\tfrac{1}{2}\left(\gamma^{2}+\ell^{2}\omega^{2}\right)+{\cal O}\big(\tfrac1r\big)\\
g_{t\varphi}&=\left(\gamma\Omega+a\omega\ell^{2}\right)\frac{r^{2}}{4\ell^{2}}+\tfrac{1}{2}\left(\gamma\Omega-a\omega\ell^{2}\right)+{\cal O}\big(\tfrac1r\big) \displaybreak[1] \\
g_{rr} & =\frac{\ell^{2}}{r^{2}}+{\cal O}\big(\tfrac1{r^3}\big)\\ 
g_{rt} & ={\cal O}\big(\tfrac1r\big)\\
g_{r\varphi} & ={\cal O}\big(\tfrac1r\big)\,. 
\end{align}
\end{subequations}
The asymptotic symmetries are then found to be spanned by 
\begin{subequations}
\label{eq:xi_mu-Asymptotic}
\begin{align}
\xi^{t} & =\frac{\eta^{+}\mathcal{J}^{-}+\eta^{-}\mathcal{J}^{+}}{\zeta^{+}\mathcal{J}^{-}+\zeta^{-}\mathcal{J}^{+}}+{\cal O}\big(\tfrac1{r^3}\big) \\
\xi^{\varphi} & =\frac{\eta^{+}\zeta^{-}-\eta^{-}\zeta^{+}}{\zeta^{+}\mathcal{J}^{-}+\zeta^{-}\mathcal{J}^{+}}+{\cal O}\big(\tfrac1{r^3}\big)\\
\xi^{r} & ={\cal O}\big(\tfrac1{r^2}\big)
\end{align}
\end{subequations}
provided that the dynamical fields transform according to \eqref{eq:deltaJmn}.

It is then simple to verify that the corresponding canonical generators reduce to \eqref{eq:Q-Hamiltonian}, which means that the asymptotic symmetry algebra coincides with the NHSA, both spanned by the affine $\hat{u}\left(1\right)$ currents with the same levels as in \eqref{eq:NHSA}.

\section{Flat space generalization}\label{se:3}

In this section we extend the results in \cite{Afshar:2016wfy} for the case of vanishing cosmological constant. In section \ref{se:3.1} we introduce soft hairy cosmology line-elements. In section \ref{se:3.2} we present near horizon boundary conditions in the Chern--Simons formulation. In section \ref{se:3.3} we derive the associated canonical charges, whose symmetry algebra we study in section \ref{se:3.4}. In section \ref{se:3.5} we construct flat space soft hair. In section \ref{se:3.6} we compare with standard asymptotically flat results and recover BMS$_3$ as a composite algebra through a Sugawara-like construction. In section \ref{se:3.7} we summarize the metric formulation in flat space. In section \ref{se:td} we solve the regularity conditions of relevance for thermodynamics of cosmological spacetimes endowed with soft hair, and explore some aspects of their entropy.

\subsection{Soft hairy cosmologies}\label{se:3.1}

The class of locally flat spacetimes endowed with non-extremal horizons we look for can be readily obtained from the limit of large AdS radius, $\ell\to\infty$, of the line element of the soft hairy black holes in \eqref{eq:tba}. The metric is then found to be given by 
\begin{align}
\extd s^{2} & =\extd r^{2}+(\Omega^{2}-a^{2}r^{2})\,\extd t^{2}+2(\Omega\gamma+a\omega r^{2})\,\extd t\extd\varphi\nonumber \\
 & +(\gamma^{2}-\omega^{2}r^{2})\,\extd\varphi^{2}\,.\label{eq:Flat-Cosmology-1}
\end{align}
In this patch of coordinates, the geometry generically describes the inner region of a class of spacetimes with a cosmological horizon located at $r=0$, and a chronological singularity at $r_{s}^{2}=\gamma^{2}/\omega^{2}$, so that the region that corresponds to $r>r_{s}$ can be excised in order to avoid closed timelike curves. The Lagrange multipliers $a$ and $\Omega$ turn out to be arbitrary fixed functions of $t$, $\varphi$, and as shown below, the functions $\omega$ and $\gamma$ are related to the global charges so that they cannot be gauged away. Local flatness implies that time evolution of the dynamical fields is determined
by the spatial derivative of the Lagrange multipliers, precisely as in Eq.~\eqref{eq:r6}. It is then worth emphasizing that the reduced phase space obtained from requiring the class of metrics in \eqref{eq:tba} to be of negative constant curvature, exactly coincides with the one obtained from demanding \eqref{eq:Flat-Cosmology-1} to be locally flat. 
Note that generic configurations are not spherically symmetric, possessing ripples that cannot be gauged away because they are characterized by global soft hair charges. Hence, these geometries are analogues of black flowers in AdS$_{3}$ discussed in section \ref{se:2.1} and therefore we shall refer to them as ``cosmological flowers''.

In order to cover a wider spacetime region that includes the cosmological horizon, it is useful to express the spacetime metric \eqref{eq:Flat-Cosmology-1} in ingoing Eddington--Finkelstein coordinates, according to
\begin{equation}
r^{2}=\frac{2}{a}\rho\qquad t=v-\frac{1}{2a}\text{log}\left(\rho\right)\label{eq:ChangeofCoordFlat}
\end{equation}
so that in a co-rotating frame ($\Omega=0$) and for constant $a$, the line element reads 
\begin{multline}
\extd s^{2} =-2a\rho\,\extd v^{2}+2\extd v\extd\rho + 4\omega\rho\extd v\extd\varphi \label{eq:Null-Metric-Flat-1}\\
 -\frac{2\omega}{a}\,\extd\varphi\extd\rho +\Big(\gamma^{2}-\frac{2\omega^{2}}{a}\rho\Big)\extd\varphi^{2}\,. 
\end{multline}
The radial coordinate now ranges as $-\infty<\rho<\rho_{s}=\frac{a}{2}r_{s}^{2}$. It is simple to verify that the change of coordinates \eqref{eq:ChangeofCoordFlat} as well as the metric in \eqref{eq:Null-Metric-Flat-1} are directly recovered in the $\ell\rightarrow\infty$ limit from \eqref{eq:ChangeofCoordAdS} and \eqref{eq:Null-AdS-metric}, respectively. 

Remarkably, the spectrum of solutions in \eqref{eq:Null-Metric-Flat-1} is regular for arbitrary functions $\omega\left(\varphi\right)$ and $\gamma\left(\varphi\right)$ (see section \ref{se:td}). Indeed, this can be directly seen for the case of constant $\omega$ and $\gamma$, which for $\omega\neq0$ describes the class of stationary spherically symmetric cosmological spacetimes discussed in \cite{Cornalba:2002fi,Cornalba:2003kd}, while for $\omega=0$ our static solution does not become singular, but it is instead given by the product of Rindler spacetime times a circle of radius $\gamma$. 

In the next section we construct a suitable set of boundary conditions that accommodates the family of locally flat solutions described here, even in the case of a generic choice of $\Omega$ and $a$.

\subsection{Chern--Simons formulation}\label{se:3.2}

We use again the Chern--Simons action \eqref{eq:r3}, now with a dimensionful Chern--Simons level $k=1/(4G)$. In the absence of a cosmological constant the connection ${\cal A}$ can be decomposed into components as 
\begin{equation}
{\cal A}=A_{L}^{n}L_{n}+A_{M}^{n}M_{n}\;,\label{eq:r28}
\end{equation}
with respect to the isl$(2)$ generators obeying the algebra
\begin{align}
[L_{n},\,L_{m}] & =(n-m)\,L_{n+m}\\{}
[L_{n},\,M_{m}] & =(n-m)\,M_{n+m}\\{}
[M_{n},\,M_{m}] & =0
\end{align}
where $n,m=0,\pm1$. From a geometric perspective the components $A_{M}$ correspond to the dreibein and the components $A_{L}$ to the (dualized) spin-connection, so that following the conventions in \cite{Barnich:2013yka,Afshar:2013bla,Afshar:2015wjm}, the line element reads 
\begin{equation}
\extd s^{2}=g_{\mu\nu}\,\extd x^{\mu}\extd x^{\nu}=-4A_{M}^{+}A_{M}^{-}+\left(A_{M}^{0}\right)^{2}\,,\label{eq:r8sm}
\end{equation}
and the nonvanishing components of the invariant bilinear form are given by
\begin{equation}
\left\langle L_{1},\,M_{-1}\right\rangle =\left\langle L_{-1},\,M_{1}\right\rangle =-2\text{\quad\quad}\left\langle L_{0},\,M_{0}\right\rangle =1\;.\label{eq:r32sm}
\end{equation}
Here, the quantities $v, \rho, \gamma, G, A_{M}$ have length dimensions one, $\varphi, \omega, \Omega, {\cal A}, L_{n}, A_{L}$ are dimensionless, and $a, k, M_{n}$ have length dimensions minus one.

The boundary conditions we propose are realized for connections of the form 
\begin{equation}
{\cal A}=b^{-1}\,\left(\extd+\frak{a}\right)\,b,\label{eq:r7sm}
\end{equation}
with the ISL$(2)$ group element 
\begin{equation}
b=\exp\Big(-\frac{1}{a}M_{1}\Big)\exp\Big(\frac{\rho}{2}M_{-1}\Big)\label{eq:r9sm}
\end{equation}
and the auxiliary connection
\begin{equation}
\frak{a}=\left(-a\,\extd v+\omega\,\extd\varphi\right)L_{0}+\left(\Omega\,\extd v+\gamma\,\extd\varphi\right)M_{0}\,.\label{eq:r21sm}
\end{equation}
Following \cite{Henneaux:2013dra}, the arbitrary functions of $v$, $\varphi$, given by $\omega$ and $\gamma$ are identified with the dynamical fields, while $\Omega$ and $a$ correspond to the Lagrange multipliers which can be assumed to be arbitrary functions of $v$, $\varphi$ that are held fixed at the boundary without variation ($\delta\Omega=\delta a=0$). The vanishing of the field strength implies (prime denotes again $\partial_\vp$ and dot $\partial_v$) 
\begin{equation}
\dot\gamma=\Omega^{\prime}\text{\quad\quad}\dot\omega=-a^{\prime}\text{\,}.\label{eq:r10sm}
\end{equation}
Note that in the particular case of $\Omega=0$ and $a$ constant, from Eqs.~\eqref{eq:r7sm} and \eqref{eq:r8sm}, one recovers the line element in \eqref{eq:Null-Metric-Flat-1}.

\subsection{Canonical charges}\label{se:3.3}

In the Hamiltonian approach \cite{Regge:1974zd}, the surface integrals associated to the variation of the canonical generators are found to be given by
\begin{equation}
\delta Q\left[\epsilon\right]=-\frac{k}{2\pi}\int\extd\varphi\left\langle \epsilon\,\delta\frak{a}_{\varphi}\right\rangle \label{eq:rc3sm}
\end{equation}
where 
\begin{equation}
\epsilon=\epsilon_{L}^{n}L_{n}+\epsilon_{M}^{n}M_{n}\;,\label{eq:r29}
\end{equation}
is an arbitrary Lie-algebra-valued parameter. Taking into account the expression for the bilinear form in \eqref{eq:r32sm}, as well as the asymptotic form of the auxiliary gauge field in \eqref{eq:r21sm}, the surface integrals in \eqref{eq:rc3sm} readily evaluate as 
\begin{equation}
\delta Q=-\frac{k}{2\pi}\int\extd\varphi\left(\epsilon_{L}^{0}\delta\gamma+\epsilon_{M}^{0}\delta\omega\right)\text{\,}.\label{eq:r25sm}
\end{equation}
These surface integrals turn out to be nontrivial for the asymptotic symmetries, which correspond to the ones that maintain the asymptotic form of $\frak{a}$ in \eqref{eq:r21sm}, i.e., those that fulfill
\begin{equation}
\delta_{\epsilon}\frak{a}=\extd\epsilon+\left[\frak{a},\epsilon\right]\label{eq:r27}
\end{equation}
up to trivial gauge transformations. The asymptotic symmetries are found to be spanned by
\begin{align}
\epsilon_{L}^{0} & =\eta_L & \epsilon_{M}^{0} & =\eta_M\label{eq:bcptsm}
\end{align}
where $\eta_L$ and $\eta_M$ are arbitrary functions of $\vp$. The free functions $\epsilon_{L,M}^{\pm}$ do not appear in the canonical charges nor in the transformation rules 
\begin{equation}
\delta\omega=\eta_L^{\prime}\text{\quad\quad}\delta\gamma=\eta_M^{\prime}\label{eq:trans}
\end{equation}
and hence they generate trivial gauge transformations. 

The canonical generators are then given by 
\begin{equation}
Q\left[\eta_L,\eta_M\right]=-\frac{k}{2\pi}\,\int\extd\varphi\left(\eta_L\,\gamma+\eta_M\,\omega\right)\,,\label{eq:rc5sm}
\end{equation}
being manifestly finite and conserved in advanced time.

\subsection{Symmetry algebra}\label{se:3.4}

Having established the canonical charges we determine now their symmetry algebra. Since their Poisson brackets fulfill $\{Q(\xi_{1}),Q(\xi_{2})\}=\delta_{\xi_{2}}Q(\xi_{1})$, it is straightforward to obtain the algebra of the canonical generators from the transformation law in \eqref{eq:trans}, which is found to coincide exactly with the one in the case of negative cosmological constant \eqref{eq:NHSA}. Indeed, expanding in Fourier modes 
\begin{equation}
J_{n}=\frac{k}{2\pi}\,\int\extd\varphi\text{\,}e^{in\varphi}\omega\text{\quad\quad}K_{n}=\frac{k}{2\pi}\,\int\extd\varphi\text{\,}e^{in\varphi}\gamma\;,\label{eq:rc6sm}
\end{equation}
the commutators fulfill 
\begin{subequations}
\label{eq:NHSAsm}
\begin{align}
[J_{n},\,J_{m}] & =[K_{n},\,K_{m}]=0 \\{}
[J_{n},\,K_{m}] & =k\,n\,\delta_{n+m,\,0}
\end{align}
\end{subequations}
so that if one changes the basis according to 
\begin{equation}
P_{0}=K_{0}\text{\,},\text{\quad}P_{n}=-i\,\frac{K_{-n}}{kn}\;\text{if }\;n\neq0\text{\,},\text{\quad}X_{n}=J_{n}\,,\label{eq:rc7sm}
\end{equation}
the canonical commutators in Casimir--Darboux coordinates in \eqref{eq:heisenberg1}, \eqref{eq:heisenberg} are recovered, precisely as in the asymptotically AdS case. As mentioned in Section \ref{se:2.3}, the asymptotic symmetries are then described by two affine $\hat{u}(1)$ current algebras with the same levels as in \eqref{eq:NHSA}.

\subsection{Flat space soft hair}\label{se:3.5}

We generalize now the discussion of section \ref{se:soft} to flat space. For simplicity let us assume a co-rotating frame ($\Omega=0$) and constant $a$. The surface integral associated to the generator of translations in advanced time is given by
\begin{equation}
H=Q[\epsilon|_{\partial_{v}}]\;,
\end{equation}
where $\epsilon|_{\partial_{v}}$ stands for the Lie-algebra-valued parameter associated to $\partial_{v}$, given by (see e.g. \cite{Witten:1988hc})
\begin{equation}
\left.\epsilon\right|_{\partial_{v}}=a_{v}=-aL_{0}\,.\label{eq:rc8sm}
\end{equation}
The Hamiltonian is then given by $H=-a\,P_{0}$, being identical to the result on AdS$_{3}$ \cite{Afshar:2016wfy} and again commutes with all canonical coordinates $X_{n},\,P_{n}$. 

We consider now all descendants $|\psi_s(\{n_i^\pm\})\rangle $ 
\begin{equation}
|\psi_s(\{n_i^\pm\})\rangle \propto \prod_{n_i^\pm>0} J_{-n_i^+} K_{-n_i^-} |\psi\rangle\label{eq:rc40sm}
\end{equation}
of some state $|\psi\rangle$ (e.g.~the vacuum state).\footnote{%
Even though the condition $n_i^\pm > 0$ suggests that the states $|\psi\rangle$ are highest weight descendants of a given vacuum state our main result that all states have the same energy eigenvalue also holds for other representations discussed in the context of flat space holography such as e.g.~induced representations \cite{Barnich:2014kra, Campoleoni:2015qrh, Campoleoni:2016vsh, Oblak:2016eij}. We thank Glenn Barnich, Blaza Oblak and Max Riegler for discussions on induced representations.
} Since $H$ commutes with all of the generators, one finds again that the energy of any soft hair descendant $|\psi_s(\{n_i^\pm\})\rangle$ coincides with the one of the original state $|\psi\rangle$. 

Hence, not only on AdS$_{3}$, but also in the flat space case, all descendants of some state turn out to possess the same energy as that state, so that this kind of excitations can be regarded again as ``soft hair'' in the sense of \cite{Hawking:2016msc}.

\subsection{Emergence of composite BMS$_{3}$ generators}\label{se:3.6}\label{se:Linking Near-Asympt}

Here we show how the BMS$_{3}$ algebra, with the precise central extension found in \cite{Barnich:2006av}, naturally emerges from composite operators of the affine $\hat{u}\left(1\right)$ currents \eqref{eq:NHSAsm} in a unique way, through an analogue of the twisted Sugawara construction. 

In order to do that, it is useful to compare the new set of boundary conditions described by \eqref{eq:r7sm}, \eqref{eq:r9sm}, \eqref{eq:r21sm} with the standard ones in \cite{Barnich:2006av}. This task can be successfully achieved only once the standard set of boundary conditions is enhanced so as to accommodate a generic choice of Lagrange multipliers as in \cite{Gary:2014ppa, Matulich:2014hea}, which here are allowed to depend on the dynamical fields. The comparison can then be explicitly carried out provided that the asymptotic behaviour in Eqs.~\eqref{eq:r7sm}, \eqref{eq:r9sm}, \eqref{eq:r21sm} is expressed in terms of the same gauge choice as in \cite{Gary:2014ppa, Matulich:2014hea}. For a generic choice of Lagrange multipliers, which are not yet specified, the asymptotic form of the connection is then given by
\begin{align}
\hat{A} & =\hat{b}^{-1}(\extd+\hat{\frak{a}})\hat{b}\qquad\hat{b}=\exp\big(\tfrac{\rho}{2}M_{-1}\big)\label{eq:Ahat-Flat}\displaybreak[1] \\
\hat{\frak{a}}_{\vp} & =L_{1}-\frac{\cM}{2}L_{-1}-\frac{\cN}{2}M_{-1}\label{eq:ahat-phi}\displaybreak[1] \\
\hat{\frak{a}}_{v} & =\mu_{M}M_{1}+\mu_LL_{1}-\mu_{M}^{\prime}M_{0}-\mu_L'L_{0}+\tfrac{1}{2}\big(\mu_{M}^{\prime\prime}\nonumber \\
 & -\cM\mu_{M}-\cN\mu_L\big)M_{-1}+\tfrac{1}{2}\big(\mu_L''-\cM\mu_L\big)L_{-1}\,,\label{eq:ahat-u}
\end{align}
where ${\cal M}$, $\cN$, $\mu_{M}$, $\mu_L$ are arbitrary functions of $v$, $\varphi$.

We then look for a permissible gauge transformation\footnote{%
In the sense of \cite{Bunster:2014mua}, a gauge transformation is dubbed permissible if it does not interfere with the asymptotic
symmetry algebra.} 
spanned by a group element $g$, that relates the auxiliary connections $\frak{a}$ in \eqref{eq:r21sm} with the auxiliary gauge field $\hat{\frak{a}}$ given by \eqref{eq:ahat-phi}, \eqref{eq:ahat-u}, i.e.,
\begin{equation}
\hat{\frak{a}}=g^{-1}(\extd+\frak{a})g\;.\label{eq:ahatTOa}
\end{equation}
The group element $g$ is found to be given by
\eq{
g = e^{y(v,\vp)M_{1}}e^{x(v,\vp)L_{1}}e^{-\tfrac{1}{2}\gamma M_{-1}}e^{-\tfrac{1}{2}\omega L_{-1}}
}{eq:gFlat}
with
\begin{subequations}
\label{eq:x-yprime}
\begin{align}
x^{\prime}&=1+\omega x & y^{\prime}&=\gamma x+\omega y\\
\partial_{v}x&=\mu_L-ax & \partial_{v}y&=\mu_{M}-ay+x\Omega
\end{align}
\end{subequations}
On-shell consistency of Eqs.~\eqref{eq:x-yprime} then implies
\begin{subequations}
\label{eq: a-Omega-To MuN-MuM}
\begin{align}
\mu_L^{\prime}-\omega\mu_L & =a\\
\mu_{M}^{\prime}-\omega\mu_{M}-\gamma\mu_L & =-\Omega
\end{align}
\end{subequations}
so that $\mu_{M}$ and $\mu_L$ not only depend on the functions $a$ and $\Omega$ that are held fixed at the boundary, but also posses a non-local dependence on the dynamical fields $\gamma$, $\omega$.

The gauge fields $\hat{\frak{a}}$ and $\frak{a}$ are then related through a permissible gauge transformation spanned by $g$ in \eqref{eq:gFlat}, provided
\begin{subequations}
\label{eq:MiuraFlat}
\begin{align}
{\cal M} & =\tfrac{1}{2}\omega^{2}+\omega^{\prime} \\
\cN & =\omega\gamma+\gamma^{\prime}\,.
\end{align}
\end{subequations}

Summarizing, the boundary conditions in \eqref{eq:r7sm}, \eqref{eq:r9sm}, \eqref{eq:r21sm}, once expressed in the gauge choice of \cite{Gary:2014ppa,Matulich:2014hea}, are described by Lagrange multipliers $\mu_{M}$, $\mu_L$ that depend non-locally on the dynamical variables $\gamma$, $\omega$ according to \eqref{eq: a-Omega-To MuN-MuM}, where $a$, $\Omega$ turn out to be fixed at the boundary without variation ($\delta a=\delta\Omega=0$). The functions ${\cal M}$, $\cN$ then depend on the global charges $\gamma$, $\omega$ according to \eqref{eq:MiuraFlat}. 

It is amusing to verify that the field equations that correspond to the local flatness of \eqref{eq:Ahat-Flat}, given by
\begin{subequations}
\begin{align}
\dot{\cal M} & =2{\cal M}\mu_L^{\prime}+{\cal M}^{\prime}\mu_L-\mu_L^{\prime\prime\prime}\\
\dot{\cN} & =2\cN\mu_L^{\prime}+\cN^{\prime}\mu_L+2{\cal M}\mu_{M}^{\prime}+{\cal M}^{\prime}\mu_{M}-\mu_{M}^{\prime\prime\prime}
\end{align}
\end{subequations}
by virtue of our boundary conditions, which in this gauge choice are expressed by \eqref{eq: a-Omega-To MuN-MuM}, \eqref{eq:MiuraFlat}, reduce to $\dot\gamma=\Omega^{\prime}$, $\dot\omega=-a^{\prime}\text{\,}$, in full agreement with \eqref{eq:r10sm}, which were readily obtained in the gauge choice of section \ref{se:3.2}.

It should also be highlighted that \eqref{eq:MiuraFlat} corresponds to a ``flat analogue'' of the twisted Sugawara construction. Indeed, one can verify that the currents ${\cal M}$, $\cN$ actually obey the BMS$_{3}$ algebra with the central extension found in \cite{Barnich:2006av}. This can be seen as follows. According to \eqref{eq:trans}, the transformation law of the dynamical fields under the affine asymptotic symmetries is given by $\delta\omega=\eta_L^{\prime}$, $\delta\gamma=\eta_M^{\prime}$, and by virtue of \eqref{eq: a-Omega-To MuN-MuM}, the relationship between the functions that span the asymptotic symmetries in both gauge choices reads
\begin{subequations}
\label{eq:ParametersFlat-ahat-a}
\begin{align}
\epsilon_{L}^{\prime}-\omega\epsilon_{L} & =-\eta_L \\
\epsilon_{M}^{\prime}-\omega\epsilon_{M}-\gamma\epsilon_{L} & =-\eta_M\;.
\end{align}
\end{subequations}
Hence, the transformation law of ${\cal M}$ and $\cN$ can be directly obtained from \eqref{eq:MiuraFlat}
\begin{subequations}
\label{eq:nolabel}
\begin{align}
\delta{\cal M} & =2{\cal M}\epsilon_{L}^{\prime}+{\cal M}^{\prime}\epsilon_{L}-\epsilon_{L}^{\prime\prime\prime}\;,\\
\delta \cN & =2\cN\epsilon_{L}^{\prime}+\cN^{\prime}\epsilon_{L}+2{\cal M}\epsilon_{M}^{\prime}+{\cal M}^{\prime}\epsilon_{M}-\epsilon_{M}^{\prime\prime\prime}\;,
\end{align}
\end{subequations}
which implies that the currents spanned by ${\cal M}$, $\cN$ fulfill the centrally-extended BMS$_{3}$ algebra.

In other words, expanding in Fourier modes, Eq.~\eqref{eq:MiuraFlat} reads
\begin{subequations}
\label{eq:whatever}
\begin{align}
M_{n} & =\frac{1}{2k}\,\sum_{p\in\mathbb{Z}}J_{n-p}J_{p}+in\,J_{n}\label{eq:Miura-Flat-ModesI}\\
L_{n} & =\frac{1}{k}\,\sum_{p\in\mathbb{Z}}J_{n-p}K_{p}+in\,K_{n}
\label{eq:Miura-Flat-ModesII}
\end{align}
\end{subequations}
so that the generators $M_{n}$, $L_{n}$ fulfill the centrally-extended BMS$_{3}$ algebra
\begin{align}
[L_{n},\,L_{m}] & =(n-m)L_{n+m}\\{}
[L_{n},\,M_{m}] & =(n-m)M_{n+m}+kn^{3}\delta_{n+m,\,0}\\{}
[M_{n},\,M_{m}] & =0\,.
\end{align}
It is then remarkable that the standard asymptotic BMS$_{3}$ algebra, which is obtained from a different set of boundary conditions, that is defined through keeping $\mu_{M}$ and $\mu_L$ to be fixed at the boundary without variation [$\mu_{M}=\bar{\mu}_{M}\left(v,\varphi\right)$, $\mu_{L}=\bar{\mu}_{L}\left(v,\varphi\right)$] \cite{Gary:2014ppa, Matulich:2014hea}, naturally emerges as a composite one in terms of the $\hat{u}\left(1\right)$ currents that correspond to the asymptotic symmetries of our boundary conditions, described by \eqref{eq: a-Omega-To MuN-MuM} and \eqref{eq:MiuraFlat}. 

We would like to emphasize that, although the currents ${\cal M}$, $\cN$ fulfill the BMS$_{3}$ algebra, their corresponding global charges actually generate the affine algebra in \eqref{eq:NHSAsm}. Indeed, by virtue of \eqref{eq:MiuraFlat} and \eqref{eq:ParametersFlat-ahat-a}, the variation of the global charges reads
\begin{align}
\delta Q & =-\frac{k}{2\pi}\,\int\extd\varphi\left(\epsilon_{M}\,\delta{\cal M}+\epsilon_{L}\,\delta\cN\right) \nonumber \\
 & =-\frac{k}{2\pi}\,\int\extd\varphi\left(\eta_L\,\delta\gamma+\eta_M\,\delta\omega\right)\;.
\end{align}
Hence, they manifestly fulfill the $\hat{u}\left(1\right)$ current algebra \eqref{eq:NHSAsm}.

\subsection{Metric formulation}\label{se:3.7}

The family of locally flat soft hairy spacetimes, once written in Gaussian normal coordinates \eqref{eq:Flat-Cosmology-1}, opens up the interesting possibility of performing a fully fledged standard canonical analysis in the inner patch, even for a generic choice of Lagrange multipliers. As in section \ref{se:III.1}, the asymptotic Killing vectors can be obtained from \eqref{eq:deltagmn}, which are found to be given by
\begin{subequations}
\label{eq:NearHorizon-Asym-Flat}
\begin{align}
\xi^{t} & =\frac{\eta_M\omega-\gamma\eta_L}{a\gamma+\Omega\omega} \\
\xi^{\varphi} & =\frac{\eta_M a+\Omega\eta_L}{a\gamma+\Omega\omega} \\
\xi^{r} & =0
\end{align}
\end{subequations}
so that the minisuperspace is preserved provided that the dynamical fields, $\gamma$ and $\omega$, transform precisely as in \eqref{eq:trans}.

It is worth pointing out that the form of the asymptotic Killing vectors in \eqref{eq:NearHorizon-Asym-Flat} agrees with the ones for the soft hairy black holes in \eqref{eq:MinisuperAKV-AdS} provided ${\cal J}^{\pm}=\gamma\pm\omega$, and $\eta_M^{\pm}=-a\pm\Omega$. Therefore, the corresponding canonical generators are directly recovered from \eqref{eq:deltaQ-H}, which are found to precisely agree with \eqref{eq:rc5sm}. Consequently, in terms of ${\cal J}^{\pm}$, their algebra manifestly acquires the form of the $\hat{u}\left(1\right)$ currents in \eqref{eq:NHSA} with levels $k=1/(4G)$.

\subsection{Thermodynamics}\label{se:td}

In order to explore the thermodynamical properties of the cosmological flowers, it is useful to consider the Euclidean continuation of the line element in the inner patch \eqref{eq:Flat-Cosmology-1}, so that the horizon is located at the origin ($r=0$) and the boundary is chosen to be at $r=r_{0}<r_{s}$. Following \cite{Bunster:2014mua}, one takes advantage of the fact that the chemical potentials manifestly appear in the metric, so that the range of the coordinates can be fixed according to the ones of a straight solid torus, being characterized by a trivial modular parameter, i.e., $0\leq\varphi<2\pi$, $0\leq\tau<\beta$, where $T=\beta^{-1}$ is the Hawking temperature. Thus, requiring regularity of the geometry around the cosmological horizon, fixes the temperature and angular chemical potential according to 
\begin{equation}
a^{2}=\frac{4\pi^{2}}{\beta^{2}}\text{\quad\quad}\Omega=0\label{eq:diagholcon}
\end{equation}
in full analogy with the black flowers \eqref{eq:cp}. It is worth emphasizing that regularity again holds independently of the value of the global charges.

Alternatively, regularity of the Euclidean configuration can be directly implemented through requiring the holonomy of the gauge fields along the thermal cycle to be trivial. One of the advantages of this procedure is that it can be carried out directly in terms of the auxiliary connections, and hence it does not depend on the radial coordinate. Therefore, the procedure holds for the inner and the outer patch provided that the orientation is suitably taken into account.

As pointed out in \cite{Matulich:2014hea}, $ISL(2,\mathbb{R})$ does not admit a suitable standard matrix representation from which the Casimir operators, and hence the invariant bilinear form \eqref{eq:r32sm}, could be recovered from the trace of a product of the generators. Therefore, regularity of the Euclidean configuration cannot be straightforwardly implemented through the diagonalization of the holonomy matrix, though it is possible to do so along the lines of \cite{Gary:2014ppa} using a 3+1 dimensional representations of $ISL(2,\mathbb{R})$. Here we implement the regularity conditions following \cite{Matulich:2014hea}:

\paragraph{(i)} One first finds a proper group element of the form $h=\text{exp}\left(\lambda^{n}M_{n}\right)$ that permits to gauge away the temporal components of the connection along $M_{n}$, so that one can consistently set the components of the dreibein along time to vanish ($A_{M\tau}=0$). Thus, the chemical potential of electric type becomes fixed, and generically can be expressed in terms of the magnetic type one and the global charges. 

\paragraph{(ii)} The remaining conditions can then be fulfilled through diagonalizing the holonomy matrix associated to the spin connection ($A_{L\tau}$) along the thermal circle, in the fundamental representation of $SL(2,\mathbb{R})$.

For the auxiliary connection $\frak{a}$ with the gauge choice in \eqref{eq:r21sm} the group element $h$ becomes trivial, so that condition (i) readily implies that $\Omega=0$. Condition (ii) then implies that exp$(-i\beta aL_{0})=-\unity$, so that 
\begin{equation}
a^{2}=\frac{4\pi^{2}}{\beta^{2}}\left(2n+1\right)^{2}\,,\label{eq:a-reg}
\end{equation}
where $n$ is an integer. The branch that is continuously connected to the cosmological flower (without conical surpluses) then corresponds to $n=0$. 

Alternatively, if one considers the auxiliary connection for the gauge choice as in \eqref{eq:ahat-u}, the group element can be chosen to be given by
\begin{equation}
h=\exp\left[-\frac{1}{2\mu_L}\left(2\mu_{M}M_{0}-\mu_{M}^{\prime}M_{-1}\right)\right]\,,
\end{equation}
and thus, condition (i) is fulfilled provided
\begin{equation}
\mu_{M}^{\prime}\mu_L^{\prime}-\mu_{M}\mu_L^{\prime\prime}-\mu_L\left(\mu_{M}^{\prime\prime}-2{\cal M}\mu_{M}-\cN\mu_L\right)=0\,.\label{eq:Holoau-1}
\end{equation}
Condition (ii) then reads exp$(i\beta\hat{\frak{a}}_{Lv})=-\unity$, which is equivalent to
\begin{equation}
\text{tr}\left(\hat{\frak{a}}_{Lv}^{2}\right)=\frac{2\pi^{2}}{\beta^{2}}\left(2n+1\right)^{2}\;,\label{eq:traceau}
\end{equation}
and evaluates as
\begin{equation}
\mu_L^{\prime2}-2\mu_L\left(\mu_L^{\prime\prime}-{\cal M}\mu_L\right)=\left(\frac{2\pi}{\beta}\left(2n+1\right)\right)^{2}\;.\label{eq:Holoau-2}
\end{equation}
Making use of the expressions that define our boundary conditions in this gauge, given by \eqref{eq: a-Omega-To MuN-MuM}, \eqref{eq:MiuraFlat}, the condition in \eqref{eq:Holoau-1} yields $\Omega=0$, while the remaining condition \eqref{eq:Holoau-2} reduces to \eqref{eq:a-reg}. 

As an additional remark of this subsection, it is worth pointing out that the regularity conditions in \eqref{eq:Holoau-1}, \eqref{eq:Holoau-2} can also be used for a wider set of boundary conditions than the ones defined through \eqref{eq: a-Omega-To MuN-MuM}, \eqref{eq:MiuraFlat}. Indeed, for the standard set of boundary conditions that can be described by fixing $\mu_{M}$ and $\mu_L$ to be constants at infinity, the conditions \eqref{eq:Holoau-1}, \eqref{eq:Holoau-2} reduce to the ones found in \cite{Gary:2014ppa, Matulich:2014hea}.

Cosmological flowers are also found to posess an entropy that depends only on the zero modes of the $\hat{u}\left(1\right)$ currents, since
\begin{equation}
S=\frac{A}{4G}=2\pi K_{0}\,.\label{eq:EntropyFlatI}
\end{equation}
We postpone a more thorough discussion of thermodynamical aspects, particularly of entropy, to section \ref{se:6}.

\section{Algebraic aspects}\label{se:algebras}

In this section we summarize for convenience and future reference the various Sugawara-like constructions that start with infinite copies of the Heisenberg algebras together with a finite number of Casimirs. Equivalently, one can start with the same number of $\hat{u}(1)$ current algebras and construct various composite algebras of interest. 
We start with the most prominent one, the Virasoro algebra, then continue with the BMS construction from the present paper, and finally list further similar constructions of interest in three dimensions (twisted warped and higher spin algebras) and four dimensions. 

Before starting we make a couple of comments on conventions. In our discussion we shall freely rescale the $\hat{u}(1)$ levels according to convenience by rescaling the corresponding generators. All commutators not displayed vanish. Finally, with the exception of the first Virasoro algebra presented below, all algebras refer to field theories defined on the plane, with standard consequences for the central terms.

\subsection{Virasoro from Heisenberg}\label{se:5.1}

From a near horizon perspective the central charge of the symmetry algebra should be independent of the AdS radius and any asymptotic behaviour of the solution. This amounts to an ${\mathcal O}(1)$ level for the $\hat u(1)$ currents
\eq{
[{\cal J}^\pm_n,\,{\cal J}^\pm_m]=\frac n2\,\delta_{n+m,\,0}\,.
}{eq:alg1}
The (untwisted) Sugawara construction 
\eq{
{\cal L}_n^\pm = \sum_{p\in\mathbb{Z}}\colon{\cal J}_{n-p}^\pm{\cal J}_p^\pm \colon
}{eq:alg2}
where $\colon\colon$ denotes normal ordering, leads to a (near horizon) Virasoro algebra of central charge 1
\eq{
[{\cal L}_n,\,{\cal L}_m] = (n-m){\cal L}_{n+m} + \frac{1}{12}\,(n^3-n)\,\delta_{n+m,\,0}\,.
}{eq:alg3}
Note that the current algebra remains untwisted here,
\eq{
[{\cal L}^\pm_n,\,{\cal J}^\pm_m]=-m{\cal J}_{n+m}\,.
}{eq:alg4}
The near horizon algebra above appeared in the construction of BTZ microstates, dubbed ``near horizon fluffs'', in \cite{Afshar:2016uax}; see the end of section \ref{se:micro} for a brief discussion of these microstates.

From an asymptotically AdS$_3$ perspective the most natural Sugawara construction of two $\hat{u}(1)$ current algebras,
\eq{
[J_n^\pm,\,J_m^\pm]=\tfrac12\,kn\,\delta_{n+m,\,0}
}{eq:alg5}
is given by a twisted one,
\eq{
L_n^\pm = \frac1k\,\sum_{p\in\mathbb{Z}} J_{n-p}^\pm J_p^\pm + in J_n^\pm\,.
}{eq:alg6}
We assumed here implicitly a large $k$ limit where normal ordering can be neglected. The associated algebra is a Virasoro algebra with the Brown--Henneaux central charge $c=6k$
\eq{
[L_n,\,L_m] = (n-m)L_{n+m} + \frac{c}{12}\,n^3\,\delta_{n+m,\,0}\,.
}{eq:alg7}
Note that the current algebra now is twisted
\eq{
[L^\pm_n,\,J^\pm_m]=-mJ_{n+m} + \tfrac i2\,k n^2 \delta_{n+m,\,0}\,.
}{eq:alg8}
The algebra above appeared in the mapping from quantities in diagonal gauge to highest weight gauge, i.e., in the mapping from ``near horizon'' variables to ``asymptotic variables'', see \cite{Afshar:2016wfy}.

\subsection{BMS$_{3}$ from Heisenberg}\label{se:5.2}

As we have shown in this work, in locally flat space the two current algebras appear naturally in algebraically off-diagonal form \eqref{eq:rc6sm}
\eq{
[J_n,\,K_m]=kn\,\de_{n+m,\,0}
}{eq:alg9}
The algebra diagonalizes by linearly combining the generators as 
\eq{
J_{\pm n}^\pm = \tfrac12\,(K_n \pm J_n)\,, 
}{eq:alg27}
but since the Fourier modes $J_n$, $K_n$ arise naturally in the analysis of the flat space canonical charges it is convenient to use them. 

From an asymptotically flat perspective the most natural Sugawara-like constructions are given by 
\eq{
L_n = \frac1k\,\sum_{p\in\mathbb{Z}} J_{n-p} K_p + in K_n
}{eq:alg10}
and
\eq{
M_n =  \frac1{2k}\,\sum_{p\in\mathbb{Z}} J_{n-p} J_p + in J_n
}{eq:alg11}
where again we refrain from introducing normal ordering since we assume large $k$. [Note, however, that the expression \eqref{eq:alg11} is already normal-ordered since the generators $J_n$ commute among themselves.] The composite generators $L_n$ and $M_n$ then obey the BMS$_3$ algebra \cite{Ashtekar:1996cd} with off-diagonal central extension \cite{Barnich:2006av}.
\begin{align}
[L_n,\,L_m] &= (n-m)\,L_{n+m}\\
[L_n,\,M_m] &= (n-m)\,M_{n+m} + k\,n^3\,\delta_{n+m,\,0} \label{eq:bms}
\end{align}
The non-vanishing commutators of the BMS$_3$ generators with the current algebra generators are given by
\begin{align}
[L_n,\,J_m] &= -mJ_{n+m} + ik\,n^2\,\de_{n+m,\,0} \\
[M_n,\,K_m] &= -mJ_{n+m} + ik\,n^2\,\de_{n+m,\,0} \\
[L_n,\,K_m] &= -mK_{n+m}
\end{align}
The algebra above appeared in the analogue of the mapping from quantities in diagonal gauge to highest weight gauge, i.e., in the mapping from ``near horizon'' variables to ``asymptotic variables'', see section \ref{se:3.6}.

\subsection{(Twisted) warped from Heisenberg}\label{se:5.3}

Previous constructions of NHSAs yielded either the centerless warped conformal algebra \cite{Donnay:2015abr} or the twisted warped conformal algebra \cite{Afshar:2015wjm}. We display now the Sugawara construction that yield the latter algebra from two $\hat{u}(1)$ current algebras. It turns out to be convenient to employ their off-diagonal version \eqref{eq:alg9}.

The twisted warped conformal algebra follows from the Sugawara-like construction
\eq{
L_n = \frac1k\,\sum_{p\in\mathbb{Z}} J_{n-p} K_p + in K_n
}{eq:alg12}
yielding
\begin{subequations}
\label{eq:alg28}
\begin{align}
[L_n,\,L_m] &= (n-m)\,L_{n+m}\\
[L_n,\,J_m] &= -mJ_{n+m} + ik\,n^2\,\de_{n+m,\,0} 
\end{align}
\end{subequations}
exactly as in the BMS$_3$ case. (The remaining part remains untwisted, $[L_n,\,K_m]=-mK_{n+m}$.) 

We close with three comments. Dropping the last term in the Virasoro generators \eqref{eq:alg12} yields the centerless version of that algebra. Introducing normal ordering in \eqref{eq:alg12} leads to a Virasoro central charge $c=2$, which is natural given that we have two $\hat{u}(1)$ current algebras.\footnote{We thank Max Riegler for discussions about this case.} The twisted centerless warped conformal algebra appears as a sector in the BMS construction above.

\subsection{Higher spin algebras from Heisenberg}\label{se:5.4}

Asymptotic symmetry algebras analogous to the spin-2 case were discovered in $sl(n)\oplus sl(n)$ higher spin theories in the principal embedding \cite{Grumiller:2016kcp}. These boundary conditions yield as many $\hat{u}(1)$ current algebras as there are elements in the Cartan subalgebra of the gauge algebra. 

For concreteness we display here the spin-3 case. The asymptotic symmetries are then spanned by four $\hat{u}(1)$ current algebras, whose levels are conveniently normalized as
\begin{align}
[J_n^\pm,\,J_m^\pm] &= \tfrac12\,kn\,\de_{n+m\,0}\\
[J_n^{(3)\,\pm},\,J_m^{(3)\,\pm}] &= \tfrac23\,kn\,\de_{n+m,\,0}
\end{align}
The twisted Sugawara construction that appears in the mapping between near horizon and asymptotic variables \cite{Grumiller:2016kcp} (again ignoring normal ordering)
\begin{align}
L_n^\pm &= \frac1k\,\sum_{p\in\mathbb{Z}} \big(J_{n-p}^\pm J_p^\pm +  \tfrac{3}{4} J_{n-p}^{(3)\,\pm} J_p^{(3)\,\pm}\big) + in J_n^\pm \\
W_n^\pm &= \frac1{k^2}\sum_{p,q\in\mathbb{Z}}\!\! \big(J_{n-p-q}^{(3)\,\pm} J_p^{(3)\,\pm} - 4J_{n-p-q}^\pm J_p^\pm \big)J_q^{(3)\,\pm} \nonumber \\
& \quad -\frac ik \sum_{p\in\mathbb{Z}} (3n-2p)J_{n-p}^{(3)\,\pm} J_p^\pm +\tfrac12 n^2 J_n^{(3)\,\pm} 
\end{align}
then yields the (semi-classical) W$_3$ algebra
\begin{align}
[L_n^\pm,\,L_m^\pm] &= (n-m)\,L_{n+m}^\pm + \tfrac12\,k\,n^3\,\de_{n+m,\,0}\\
[L_n^\pm,\,W_m^\pm] &= (2n-m)\,W^\pm_{n+m}\\
[W_n^\pm,\,W_m^\pm] &= \tfrac13(n-m)(2n^2+2m^2-nm)L_{n+m}^\pm \nonumber \\
&\!\!\!\!\!\!\!\!\!\!\!\!\!\!\!\! + \frac{16}{3k}(n-m) \sum_{p\in\mathbb{Z}} L_{n+m-p}^\pm L_p^\pm + \tfrac16 k n^5\,\de_{n+m,0} 
\end{align}
whose generators have the following commutations relations with the $\hat{u}(1)$ currents.
\begin{align}
[L_n^\pm,\,J_m^\pm] &= -mJ_{n+m}^\pm  + \tfrac i2\,k n^2 \delta_{n+m,\,0}\\
[L_n^\pm,\,J_m^{(3)\,\pm}] &=  -mJ_{n+m}^{(3)\,\pm} \displaybreak[1] \\
[W_n^\pm,\,J_m^\pm] &= \frac{4m}{k}\,\sum_{p\in\mathbb{Z}} J_{n+m-p}^\pm J_p^{(3)\,\pm}\nonumber\\
&\quad +\tfrac i2\,m(3n+2m)\,J_{n+m}^{(3)\,\pm} \displaybreak[1]\\
[W_n^\pm,\,J_m^{(3)\,\pm}] &= \frac{8m}{3k}\sum_{p\in\mathbb{Z}}\big(J_{n+m-p}^\pm J^\pm_p - \tfrac34\,J_{n+m-p}^{(3)\,\pm}J_p^{(3)\,\pm}\big) \nonumber\\
&\! +\frac{2i}{3}m(n-2m)J_{n+m}^\pm + \tfrac13kn^3\de_{n+m,0}
\end{align}

Constructions analogous the the spin-3 case reviewed above work also for higher spins, see \cite{Grumiller:2016kcp} for more details.

\subsection{Kerr near horizon algebra from Heisenberg}\label{se:5.5}

All the algebraic constructions above were attached to three-dimensional theories of gravity or higher spin theories. It is physically interesting to inquire about similar algebraic constructions in higher dimensions. We summarize here the construction in four-dimensional flat space discovered recently \cite{Afshar:2016uax}, with possible applications to non-extremal Kerr black holes.

Starting with four $\hat{u}(1)$ current algebras
\eq{
[J_n^\pm,\,J_m^\pm] = -[K_n^\pm,\,K_m^\pm] = \tfrac12\,n\,\de_{n+m,\,0}
}{eq:alg13}
the Sugawara-like constructions
\begin{align}
Y_n^\pm &= \sum_{p\in\mathbb{Z}} (J_{n-p}^\pm + K_{n-p}^\pm)(J_p^\pm - K_p^\pm) \\
T_{(n,\,m)} &= (J_n^++K_n^+)(J_m^-+K_m^-)
\end{align}
yield the four-dimensional near horizon algebra of \cite{Donnay:2015abr}
\begin{align}
[Y_n^\pm,\,Y^\pm_m] &= (n-m)\,Y^\pm_{n+m}\\
[Y_l^+,\,T_{(n,\,m)}] &= -n T_{(n+l,\,m)}\\
[Y^-_l,\,T_{(n,\,m)}] &= -m T_{(n,\,m+l)} \,.
\end{align}
This is a strong algebraic hint that our soft Heisenberg hair discussion generalizes to four dimensional (non-extremal) Kerr black holes.

\section{Entropy}\label{se:6}

The entropy associated with our near horizon metrics can be calculated in numerous ways, which we shall do in this section. We broadly classify the calculations into macroscopic, microscopic and thermodynamic calculations. All our results turn out to agree with each other. Both for the locally AdS and the locally flat case entropy is given by 
\begin{equation}
S=\frac{A}{4G}=\beta |E|=2\pi\big(J_{0}^{+}+J_{0}^{-}\big)\label{eq:entropy0}
\end{equation}
where the first equality gives the macroscopic result (with $A=\oint\gamma\extd\vp$ and $G$ is Newton's constant), the second the thermodynamic result [with $\beta=1/T=2\pi/a$ and $|E|=a(J_{0}^{+}+J_{0}^{-})$] and the last the microscopic result.

We review first the macroscopic derivation of entropy in section \ref{se:macro} and then address microscopic derivations in section \ref{se:micro}. In section \ref{se:flat} we generalize our results to flat space cosmological flowers. Finally, we discuss and speculate on some of the unusual thermodynamical aspects of our theory in section \ref{se:thermo}.

\subsection{Macroscopic entropy}\label{se:macro}

In this subsection we determine the entropy macroscopically. The Bekenstein--Hawking formula 
\begin{equation}
S_{\text{BH}}=\frac{A}{4G}\label{eq:SBH}
\end{equation}
can be obtained by a number of methods, like Wald's \cite{Wald:1993nt,Iyer:1994ys} or Solodukhin's \cite{Solodukhin:1994yz}. We do not display these standard derivations and instead derive macroscopic entropy here in the Chern--Simons formulation.

The general result for the entropy for gravity theories in the Chern--Simons formulation yields 
\begin{align}
S_{\textrm{\tiny CS}}&=\frac{k}{2\pi}\,\int\extd\vp\wedge\extd\tau\,\langle {{\cal A}_{\tau}\cal A}_{\vp}\rangle \nonumber\\
&=-\frac{k}{2\pi}\,\beta\,\sigma\,\oint\extd\vp\,\langle {{\cal A}_{\tau}\cal A}_{\vp}\rangle \label{eq:SCS}\\
&=-\beta\,\sigma\big(J_{0}^{+}\zeta^{+}+J_{0}^{-}\zeta^{-}\big)=2\pi\,\big(J_{0}^{+}+J_{0}^{-}\big)\,.\nonumber
\end{align}
The first equality follows from the general discussion reviewed e.g.~in \cite{Bunster:2014mua} (see also Refs. \cite{Perez:2012cf,Perez:2013xi,deBoer:2013gz}). The second equality follows from our definitions and results in sections \ref{se:2} and \ref{se:3}. As explained in \cite{Matulich:2014hea}, $\sigma$ stands for the orientation of the torus, which for black flowers on AdS$_{3}$ ($\sigma=1$) turns out to be the opposite of soft hairy cosmologies on flat space ($\sigma=-1$). The last equality is a consequence of the regularity conditions that dictate $\zeta^{\pm}=-2\pi \sigma\beta^{-1}$. The result \eqref{eq:SCS} coincides with the last expression in \eqref{eq:entropy0}.

\subsection{Microscopic entropy}\label{se:micro}

For zero-mode solutions like the BTZ black hole there is a number of different microstate countings that yield the correct result for the entropy. Each is based on a different symmetry algebra that is composite in terms of our asymptotic symmetry generators $J_{n}^{\pm}$ fulfilling the $\hat{u}\left(1\right)$ currents algebras in \eqref{eq:NHSA}. However, a naive standard microstate counting does not work for generic black flowers. We show now why this is the case.

Even though black flowers are generically endowed with all of the possible $\hat{u}\left(1\right)$ global charges, according to \eqref{eq:SCS}, their entropy only depends on the zero modes. Besides, in the spherically symmetric case (BTZ black hole), by virtue of the (twisted) Sugawara construction given by \eqref{eq:alg6} \cite{Afshar:2016wfy}, the Virasoro generators $L_{n}^{\pm}$ can be obtained from the $\hat{u}(1)$ ones $J_{n}^{\pm}$, so that \eqref{eq:SCS} reduces to the known result for the black hole entropy\footnote{In the following quantities like $L_{0}$ or $J_{0}^{\pm}$ often refer to vacuum expectation values and not to the corresponding operators; the meaning should always be clear from the context.} \cite{Bloete:1986qm,Cardy:1986ie,Strominger:1997eq}
\begin{equation}
S=S_{\text{Cardy}}=2\pi\sqrt{\frac{c}{6}\,L_{0}^{+}}+2\pi\sqrt{\frac{c}{6}\,L_{0}^{-}}\,.\label{eq:cardy}
\end{equation}
However, as it can be directly seen from the map in \eqref{eq:alg6}, a naive direct application of \eqref{eq:cardy} to the case of a generic soft hairy black hole, for which $J_n^{\pm}\neq$0, clearly does not lead to the right result because the entropy would manifestly acquire an explicit dependence on soft hair charges. To pinpoint the problematic issue with these states we consider a specific one defined by 
\eq{
|B\rangle = J_{-1}^+ |\textrm{BTZ}\rangle
}{eq:angelinajolie} 
where the state $|\textrm{BTZ}\rangle$ obeys the highest weight conditions $J_{n}^{\pm}|\textrm{BTZ}\rangle=0$ for all positive $n$. Algebraically, this seems to make sense since BTZ black holes could be defined by the conditions \cite{Sheikh-Jabbari:2016npa} $\langle\textrm{BTZ}|J_n|\textrm{BTZ}\rangle=0$, for all $n\neq0$, while the vacuum expectation values of $J_{0}^{\pm}$ are related to mass and angular momentum, through $L_{0}^{\pm}=k^{-1}(J_{0}^{\pm})^2$. The key observation is that the state $|B\rangle$ does not obey all the highest weight conditions since particularly $J_{1}^{+}|B\rangle\neq0$. As a consequence, it does neither obey all the Virasoro highest weight conditions since particularly $L_{1}^{+}|B\rangle\neq0$. This prevents us from doing the usual Cardy counting for generic black flowers. Similar remarks apply to countings based on other compositions of Heisenberg algebras, like the BMS$_{3}$ algebra or the warped conformal algebra (see section \ref{se:algebras}).

One may then wonder how the Cardy formula could be suitably modified or extended so as to reproduce the right semiclassical result for the black flower entropy in \eqref{eq:SCS}. Indeed, such a possibility exists and invokes a microstate counting that relies on an anisotropic extension of $S$ modular invariance, without the need of explicitly identifying any microstates. It is based on recent results for theories with anisotropic Lifshitz scaling in two spacetime dimensions, $t\to \lambda^z t$, $\vp\to\lambda\vp$, whose left and right movers are assumed to be decoupled and characterized by the same dynamical exponent $z$ \cite{Gonzalez:2011nz,Perez:2016vqo}. At fixed values of left and right energies $\Delta_{\pm}\gg\Delta_{0}^{\pm}\left[z\right]$, the asymptotic growth of the number of states is given by 
\begin{equation}
S=2\pi(1+z)\sum_{\pm}\Delta_{\pm}^{1/(1+z)}\exp\left(\tfrac{z}{1+z}\,\ln\,\left(\Delta_{0}^{\pm}\left[1/z\right]/z\right)\right)\,,\label{eq:SKdV}
\end{equation}
where the spectrum is assumed to possess a gap and $-\Delta_{0}^{\pm}\left[z\right]$ stand for the ground state energy of left and right movers. Note that for the isotropic case $z=1$ the Cardy formula \eqref{eq:cardy} is recovered, upon identifying $L_0^\pm = \Delta_\pm$ and $\Delta_0^\pm = c/24$. As we shall see in the next paragraph, our case corresponds to $z=0$.

The reason why we can apply \eqref{eq:SKdV} to black flowers is that the boundary conditions in \cite{Afshar:2016wfy} containing soft hairy black holes belong to a class of boundary conditions on AdS$_3$, for which the ``boundary gravitons'' obey the field equations of the $n^{\textrm{th}}$ KdV hierarchy, extended to fractional $n$, for the special case  $n=-\frac{1}{2}$ \cite{Perez:2016vqo}. For a generic value of $n$, left and right movers then possess a dynamical exponent given by $z=2n+1$, and the ground state energies that correspond to the ones of global AdS$_3$ were found to be given by
\begin{equation}
-\Delta_{0}^{\pm}\left[z\right]=\frac{k}{2}\,\frac{1}{1+z}\,\left(-1\right)^{(1+z)/2}\,.\label{eq:SKdV2}
\end{equation}
Remarkably, the asymptotic growth of the number of states for the set of boundary conditions in \cite{Afshar:2016wfy} can then be obtained from \eqref{eq:SKdV} and \eqref{eq:SKdV2} in the limit $z\to0$, so that the entropy reduces to
\begin{equation}
S\big|_{z=0}=2\pi\Delta_{+}+2\pi\Delta_{-}\,,\label{eq:SKdV4}
\end{equation}
which precisely agrees with the last expression in \eqref{eq:entropy0} for a generic soft hairy black hole upon identifying $\Delta_{\pm}=J_{0}^{\pm}$. 

Although the left- and right-energies $-\Delta_0^\pm$ of global AdS$_3$ drop out in our entropy formula \eqref{eq:SKdV4}, it is still of interest to compare the general result \eqref{eq:SKdV2} for $z=0$ with the corresponding result within the theory specified by the boundary conditions reviewed in section \ref{se:2}. At first glance this is problematic, since for fixed (real, positive) Rindler acceleration none of our states is maximally symmetric. However, by analytic continuation to complex values we obtain a maximally symmetric line-element \eqref{eq:r5} with the choices\footnote{%
This is perhaps seen most easily by comparing with the near horizon line-element for $\Omega=0$, $\extd s_{\textrm{\tiny nh}}^{2}=-r^{2}a^{2}\,\extd t^{2}+\extd r^{2}+\ga^{2}\extd\vp^{2}+\dots$ Then it is evident that choosing imaginary $a$ and $\ga$ as in \eqref{eq:groundstate} effectively exchanges $\vp$ and $t$. At the ``self-dual point'' where $\vp$ and $t$ are $2\pi$ periodic regularity at the horizon requires $a=\pm i$, while the Killing vector analysis in appendix \ref{se:KV} leads to the result $J_{0}^{+}=J_{0}^{-}=\pm ik/2$.} 
\begin{equation}
a=\pm i\quad\quad\Omega=0\quad\quad J_{0}^{+}=J_{0}^{-}=\pm i\,\frac{k}{2}\,,\label{eq:groundstate}
\end{equation}
so that left and right ground state energies $-\Delta_{0}^{\pm}\left[z\right]$ in \eqref{eq:SKdV2} exactly coincide with $J_{0}^{\pm}$ in \eqref{eq:groundstate} for $z=0$. Thus, we have the curious situation that our physical spectrum (real, positive values of $a$ and $J_{0}^{\pm}$) is gapped from the ground state by an imaginary amount. The same feature was found previously in Rindleresque holography \cite{Afshar:2015wjm}.

Besides a traditional microstate counting, which we have achieved above, it is also of interest to explicitly identify the black hole microstates. Exploiting the near horizon symmetry algebra and both Sugawara constructions for asymptotic ($L_n^\pm$) and near horizon (${\cal L}_n^\pm$) Virasoro algebras summarized in section \ref{se:5.1}, as well as the working hypothesis ${\cal L}_0 = c L_0$, a proposal for these microstates was presented recently in \cite{Afshar:2016uax}. It was found that soft hair descendants fall into two classes, the ``horizon fluffs'', which obey all Virasoro highest weight conditions, and the remaining ones, which violate some of the Virasoro highest weight conditions. In terms of near horizon generators ${\cal J}_n^\pm$ the microstates $|{\cal B}\rangle$ of a BTZ black hole with energies $L_0^\pm$ are given by all states of the form
\eq{
|{\cal B}\rangle \sim \prod {\cal J}_{-n_i^+}^+ \,{\cal J}_{-n_i^-}^- |0\rangle 
}{eq:fluff1}
subject to the conditions $\langle{\cal B}|({\cal J}_0^{\pm})^2|{\cal B}\rangle = c L_0^\pm$ and $\langle{\cal B}|{\cal J}_{cn}|{\cal B}\rangle=0$ for all $n\neq 0$.
Our soft hair generators $J_n^\pm$ are related to these near horizon generators through $J_n^\pm = {\cal J}^\pm_{cn}/\sqrt{6}$, where the central charge $c=6k$ is assumed to be a (large) integer. This means that the microstates \eqref{eq:fluff1} are highest weight states  with respect to soft hair generators, $J_n^\pm|{\cal B}\rangle=0$ for $n>0$, but they are not highest weight states with respect to the near horizon generators ${\cal J}_n^\pm$. As shown in \cite{Afshar:2016uax}, the degeneracy of the microstates \eqref{eq:fluff1} in the classical (large $c$) limit correctly accounts for the Bekenstein--Hawking entropy of BTZ black holes. By the same arguments it works also for arbitrary black flowers generated by acting with arbitrary combinations of $J_{-n^\pm}^\pm$ on BTZ.

\subsection{Flat space entropy}\label{se:flat}

It is known that the entropy of flat space cosmologies \cite{Barnich:2012xq, Bagchi:2012xr} can be obtained from the inner horizon AdS entropy by suitable rescalings with the AdS radius \cite{Riegler:2014bia, Fareghbal:2014qga}. In terms of Cardy-like formulas the main change between outer and inner horizon entropies is a relative sign change between the two additive terms in the corresponding entropy formula. As explained in \cite{Matulich:2014hea} this is because the solid torus possesses a reversed orientation as compared with the black hole. In our case this would change the last expression in \eqref{eq:entropy0} to $S_{\textrm{\tiny inner}}=2\pi(J_0^+ - J_0^-)$. Note also that the automorphism $J^-_n\to - J_n^-$ of our NHSA allows to redefine the generators such that our result \eqref{eq:entropy0} still holds for inner horizons! Thus, we shall always assume that the $\hat u(1)$ current algebra generators are defined with suitable signs so that the last expression in \eqref{eq:entropy0} is valid.

An alternative way to obtain the entropy of cosmological flowers, as described above, is performing the usual \.In\"on\"u--Wigner contraction from AdS to flat space by rescaling suitably some generators with the AdS radius $\ell$.
\begin{subequations}
\label{eq:IW}
\begin{align}
L_n &= L_n^+ - L_{-n}^-\\
M_n &= \frac1\ell\big(L_n^+ + L_{-n}^-\big)\\
K_n &= J_n^+ + J_{-n}^- \\
J_n &= \frac1\ell \big( J_n^+ - J_{-n}^-\big)
\end{align}
\end{subequations}
Note that the relative signs chosen in $K_n$ and $J_n$ are in accordance with the automorphism mentioned above. These definitions are compatible with all the algebras in our paper and with the respective Sugawara-like constructions. 

To see this explicitly it is important to recall the differences between the flat space Chern--Simons level $k_{\textrm{\tiny flat}}=1/(4G)=:k$ and the AdS Chern--Simons level $k_{\textrm{\tiny AdS}}=\ell k$. The NHSAs \eqref{eq:NHSA} and \eqref{eq:NHSAsm} are then compatible with the contraction \eqref{eq:IW}. At finite (but large) $\ell$ the twisted Sugawara construction \eqref{eq:alg9} and the relations above (together with their inversion, $J_{\pm n}^\pm=(K_n\pm\ell J_n)/2$) yield
\begin{align}
L_n &= \frac{1}{\ell k}\sum_{p\in\mathbb{Z}}\big(J^+_{n-p}J_p^+ - J_{-n-p}^-J_p^-\big) +in\big(J_n^+ + J_{-n}^-\big) \nonumber \\
&= \frac1k\sum_{p\in\mathbb{Z}}J_{n-p}K_p + inK_n \\
M_n &= \frac{1}{\ell^2 k}\sum_{p\in\mathbb{Z}}\big(J^+_{n-p}J_p^+ + J_{-n-p}^-J_p^-\big) +\frac{in}{\ell}\,\big(J_n^+ - J_{-n}^-\big) \nonumber \\
&= \frac1{2k}\sum_{p\in\mathbb{Z}}J_{n-p}J_p + inJ_n + {\cal O}(1/\ell)\,.
\end{align}
In the $\ell\to\infty$ limit the results above coincide precisely with \eqref{eq:whatever}.

This means that the entropy of soft hairy cosmological spacetime can be obtained purely algebraically as 
\eq{
S = 2\pi\big(J_0^+ + J_0^-\big) = 2\pi K_0 = \frac{A}{4G} = -\beta E\,
}{eq:entropy42}
in agreement with \eqref{eq:SCS}.
The first equality is the inner horizon entropy in AdS, taking into account our automorphism above. 
The second equality is the flat space entropy and follows from the definitions \eqref{eq:IW}. The third equality follows from the relation \eqref{eq:rc6sm}. The final equality follows from the discussion in section \ref{se:3.5}. The sign concurs with a corresponding sign flip in inner horizon thermodynamics \cite{Castro:2012av, Detournay:2012ug}.

As a consequence of our contraction procedure above all macroscopic and microscopic formulas obtained for the entropy in AdS can be used to recover the corresponding flat space results. As an example let us quote the flat space Cardy formula \cite{Barnich:2012xq, Bagchi:2012xr} for flat space cosmologies, 
\eq{
S_{\textrm{\tiny FSC}} = 2\pi |L_0| \sqrt{\frac{c_M}{2M_0}}\,.
}{eq:entropy26}
Here $L_0$ and $M_0$ are the vacuum expectation values of the BMS zero-mode generators and $c_M$ is the coefficient in the anomalous term of the mixed BMS$_3$ commutator \eqref{eq:bms}, $c_M=k$. Inserting into \eqref{eq:entropy26} our results for the contraction above then yields
\eq{
S_{\textrm{\tiny FSC}} = 2\pi \frac1k |J_0| K_0 \sqrt{\frac{k^2}{J_0^2}} = 2\pi K_0
}{eq:entropy25}
which agrees with our general result \eqref{eq:entropy42}. However, it should be highlighted that, as in the case of soft hairy black holes on AdS$_{3}$, a naive direct application of \eqref{eq:entropy26} to the case of cosmological flowers does not yield the correct result (see section \ref{se:macro}). In this sense, we expect that there is an analogous flat space contraction of the microscopic entropy formula \eqref{eq:SKdV}.

\subsection{Thermodynamic entropy}\label{se:thermo}

We conclude this section with some discussion on peculiar thermodynamical features associated with our boundary conditions, as well as some related speculations.

Our theory is singled out from a thermodynamic perspective by the independence of temperature from any of the extensive variables (energy, angular momentum or entropy). We stress that this is a highly unusual property, not just from a thermodynamical perspective but also from a gravitational one. In any other gravitational setup we are aware of the temperature associated with some regular black hole solution generally depends on parameters like mass or angular momentum, the canonical example being Hawking's temperature law $T\propto 1/M$ for Schwarzschild black holes of mass $M$. 

We believe that the property that temperature does not depend on the state is key in making our boundary conditions so suitable for microscopic purposes, since we can put the theory at any temperature we like and still all states in the theory (with real vacuum expectation values $J_0^\pm$) are regular.

Clearly, our setup is thermodynamically an extremely simple system. This is a direct consequence of our assumption that Rindler acceleration (which determines temperature) is a state-independent quantity.

This means that the first law 
\begin{equation}
\extd E=T\text{\,}\extd S\label{eq:entropy3}
\end{equation}
can be integrated trivially since $T$ does not depend on $S$. Integrating the first law \eqref{eq:entropy3} yields the thermodynamic result displayed in \eqref{eq:entropy0}. We can rephrase the statement above in terms of the Gibbs--Duhem relation, which states that the total Legendre transformation with respect to all pairs of intensive/extensive variables of the internal energy vanishes. In our case the only extensive variable on which internal energy depends is the entropy $E=E(S)$. This means that the Gibbs--Duhem relation in our case reads $E-TS=0$. From the Gibbs--Duhem relation we then recover the result \eqref{eq:entropy0}. Another way to read the Gibbs--Duhem relation is to observe that Helmholtz free energy vanishes in our theory. 
\begin{equation}
F=-T\,\ln\,Z_{c}=0\label{eq:entropy7}
\end{equation}
This means that the canonical partition function $Z_{c}$ must equal
to unity. 
\begin{equation}
Z_{c}=1=\text{Tr}e^{-\beta E}=e^{-\beta E}\text{Tr}\unity\label{eq:entropy4}
\end{equation}
From the identities \eqref{eq:entropy4} we learn that the microcanonical
partition function $Z$ is given by 
\begin{equation}
Z=\text{Tr}{\unity}=e^{\beta E}=W\label{eq:entropy5}
\end{equation}
which, after taking the logarithm, yields again the entropy \eqref{eq:entropy0}.

The first equality \eqref{eq:entropy5} provides a purely thermodynamical glimpse into a microstate counting and the associated difficulties: we would need to count all the states $W$ in the Hilbert space that have the same energy and could then use Boltzmann's formula, 
\begin{equation}
S_{\text{Boltzmann}}=\ln W\label{eq:entropy6}
\end{equation}
to obtain again the entropy \eqref{eq:entropy0}. However, as we have seen above there are infinitely many soft hair states with the same energy, so that a naive microscopic evaluation of the entropy leads to a meaningless divergent result. What we need, therefore, is some cutoff on the amount of soft hair that a horizon can have, depending on its area. A proposal to obtain such a cutoff in a controlled way was described in \cite{Afshar:2016uax}.

Here, we use a shortcut from heuristic arguments. Rewriting \eqref{eq:entropy6} as 
\begin{equation}
W=e^{S_{\text{Boltzmann}}}=e^{f\left[J_{0}^{+},J_{0}^{-}\right]}=e^{f\left[J_{0}^{+}+J_{0}^{-}\right]}=e^{\alpha\left(J_{0}^{+}+J_{0}^{-}\right)}\label{eq:entropy8}
\end{equation}
the first equality is evident. The second equality uses the fact that entropy can only depend on the zero mode charges $J_{0}^{\pm}$ through some arbitrary function $f$. The third equality invokes parity invariance to simplify the function $f$ to a function of the sum of the zero mode charges, $f[J_{0}^{+}+J_{0}^{-}]$. The last equality then uses extensitivity of the entropy to argue that the function $f$ must be linear in the charges (and homogeneous). Thus, what remains to be determined in this Boltzmann-inspired counting is the constant $\alpha>0$. Assuming that global AdS$_3$ has no entropy, i.e., $W_{\textrm{\tiny AdS}}=1$, and using the result \eqref{eq:groundstate} we conclude 
\begin{equation}
\alpha=\frac{2\pi n}{k}\text{\quad\quad}n\in\mathbb{Z}^{+}\,.\label{eq:entropy9}
\end{equation}
Since entropy should be linear in the inverse Newton constant $k$ and $J_{0}^{\pm}$ is also linear in $k$, we know that $n$ in \eqref{eq:entropy9} must be a multiple of $k$ and $k$ must be quantized in the integers. Assuming that the smallest positive number $\alpha$ compatible with all requirements spelled out above is the correct choice, we then obtain 
\begin{equation}
\alpha=2\pi\label{eq:entropy10}
\end{equation}
and consequently 
\begin{equation}
W=e^{S_{\text{Boltzmann}}}=e^{2\pi\left(J_{0}^{+}+J_{0}^{-}\right)}\label{eq:entropy11}
\end{equation}
which is again compatible with \eqref{eq:entropy0}.

\section{Comparison to other approaches}\label{se:42}

\subsection{General remarks}

In the past year, three different near horizon boundary conditions were proposed \cite{Donnay:2015abr,Afshar:2015wjm,Afshar:2016wfy}, some of which can be interpreted as modified asymptotically AdS boundary conditions. However, these boundary conditions appear more natural when expanding the metric in a near horizon approximation like \eqref{eq:r2}. While imposing boundary conditions at some arbitrary locus inside the manifold may lack physical motivation, as discussed in the introduction the imposition of the existence of a horizon (and some smooth neighborhood around it) has a clear physical motivation, namely to ensure that the state space considered by a theory subject to these boundary conditions has exclusively states with a horizon. Even under these premises and restricting to Einstein gravity in three spacetime dimensions (for the time being with negative cosmological constant) there is still a number of choices one can make.

The most important choice is to fix the Rindler acceleration $a$ in \eqref{eq:r2} either as a state-independent quantity (or, equivalently, as chemical potential/source from a holographic view point) or as a state-dependent quantity (depending on the charges or vacuum expectation values from a holographic view point). We mentioned already that we make the former choice, which implies that all states in the theory have the same temperature 
\eq{ 
T = \frac{a}{2\pi}\,. 
}{eq:unruh}
Since typically different states have different temperatures --- for instance, the BTZ black holes for different masses and angular momenta have different temperatures --- we elaborate now why we make this choice.

First of all, the alternative is difficult to implement. One obstacle is that the near horizon line-element \eqref{eq:r2} is invariant under rescalings of Rindler acceleration with a simultaneous rescaling of the coordinates, 
\eq{ 
a\to \lambda a\qquad\rho\to\lambda\rho\qquad v\to v/\lambda 
}{eq:r30} 
which means that there is no operational meaning to a statement like ``the Rindler acceleration is 42'' since we can always rescale Rindler acceleration to unity using \eqref{eq:r30}. Therefore, we need to break the scaling symmetry \eqref{eq:r30} if we want to make sense of a state-dependent Rindler acceleration. A previous work \cite{Afshar:2015wjm} achieved this through periodically identifying advanced/retarded time, $v\to v+2\pi L$, with some length scale $L$ that breaks the invariance \eqref{eq:r30}. While this compactification of a lightlike direction had some additional advantages, the physical interpretation of this setup and the dual field theory (if there is any) is difficult. Another difficulty is that solving the Einstein equations in the near horizon approximation with an arbitrary function $a(v,\,\vp)$ leads to the conditions that $a$ can depend on $v$ only, which means that either $a$ is pure gauge (and hence state-independent) or that the associated charges are not conserved in advanced time.

The second reason to opt for fixed Rindler acceleration is that it automatically fixes the scale invariance \eqref{eq:r30}. Finally, and perhaps most importantly, a good physical reason to fix the temperature for all states to be the same is that the type of questions one would like to ask in the near horizon setup are questions about one fixed macrostate --- a black hole or cosmological spacetime --- with fixed temperature. For instance, the type of question we want our near horizon theory to be able to answer is ``given a BTZ black hole with temperature $T$, what is the number of microstates that contribute to the ensemble that describes this macrostate''.

We address now in more detail relations and differences to previous approaches.

\subsection{Comparison with \cite{Donnay:2015abr}}

Donnay, Giribet, Gonz\'alez and Pino \cite{Donnay:2015abr} formulated near horizon boundary conditions in three-dimensional AdS and flat space, as well as in four-dimensional flat space. Their NHSA differs from ours, though it is also possible to obtain it as a composite algebra through a Sugawara-like construction, see section \ref{se:algebras}. More precisely, they obtained in three dimensions an untwisted version of the warped CFT algebra \eqref{eq:alg28}.

Since their paper uses Gaussian null coordinates like in \eqref{eq:Null-AdS-metric} we employ these coordinates for comparison. Both their boundary conditions and ours are preserved by asymptotic Killing vectors of the form
\eq{
\xi^v = T(\vp) + \dots\qquad \xi^\vp = Y(\vp) + \dots\qquad \xi^\rho = \dots
}{eq:dggp1}
where the ellipsis refers to subleading terms that do not contribute to the canonical boundary charges. 

The key difference between their boundary conditions and ours is that in their case $T$ and $Y$ are state-independent, whereas in our case the variations of $\gamma,\omega$ are state-independent.
\begin{equation}
\delta_{\xi}\gamma=\tfrac{\ell}{2}\left(\eta^{+}-\eta^{-}\right)^{\prime}\text{\quad\quad}\delta_{\xi}\omega=\tfrac{1}{2}\left(\eta^{+}+\eta^{-}\right)^{\prime}\label{eq:AKVs2}
\end{equation}
This implies that the state-dependent functions $\gamma$ and $\omega$ transform under the symmetry generators \eqref{eq:dggp1} as 
\begin{equation}
\delta_{\xi}\gamma=\left(\gamma\,Y\right)^{\prime}+\Omega\,T^{\prime}\text{\,}\text{\,},\text{\quad}\delta_{\xi}\omega=\left(\omega\,Y\right)^{\prime}-a\,T^{\prime}\,.\label{eq:AKVs1}
\end{equation}
and that the functions $T$ and $Y$ can be expressed as 
\begin{equation}
T=\frac{{\cal J}^{+}\eta^{-}+{\cal J}^{-}\eta^{+}}{{\cal J}^{+}\zeta^{-}+{\cal J}^{-}\zeta^{+}}\text{\quad\quad}Y=\frac{\zeta^{-}\eta^{+}-\zeta^{+}\eta^{-}}{{\cal J}^{+}\zeta^{-}+{\cal J}^{-}\zeta^{+}}\,.\label{eq:AKVs3}
\end{equation}
As it must be, the results \eqref{eq:AKVs3} concur with \eqref{eq:xi-mu}.

\subsection{Relationship to other work}

The soft hair proposal by Hawking, Perry and Strominger \cite{Hawking:2016msc} and related work \cite{Strominger:2014pwa, Flanagan:2015pxa, Compere:2016hzt} has engendered a lot of research in the past year, see \cite{Penna:2015gza, Hooft:2016itl, Averin:2016ybl, Compere:2016jwb, Compere:2016hzt, Sheikh-Jabbari:2016unm, Sheikh-Jabbari:2016lzm, Blau:2016juv, Eling:2016xlx, Ellis:2016atb, Giddings:2016plq, Compere:2016gwf, Muck:2016stv, Hotta:2016qtv, Averin:2016hhm, Mirbabayi:2016axw, Donnay:2016ejv, Sheikh-Jabbari:2016npa, Cai:2016idg, Banerjee:2016nio, Setare:2016qob} for some selected references and \cite{Hawking:2016sgy} for their most recent work.

Our boundary conditions in \cite{Afshar:2016wfy} provide a concrete realization of their proposal for soft hairy black holes in three-dimensional general relativity with negative cosmological constant, albeit with a symmetry algebra that was not anticipated in \cite{Hawking:2016msc}, namely two $\hat{u}(1)$ current algebras or, equivalently, infinitely many Heisenberg algebras supplemented by two Casimirs. It is then fair to ask how sensitive to the chosen model are our conclusions about the symmetry algebra, soft Heisenberg hair, entropy, etc.

In fact, we are convinced that our conclusions are fairly universal, in particular that the near horizon symmetries are governed by infinite copies of the Heisenberg algebra, together with some Casimirs. In the present work we have provided evidence for this universality by showing that also horizons in flat space lead to the same symmetry algebra in three-dimensional general relativity. There is, however, already substantial additional evidence for universality that has appeared while this paper was in progress.

It has been shown that Heisenberg algebras arise also as NHSAs for BTZ black holes in Chern--Simons like theories of gravity, i.e., in theories that go beyond general relativity by including higher derivative corrections \cite{Setare:2016vhy}. Our conclusions were extended in \cite{Grumiller:2016kcp} to a specific class of higher spin theories. Perhaps the most remarkable aspect of these results is that the entropy law $S=2\pi(J_0^+ + J_0^-)$ remains true in higher spin theories and reproduces the (fairly complicated) known results for entropy expressed in terms of the global charges. This provides further evidence for the expectation that the near horizon theory is extremely simple, far simpler than the asymptotic one. It was algebraically shown in \cite{Afshar:2016uax} that Heisenberg algebras arise also in the near horizon approach to general relativity in four spacetime dimensions, see section \ref{se:5.5}. 

While there are certainly more generalizations one could envisage (we mention some of them in the concluding section \ref{se:7}), in our opinion the diverse generalizations obtained so far --- higher derivative interactions, higher spins, higher dimension --- together with the results of the present work provide strong evidence for the universality of asymptotic or near horizon symmetry algebras in terms of Heisenberg algebras. 

Let us finally point out that our boundary conditions are a special case of the recently proposed general boundary conditions in AdS$_3$ \cite{Grumiller:2016pqb}. Our symmetry algebra is a subalgebra of the general asymptotic symmetry algebra found therein, which consists of two $sl(2)$ current algebras with non-vanishing levels.

\section{Conclusions}\label{se:7}

In this work we summarized and expanded upon the boundary conditions in three dimensional Einstein gravity of \cite{Afshar:2016wfy}. We have presented and discussed the metric formulation of the boundary conditions and generalized the asymptotically AdS result to asymptotically flat spacetimes. We have found that the asymptotic/near horizon symmetry algebra is independent from the curvature radius of the spacetime and hence equivalent in both asymptotically AdS and flat spacetimes. It consists of infinite copies of the Heisenberg algebra and two Casimirs, $X_{0}$, $P_{0}$. 
\begin{align}
\left[X_{n},\,X_{m}\right] & =\left[P_{n},\,P_{m}\right]=\left[X_{0},\,P_{n}\right]=\left[P_{0},\,X_{n}\right]=0\nonumber \\
\left[X_{n},\,P_{m}\right] & =i\delta_{n,m}\quad\textrm{if}\;n\neq0
\end{align}
Our results, in particular the independence of the symmetry algebra from the cosmological constant, support the interpretation that our boundary conditions describe near horizon physics.

In section \ref{se:6} we derived entropy macroscopically, thermodynamically and microscopically, and found 
\eq{
S =2\pi(J_{0}^{+}+J_{0}^{-}) \,,
}{eq:lalapetz}
compatible with the Bekenstein--Hawking law.
Remarkably, the result for entropy \eqref{eq:lalapetz} remains true in generalizations to higher spin theories \cite{Grumiller:2016kcp} and is thus fairly universal.

What is still missing in the present work is a true Heisenberg counting in locally flat space that directly uses the Heisenberg algebra or, equivalently, the $\hat{u}(1)$ current algebras \eqref{eq:NHSAsm} plus suitable additional information that provides a controlled cutoff on the soft hair spectrum. Recently, a proposal for  such a counting was performed in locally AdS$_3$ \cite{Afshar:2016uax} by using the observation of \cite{Banados:1998wy} that the Virasoro algebra with (integer) central charge $c$ can be understood as a subalgebra of the Virasoro algebra with central charge equal to unity. The $c=1$ Virasoro algebra naturally arises from the normal ordered Sugawara construction of the $\hat{u}(1)$ NHSAs (see section \ref{se:5.1}). It was then proposed that the black hole microstates correspond to all the $\hat{u}(1)$ descendants of the vacuum which lead to the same expectation value of the asymptotic $c=3\ell/(2G)$ Virasoro charges. A counting of states in this proposal was found to agree with the Bekenstein--Hawking entropy. It would be very interesting to generalize those results to the case of asymptotically flat spacetimes and, beyond Einstein gravity, to flat space chiral gravity \cite{Bagchi:2012yk}, and to other higher derivative or higher spin theories of gravity.

Further generalizations of our approach are possible and would be useful to pursue in order to check the (possible limits of) universality of our results for the NHSA and the entropy \eqref{eq:lalapetz}. For instance, one could consider black hole solutions in higher derivative theories of gravity that are not locally maximally symmetric and check whether again the the near horizon symmetries turn out to be governed by $\hat u(1)$ current algebras, and also if entropy again is given by the simple result \eqref{eq:lalapetz}. Another interesting generalization is to consider supersymmetry or flat space higher spin gravity \cite{Afshar:2013vka, Gonzalez:2013oaa}. Moreover, it would also be of interest to try to extract the extremal limit from our non-extremal boundary conditions and to compare with the vast literature on microscopic state counting for extremal black holes.

A potentially highly rewarding direction of future research is to generalize the construction presented here to higher dimensions and specifically to four dimensions. Although the use of Chern--Simons theory is limited to three dimensions, we expect the appearance of soft conserved charges in the near horizon limit to hold in higher dimensions as well, given the algebraic observations in section \ref{se:5.5}. This might provide a fruitful new direction to elucidate the peculiar properties of black hole physics, in particular of non-extremal Kerr black holes in general relativity.

\acknowledgments

We thank St\'ephane Detournay for collaboration at the early stages of this project.

We are grateful to Glenn Barnich, Jan de Boer, Steve Carlip, Borun Chowdhury, Laura Donnay, Oscar Fuentealba, Gaston Giribet, Hern\'an Gonz\'alez, Javier Matulich, Miguel Pino, Stefan Prohazka, Max Riegler, Jakob Salzer, Friedrich Sch\"oller, Shahin Sheikh-Jabbari and Cedric Troessaert for discussions.
 
We acknowledge the scientific atmosphere of the workshops ``Topics in three dimensional Gravity'' in March 2016 at the Abdus Salam ICTP, Trieste, Italy, the workshop ``Flat Holography'' in April 2016 at the Simons Center, Stony Brook, New York, the ERC and Solvay Workshop ``Holography for Black Holes and Cosmology'' at Brussels University in May 2016, and the MIAPP programme ``Higher-Spin Theory and Duality'' in May 2016 at the Munich Institute for Astro- and Particle Physics, Germany. HA wishes to thank CERN TH-division for hospitality and support during his research visit in September 2016 when this project was completed. DG also acknowledges the hospitality at CECS in February and November 2016, where this project was started and finished, respectively.

HA was supported by the Iranian National Science Foundation (INSF). DG was supported by the Austrian Science Fund (FWF), projects P~27182-N27 and P~28751-N27, and by CECs. WM was supported by the FWF project P~27182-N27. The work of AP, DT and RT is partially funded by Fondecyt grants 11130260, 11130262, 1130658, 1161311. The Centro de Estudios Cient\'ificos (CECs) is funded by the Chilean Government through the Centers of Excellence Base Financing Program of Conicyt.

\appendix

\begin{widetext}

\section{Metric in Eddington--Finkelstein gauge}\label{app:A}

In \cite{Afshar:2016wfy} we presented the locally AdS$_3$ metric in Eddington--Finkelstein gauge for constant Rindler acceleration, $a$ and vanishing rotation parameter $\Omega=0$. Here we generalize these results in the same gauge to arbitrary functions $a(v,\vp)$, $\Om(v,\vp)$.

We start with the connection \eqref{eq:r7} with the group element
\eq{
b_{\pm}=\exp\left(\pm\frac{1}{\ell\zeta^{\pm}}L_{1}\right)\exp\left(\pm\frac{\rho}{2\ell}L_{-1}\right)
}{eq:app1}
and the same choice for $\frak a^\pm$ as in the main text, see \eqref{eq:r21} but with $t$ replaced with $v$. The full connection then is given by
\begin{multline}
A^{\pm}=\frak{a}^{\pm}\pm\left(\frac{\extd\rho}{2\ell}L_{-1}-\frac{\extd\zeta^{\pm}}{\ell{\zeta^{\pm}}^{2}}L_{1}\right)-\left(\pm\frac{\extd v}{\ell} +\frac{\mathcal{J}_{\pm}}{\ell\zeta^{\pm}}\extd\varphi\right)L_{1}
+\frac{\rho}{2\ell}\left(\pm\zeta^{\pm}\extd v+\mathcal{J}_{\pm}\extd\varphi\right)L_{-1}-\frac{\rho}{\ell}\left(\frac{\extd v}{\ell}\pm\frac{\mathcal{J}_{\pm}}{\ell\zeta^{\pm}}\extd\varphi+\frac{\extd\zeta^{\pm}}{\ell{\zeta^{\pm}}^{2}}\right)L_{0} \\
-\frac{\rho^{2}}{4\ell^2}\left(\pm\frac{\extd v}{\ell}+\frac{\mathcal{J}_{\pm}}{\ell\zeta^{\pm}}\extd\varphi\pm\frac{\extd\zeta^{\pm}}{\ell{\zeta^{\pm}}^{2}}\right)L_{-1}
\label{eq:bigAsm}
\end{multline}
where the second term comes from $b_{\pm}^{-1}\extd b_{\pm}$ and the rest from applying the Baker--Campbell--Hausdorff formula, $b_{\pm}^{-1}\frak{a}^{\pm}b_{\pm}=\frak{a}^{\pm}+\dots$, and the ellipsis denotes single and double commutator terms (after two commutators the otherwise infinite series truncates). 

The most general solution of the Einstein equations \eqref{eq:EEqs} obeying our boundary conditions in Eddington--Finkelstein gauge reads 
\eq{
\extd s^{2} =\left(\Omega^{2}-2a\rho f_{1}\right)\,\extd v^{2}+2f_{2}\extd v\extd\rho+2\big(\Omega\gamma+2\tilde{\omega}\rho 
  h_{2}\big)\,\extd v\extd\varphi - 2\frac{\tilde{\omega}}{a}h_{1}\extd\varphi\extd\rho + \Big[\gamma^2+\frac{2\rho}{a}\Big(\frac{\gamma^2}{\ell^2}-\tilde{\omega}^{2}\big(1-\frac{\Omega}{a\ell}\big)^2\Big)h_3\Big] \,\extd\varphi^{2}
}{eq:r5}
We display below the functions $f_{1},f_{2}$ and $h_{1},h_{2},h_{3}$ appearing in the locally AdS$_{3}$ line-element \eqref{eq:r5}. 
\begin{align}
f_{1} & =f_{2}+\frac{\Omega}{a\ell}\dot{W}+F_{+}F_{-}\,\frac{\rho}{2a\ell^{2}}\label{eq:r17}\\
f_{2} & =\frac{F_{+}+F_{-}}{2}\\
h_{1} & =1-\frac{a}{\tilde{\omega}}A^{\prime}\\
h_{2} & =\frac{h_{1}+f_{2}}{2}-\frac{\Omega}{2\tilde{\omega}\ell}\Big(W^{\prime}+\frac{a\gamma+\tilde{\omega}\Omega}{a^{2}\ell}\dot{A}-\frac{\gamma}{\Omega}\dot{W}\Big)+\Big[h_{1}f_{2}+\frac{a}{\tilde{\omega}}\dot{W}\Big(W^{\prime}-\frac{a\gamma+\tilde{\omega}\Omega}{a^{2}\ell}\Big)\Big]\frac{\rho}{2a\ell^{2}}
\\
h_{3} & =\frac{H_{+}+H_{-}}{2}-\frac{\Omega(H_{+}-H_{-})}{2a\ell}+(H_{+}H_{-})\,\frac{\rho}{2a\ell^{2}}
\end{align}
Additionally we have defined 
\begin{align}
\tilde{\omega} & =\frac{\omega a^{2}+\gamma a\Omega\ell^{-2}}{a^{2}-\Omega^{2}\ell^{-2}}\label{eq:omega}\\
F_{\pm} & =1-\frac{\dot{a}\pm\dot{\Omega}\ell^{-1}}{(a\pm\Omega\ell^{-1})^{2}}\\
A & =\frac{a}{a^{2}-\Omega^{2}\ell^{-2}}\\
W & =\frac{\Omega\ell^{-1}}{a^{2}-\Omega^{2}\ell^{-2}}\\
H_{\pm} & =1+\frac{(\pm a^{\prime}-\Omega^{\prime}\ell^{-1})a^{2}}{(a\mp\Omega\ell^{-1})^{2}[(\pm a+\Omega\ell^{-1})\tilde{\omega}+a\gamma\ell^{-1}]}\,.
\end{align}
The quantities $f_{i}$ depend only on $a$, $\Omega$, while $h_{i}$ depend additionally on $\gamma$ and $\omega$. Prime (dot) denotes $\partial_{\vp}$ ($\partial_{v}$). 

The chemical potential $\Omega$ generates a rotating frame. If it vanishes, $\Omega=0$, we get considerable simplifications in the functions appearing in the line-element \eqref{eq:r5}. 
\begin{align}
f_{1}\big|_{\Omega=0} & =f_{2}+f_{2}^{2}\frac{\rho}{2a\ell^{2}}\\
f_{2}\big|_{\Omega=0} & =1-\frac{\dot{a}}{a^{2}}\\
h_{1}\big|_{\Omega=0} & =1+\frac{a^{\prime}}{a\omega}\\
h_{2}\big|_{\Omega=0} & =\frac{h_{1}+f_{2}}{2}+h_{1}f_{2}\,\frac{\rho}{2a\ell^{2}}\\
h_{3}\big|_{\Omega=0} & =\frac{1}{2}(H_{+}+H_{-})+(H_{+}H_{-})\,\frac{\rho}{2a\ell^{2}}
\end{align}
where $H_{\pm}=1+a^{\prime}/[a(\omega\pm\gamma\ell^{-1})]$.

If $a$ is constant it can be interpreted as Rindler acceleration. For constant chemical potentials the line-element \eqref{eq:r5} with \eqref{eq:omega} and \eqref{eq:r17} simplifies to 
\begin{multline}
\extd s^{2}\big|_{\Omega,a={\rm const.}}=\big(\Omega^{2}-2a\rho f\big)\,\extd v^{2}  + 2\extd v\extd\rho + 2\Big[\Omega\gamma+2\tilde{\omega}\rho f\Big]\extd v\extd\varphi \\
- 2\frac{\tilde{\omega}}{a}\,\extd\varphi\extd\rho + \big[\gamma^{2}+\frac{2\rho}{a\ell^{2}}\big(\gamma^{2}-\tilde{\omega}^{2}(1-\Om/(a\ell))^{2}\big)f\big]\,\extd\varphi^{2}\label{eq:r5a}
\end{multline}
where $\gamma$ and $\omega$ depend now on $\vp$, only, and all functions $f_{i}$ and $h_{i}$ above simplify, either to $f_{2}=h_{1}=1$ or to the single function $f_{1}=h_{2}=h_{3}
=1+\rho/(2a\ell^{2})$. 
If additionally $\Omega=0$ then we recover the result displayed in the main text, see \eqref{eq:Null-AdS-metric}.
\end{widetext}

\section{Killing vectors}\label{se:KV}

We consider here the six local Killing vectors for simple solutions with constant $\gamma, a$ and vanishing $\omega=\Omega=0$, describing static BTZ black holes in Eddington-Finkelstein coordinates \eqref{eq:Null-AdS-metric}. The main goal of this appendix is to find conditions for which these local Killing vectors remain well-defined globally. 

We obtain the six local Killing vectors
\begin{align}
\xi_{1} & =\partial_{v}\\
\xi_{2} & =\partial_{\vp}\displaybreak[1]\\
\xi_{3,4} & =e^{av\pm{\cal A}}\big( \partial_{v} \pm \frac{2a^{3}\ell^{3}+a^{2}\ell\rho}{\ga(a^{2}\ell^{2}+a\rho)}\,\partial_{\vp} 
 -(2a^{2}\ell^{2}+a\rho)\,\partial_{\rho}\big)\displaybreak[1]\\
\xi_{5,6} & =e^{-av\pm{\cal A}}\big(\partial_{v} \mp \frac{a^{2}\ell\rho}{\ga(a^{2}\ell^{2}+a\rho)}\,\partial_{\vp}+a\rho\,\partial_{\rho}\big)
\end{align}
where ${\cal A}=\vp\,\ga/\ell$.

If $v$ has an imaginary periodicity then these Killing vectors can be globally regular only if $v\sim v+i2\pi/a$, which is indeed the identification induced by \eqref{eq:cp}. Moreover, periodicity in the angular coordinate, $\vp\sim\vp+2\pi$, means that the Killing vectors above are globally regular only for 
\eq{ 
\gamma = \pm in\, \ell\qquad n \in \mathbb{Z}+ 
}{eq:gs} 
For $n=1$ we obtain the ground state solution, see the discussion in section \ref{se:micro}. Therefore, the ground state solution is maximally symmetric, as may be expected on general grounds. The fact that our ground state solution is gapped from the physical spectrum by an imaginary amount is interesting and has been seen already in a previous Rindleresque construction \cite{Afshar:2015wjm}.

Interestingly, only the local Killing vectors $\xi_{1}$ and $\xi_{2}$ are compatible with the fall-off conditions of asymptotic Killing vectors \eqref{eq:dggp1}. They correspond precisely to the generators of the ``wedge subalgebra'' of our symmetry algebra \eqref{eq:NHSA}.


\begin{thebibliography}{86}%
\makeatletter
\providecommand \@ifxundefined [1]{%
 \@ifx{#1\undefined}
}%
\providecommand \@ifnum [1]{%
 \ifnum #1\expandafter \@firstoftwo
 \else \expandafter \@secondoftwo
 \fi
}%
\providecommand \@ifx [1]{%
 \ifx #1\expandafter \@firstoftwo
 \else \expandafter \@secondoftwo
 \fi
}%
\providecommand \natexlab [1]{#1}%
\providecommand \enquote  [1]{``#1''}%
\providecommand \bibnamefont  [1]{#1}%
\providecommand \bibfnamefont [1]{#1}%
\providecommand \citenamefont [1]{#1}%
\providecommand \href@noop [0]{\@secondoftwo}%
\providecommand \href [0]{\begingroup \@sanitize@url \@href}%
\providecommand \@href[1]{\@@startlink{#1}\@@href}%
\providecommand \@@href[1]{\endgroup#1\@@endlink}%
\providecommand \@sanitize@url [0]{\catcode `\\12\catcode `\$12\catcode
  `\&12\catcode `\#12\catcode `\^12\catcode `\_12\catcode `\%12\relax}%
\providecommand \@@startlink[1]{}%
\providecommand \@@endlink[0]{}%
\providecommand \url  [0]{\begingroup\@sanitize@url \@url }%
\providecommand \@url [1]{\endgroup\@href {#1}{\urlprefix }}%
\providecommand \urlprefix  [0]{URL }%
\providecommand \Eprint [0]{\href }%
\providecommand \doibase [0]{http://dx.doi.org/}%
\providecommand \selectlanguage [0]{\@gobble}%
\providecommand \bibinfo  [0]{\@secondoftwo}%
\providecommand \bibfield  [0]{\@secondoftwo}%
\providecommand \translation [1]{[#1]}%
\providecommand \BibitemOpen [0]{}%
\providecommand \bibitemStop [0]{}%
\providecommand \bibitemNoStop [0]{.\EOS\space}%
\providecommand \EOS [0]{\spacefactor3000\relax}%
\providecommand \BibitemShut  [1]{\csname bibitem#1\endcsname}%
\let\auto@bib@innerbib\@empty
\bibitem [{\citenamefont {Donnay}\ \emph
  {et~al.}(2016{\natexlab{a}})\citenamefont {Donnay}, \citenamefont {Giribet},
  \citenamefont {Gonzalez},\ and\ \citenamefont {Pino}}]{Donnay:2015abr}%
  \BibitemOpen
  \bibfield  {author} {\bibinfo {author} {\bibfnamefont {L.}~\bibnamefont
  {Donnay}}, \bibinfo {author} {\bibfnamefont {G.}~\bibnamefont {Giribet}},
  \bibinfo {author} {\bibfnamefont {H.~A.}\ \bibnamefont {Gonzalez}}, \ and\
  \bibinfo {author} {\bibfnamefont {M.}~\bibnamefont {Pino}},\ }\href {\doibase
  10.1103/PhysRevLett.116.091101} {\bibfield  {journal} {\bibinfo  {journal}
  {Phys. Rev. Lett.}\ }\textbf {\bibinfo {volume} {116}},\ \bibinfo {pages}
  {091101} (\bibinfo {year} {2016}{\natexlab{a}})},\ \Eprint
  {http://arxiv.org/abs/1511.08687} {arXiv:1511.08687 [hep-th]} \BibitemShut
  {NoStop}%
\bibitem [{\citenamefont {Rindler}(1966)}]{Rindler:1966zz}%
  \BibitemOpen
  \bibfield  {author} {\bibinfo {author} {\bibfnamefont {W.}~\bibnamefont
  {Rindler}},\ }\href {\doibase 10.1119/1.1972547} {\bibfield  {journal}
  {\bibinfo  {journal} {Am.J.Phys.}\ }\textbf {\bibinfo {volume} {34}},\
  \bibinfo {pages} {1174} (\bibinfo {year} {1966})}\BibitemShut {NoStop}%
\bibitem [{\citenamefont {Carlip}(1999)}]{Carlip:1999cy}%
  \BibitemOpen
  \bibfield  {author} {\bibinfo {author} {\bibfnamefont {S.}~\bibnamefont
  {Carlip}},\ }\href@noop {} {\bibfield  {journal} {\bibinfo  {journal} {Class.
  Quant. Grav.}\ }\textbf {\bibinfo {volume} {16}},\ \bibinfo {pages} {3327}
  (\bibinfo {year} {1999})},\ \Eprint {http://arxiv.org/abs/gr-qc/9906126}
  {gr-qc/9906126} \BibitemShut {NoStop}%
\bibitem [{\citenamefont {Afshar}\ \emph
  {et~al.}(2016{\natexlab{a}})\citenamefont {Afshar}, \citenamefont
  {Detournay}, \citenamefont {Grumiller}, \citenamefont {Merbis}, \citenamefont
  {Perez}, \citenamefont {Tempo},\ and\ \citenamefont
  {Troncoso}}]{Afshar:2016wfy}%
  \BibitemOpen
  \bibfield  {author} {\bibinfo {author} {\bibfnamefont {H.}~\bibnamefont
  {Afshar}}, \bibinfo {author} {\bibfnamefont {S.}~\bibnamefont {Detournay}},
  \bibinfo {author} {\bibfnamefont {D.}~\bibnamefont {Grumiller}}, \bibinfo
  {author} {\bibfnamefont {W.}~\bibnamefont {Merbis}}, \bibinfo {author}
  {\bibfnamefont {A.}~\bibnamefont {Perez}}, \bibinfo {author} {\bibfnamefont
  {D.}~\bibnamefont {Tempo}}, \ and\ \bibinfo {author} {\bibfnamefont
  {R.}~\bibnamefont {Troncoso}},\ }\href {\doibase 10.1103/PhysRevD.93.101503}
  {\bibfield  {journal} {\bibinfo  {journal} {Phys. Rev.}\ }\textbf {\bibinfo
  {volume} {D93}},\ \bibinfo {pages} {101503} (\bibinfo {year}
  {2016}{\natexlab{a}})},\ \Eprint {http://arxiv.org/abs/1603.04824}
  {arXiv:1603.04824 [hep-th]} \BibitemShut {NoStop}%
\bibitem [{\citenamefont {Hawking}\ \emph
  {et~al.}(2016{\natexlab{a}})\citenamefont {Hawking}, \citenamefont {Perry},\
  and\ \citenamefont {Strominger}}]{Hawking:2016msc}%
  \BibitemOpen
  \bibfield  {author} {\bibinfo {author} {\bibfnamefont {S.~W.}\ \bibnamefont
  {Hawking}}, \bibinfo {author} {\bibfnamefont {M.~J.}\ \bibnamefont {Perry}},
  \ and\ \bibinfo {author} {\bibfnamefont {A.}~\bibnamefont {Strominger}},\
  }\href {\doibase 10.1103/PhysRevLett.116.231301} {\bibfield  {journal}
  {\bibinfo  {journal} {Phys. Rev. Lett.}\ }\textbf {\bibinfo {volume} {116}},\
  \bibinfo {pages} {231301} (\bibinfo {year} {2016}{\natexlab{a}})},\ \Eprint
  {http://arxiv.org/abs/1601.00921} {arXiv:1601.00921 [hep-th]} \BibitemShut
  {NoStop}%
\bibitem [{\citenamefont {Hawking}\ \emph
  {et~al.}(2016{\natexlab{b}})\citenamefont {Hawking}, \citenamefont {Perry},\
  and\ \citenamefont {Strominger}}]{Hawking:2016sgy}%
  \BibitemOpen
  \bibfield  {author} {\bibinfo {author} {\bibfnamefont {S.~W.}\ \bibnamefont
  {Hawking}}, \bibinfo {author} {\bibfnamefont {M.~J.}\ \bibnamefont {Perry}},
  \ and\ \bibinfo {author} {\bibfnamefont {A.}~\bibnamefont {Strominger}},\
  }\href@noop {} {\  (\bibinfo {year} {2016}{\natexlab{b}})},\ \Eprint
  {http://arxiv.org/abs/1611.09175} {arXiv:1611.09175 [hep-th]} \BibitemShut
  {NoStop}%
\bibitem [{\citenamefont {Grumiller}\ \emph {et~al.}(2016)\citenamefont
  {Grumiller}, \citenamefont {Perez}, \citenamefont {Prohazka}, \citenamefont
  {Tempo},\ and\ \citenamefont {Troncoso}}]{Grumiller:2016kcp}%
  \BibitemOpen
  \bibfield  {author} {\bibinfo {author} {\bibfnamefont {D.}~\bibnamefont
  {Grumiller}}, \bibinfo {author} {\bibfnamefont {A.}~\bibnamefont {Perez}},
  \bibinfo {author} {\bibfnamefont {S.}~\bibnamefont {Prohazka}}, \bibinfo
  {author} {\bibfnamefont {D.}~\bibnamefont {Tempo}}, \ and\ \bibinfo {author}
  {\bibfnamefont {R.}~\bibnamefont {Troncoso}},\ }\href {\doibase
  10.1007/JHEP10(2016)119} {\bibfield  {journal} {\bibinfo  {journal} {JHEP}\
  }\textbf {\bibinfo {volume} {10}},\ \bibinfo {pages} {119} (\bibinfo {year}
  {2016})},\ \Eprint {http://arxiv.org/abs/1607.05360} {arXiv:1607.05360
  [hep-th]} \BibitemShut {NoStop}%
\bibitem [{\citenamefont {Barnich}\ \emph {et~al.}(2016)\citenamefont
  {Barnich}, \citenamefont {Troessaert}, \citenamefont {Tempo},\ and\
  \citenamefont {Troncoso}}]{Barnich:2015dvt}%
  \BibitemOpen
  \bibfield  {author} {\bibinfo {author} {\bibfnamefont {G.}~\bibnamefont
  {Barnich}}, \bibinfo {author} {\bibfnamefont {C.}~\bibnamefont {Troessaert}},
  \bibinfo {author} {\bibfnamefont {D.}~\bibnamefont {Tempo}}, \ and\ \bibinfo
  {author} {\bibfnamefont {R.}~\bibnamefont {Troncoso}},\ }\href {\doibase
  10.1103/PhysRevD.93.084001} {\bibfield  {journal} {\bibinfo  {journal} {Phys.
  Rev.}\ }\textbf {\bibinfo {volume} {D93}},\ \bibinfo {pages} {084001}
  (\bibinfo {year} {2016})},\ \Eprint {http://arxiv.org/abs/1512.05410}
  {arXiv:1512.05410 [hep-th]} \BibitemShut {NoStop}%
\bibitem [{\citenamefont {Ba\~nados}\ \emph {et~al.}(1992)\citenamefont
  {Ba\~nados}, \citenamefont {Teitelboim},\ and\ \citenamefont
  {Zanelli}}]{Banados:1992wn}%
  \BibitemOpen
  \bibfield  {author} {\bibinfo {author} {\bibfnamefont {M.}~\bibnamefont
  {Ba\~nados}}, \bibinfo {author} {\bibfnamefont {C.}~\bibnamefont
  {Teitelboim}}, \ and\ \bibinfo {author} {\bibfnamefont {J.}~\bibnamefont
  {Zanelli}},\ }\href@noop {} {\bibfield  {journal} {\bibinfo  {journal} {Phys.
  Rev. Lett.}\ }\textbf {\bibinfo {volume} {69}},\ \bibinfo {pages} {1849}
  (\bibinfo {year} {1992})},\ \Eprint {http://arxiv.org/abs/hep-th/9204099}
  {hep-th/9204099} \BibitemShut {NoStop}%
\bibitem [{\citenamefont {Ba\~nados}\ \emph {et~al.}(1993)\citenamefont
  {Ba\~nados}, \citenamefont {Henneaux}, \citenamefont {Teitelboim},\ and\
  \citenamefont {Zanelli}}]{Banados:1992gq}%
  \BibitemOpen
  \bibfield  {author} {\bibinfo {author} {\bibfnamefont {M.}~\bibnamefont
  {Ba\~nados}}, \bibinfo {author} {\bibfnamefont {M.}~\bibnamefont {Henneaux}},
  \bibinfo {author} {\bibfnamefont {C.}~\bibnamefont {Teitelboim}}, \ and\
  \bibinfo {author} {\bibfnamefont {J.}~\bibnamefont {Zanelli}},\ }\href@noop
  {} {\bibfield  {journal} {\bibinfo  {journal} {Phys. Rev.}\ }\textbf
  {\bibinfo {volume} {D48}},\ \bibinfo {pages} {1506} (\bibinfo {year}
  {1993})},\ \Eprint {http://arxiv.org/abs/gr-qc/9302012} {gr-qc/9302012}
  \BibitemShut {NoStop}%
\bibitem [{\citenamefont {Achucarro}\ and\ \citenamefont
  {Townsend}(1986)}]{Achucarro:1987vz}%
  \BibitemOpen
  \bibfield  {author} {\bibinfo {author} {\bibfnamefont {A.}~\bibnamefont
  {Achucarro}}\ and\ \bibinfo {author} {\bibfnamefont {P.~K.}\ \bibnamefont
  {Townsend}},\ }\href {\doibase 10.1016/0370-2693(86)90140-1} {\bibfield
  {journal} {\bibinfo  {journal} {Phys. Lett.}\ }\textbf {\bibinfo {volume}
  {B180}},\ \bibinfo {pages} {89} (\bibinfo {year} {1986})}\BibitemShut
  {NoStop}%
\bibitem [{\citenamefont {Witten}(1988)}]{Witten:1988hc}%
  \BibitemOpen
  \bibfield  {author} {\bibinfo {author} {\bibfnamefont {E.}~\bibnamefont
  {Witten}},\ }\href {\doibase 10.1016/0550-3213(88)90143-5} {\bibfield
  {journal} {\bibinfo  {journal} {Nucl. Phys.}\ }\textbf {\bibinfo {volume}
  {B311}},\ \bibinfo {pages} {46} (\bibinfo {year} {1988})}\BibitemShut
  {NoStop}%
\bibitem [{\citenamefont {Brown}\ and\ \citenamefont
  {Henneaux}(1986)}]{Brown:1986nw}%
  \BibitemOpen
  \bibfield  {author} {\bibinfo {author} {\bibfnamefont {J.~D.}\ \bibnamefont
  {Brown}}\ and\ \bibinfo {author} {\bibfnamefont {M.}~\bibnamefont
  {Henneaux}},\ }\href@noop {} {\bibfield  {journal} {\bibinfo  {journal}
  {Commun. Math. Phys.}\ }\textbf {\bibinfo {volume} {104}},\ \bibinfo {pages}
  {207} (\bibinfo {year} {1986})}\BibitemShut {NoStop}%
\bibitem [{\citenamefont {Comp{\`e}re}\ \emph {et~al.}(2013)\citenamefont
  {Comp{\`e}re}, \citenamefont {Song},\ and\ \citenamefont
  {Strominger}}]{Compere:2013bya}%
  \BibitemOpen
  \bibfield  {author} {\bibinfo {author} {\bibfnamefont {G.}~\bibnamefont
  {Comp{\`e}re}}, \bibinfo {author} {\bibfnamefont {W.}~\bibnamefont {Song}}, \
  and\ \bibinfo {author} {\bibfnamefont {A.}~\bibnamefont {Strominger}},\
  }\href {\doibase 10.1007/JHEP05(2013)152} {\bibfield  {journal} {\bibinfo
  {journal} {JHEP}\ }\textbf {\bibinfo {volume} {1305}},\ \bibinfo {pages}
  {152} (\bibinfo {year} {2013})},\ \Eprint {http://arxiv.org/abs/1303.2662}
  {arXiv:1303.2662 [hep-th]} \BibitemShut {NoStop}%
\bibitem [{\citenamefont {P{\'e}rez}\ \emph {et~al.}(2016)\citenamefont
  {P{\'e}rez}, \citenamefont {Tempo},\ and\ \citenamefont
  {Troncoso}}]{Perez:2016vqo}%
  \BibitemOpen
  \bibfield  {author} {\bibinfo {author} {\bibfnamefont {A.}~\bibnamefont
  {P{\'e}rez}}, \bibinfo {author} {\bibfnamefont {D.}~\bibnamefont {Tempo}}, \
  and\ \bibinfo {author} {\bibfnamefont {R.}~\bibnamefont {Troncoso}},\ }\href
  {\doibase 10.1007/JHEP06(2016)103} {\bibfield  {journal} {\bibinfo  {journal}
  {JHEP}\ }\textbf {\bibinfo {volume} {06}},\ \bibinfo {pages} {103} (\bibinfo
  {year} {2016})},\ \Eprint {http://arxiv.org/abs/1605.04490} {arXiv:1605.04490
  [hep-th]} \BibitemShut {NoStop}%
\bibitem [{\citenamefont {Troessaert}(2013)}]{Troessaert:2013fma}%
  \BibitemOpen
  \bibfield  {author} {\bibinfo {author} {\bibfnamefont {C.}~\bibnamefont
  {Troessaert}},\ }\href {\doibase 10.1007/JHEP08(2013)044} {\bibfield
  {journal} {\bibinfo  {journal} {JHEP}\ }\textbf {\bibinfo {volume} {08}},\
  \bibinfo {pages} {044} (\bibinfo {year} {2013})},\ \Eprint
  {http://arxiv.org/abs/1303.3296} {arXiv:1303.3296 [hep-th]} \BibitemShut
  {NoStop}%
\bibitem [{\citenamefont {Grumiller}\ and\ \citenamefont
  {Riegler}(2016)}]{Grumiller:2016pqb}%
  \BibitemOpen
  \bibfield  {author} {\bibinfo {author} {\bibfnamefont {D.}~\bibnamefont
  {Grumiller}}\ and\ \bibinfo {author} {\bibfnamefont {M.}~\bibnamefont
  {Riegler}},\ }\href {\doibase 10.1007/JHEP10(2016)023} {\bibfield  {journal}
  {\bibinfo  {journal} {JHEP}\ }\textbf {\bibinfo {volume} {10}},\ \bibinfo
  {pages} {023} (\bibinfo {year} {2016})},\ \Eprint
  {http://arxiv.org/abs/1608.01308} {arXiv:1608.01308 [hep-th]} \BibitemShut
  {NoStop}%
\bibitem [{\citenamefont {Henneaux}\ \emph {et~al.}(2002)\citenamefont
  {Henneaux}, \citenamefont {Martinez}, \citenamefont {Troncoso},\ and\
  \citenamefont {Zanelli}}]{Henneaux:2002wm}%
  \BibitemOpen
  \bibfield  {author} {\bibinfo {author} {\bibfnamefont {M.}~\bibnamefont
  {Henneaux}}, \bibinfo {author} {\bibfnamefont {C.}~\bibnamefont {Martinez}},
  \bibinfo {author} {\bibfnamefont {R.}~\bibnamefont {Troncoso}}, \ and\
  \bibinfo {author} {\bibfnamefont {J.}~\bibnamefont {Zanelli}},\ }\href
  {\doibase 10.1103/PhysRevD.65.104007} {\bibfield  {journal} {\bibinfo
  {journal} {Phys. Rev.}\ }\textbf {\bibinfo {volume} {D65}},\ \bibinfo {pages}
  {104007} (\bibinfo {year} {2002})},\ \Eprint
  {http://arxiv.org/abs/hep-th/0201170} {arXiv:hep-th/0201170} \BibitemShut
  {NoStop}%
\bibitem [{\citenamefont {Henneaux}\ \emph {et~al.}(2004)\citenamefont
  {Henneaux}, \citenamefont {Martinez}, \citenamefont {Troncoso},\ and\
  \citenamefont {Zanelli}}]{Henneaux:2004zi}%
  \BibitemOpen
  \bibfield  {author} {\bibinfo {author} {\bibfnamefont {M.}~\bibnamefont
  {Henneaux}}, \bibinfo {author} {\bibfnamefont {C.}~\bibnamefont {Martinez}},
  \bibinfo {author} {\bibfnamefont {R.}~\bibnamefont {Troncoso}}, \ and\
  \bibinfo {author} {\bibfnamefont {J.}~\bibnamefont {Zanelli}},\ }\href
  {\doibase 10.1103/PhysRevD.70.044034} {\bibfield  {journal} {\bibinfo
  {journal} {Phys. Rev.}\ }\textbf {\bibinfo {volume} {D70}},\ \bibinfo {pages}
  {044034} (\bibinfo {year} {2004})},\ \Eprint
  {http://arxiv.org/abs/hep-th/0404236} {arXiv:hep-th/0404236 [hep-th]}
  \BibitemShut {NoStop}%
\bibitem [{\citenamefont {Grumiller}\ and\ \citenamefont
  {Johansson}(2009)}]{Grumiller:2008es}%
  \BibitemOpen
  \bibfield  {author} {\bibinfo {author} {\bibfnamefont {D.}~\bibnamefont
  {Grumiller}}\ and\ \bibinfo {author} {\bibfnamefont {N.}~\bibnamefont
  {Johansson}},\ }\href {\doibase 10.1142/S0218271808014096} {\bibfield
  {journal} {\bibinfo  {journal} {Int. J. Mod. Phys.}\ }\textbf {\bibinfo
  {volume} {D17}},\ \bibinfo {pages} {2367} (\bibinfo {year} {2009})},\ \Eprint
  {http://arxiv.org/abs/0808.2575} {arXiv:0808.2575 [hep-th]} \BibitemShut
  {NoStop}%
\bibitem [{\citenamefont {Henneaux}\ \emph {et~al.}(2009)\citenamefont
  {Henneaux}, \citenamefont {Martinez},\ and\ \citenamefont
  {Troncoso}}]{Henneaux:2009pw}%
  \BibitemOpen
  \bibfield  {author} {\bibinfo {author} {\bibfnamefont {M.}~\bibnamefont
  {Henneaux}}, \bibinfo {author} {\bibfnamefont {C.}~\bibnamefont {Martinez}},
  \ and\ \bibinfo {author} {\bibfnamefont {R.}~\bibnamefont {Troncoso}},\
  }\href {\doibase 10.1103/PhysRevD.79.081502} {\bibfield  {journal} {\bibinfo
  {journal} {Phys. Rev.}\ }\textbf {\bibinfo {volume} {D79}},\ \bibinfo {pages}
  {081502R} (\bibinfo {year} {2009})},\ \Eprint
  {http://arxiv.org/abs/0901.2874} {arXiv:0901.2874 [hep-th]} \BibitemShut
  {NoStop}%
\bibitem [{\citenamefont {Oliva}\ \emph {et~al.}(2009)\citenamefont {Oliva},
  \citenamefont {Tempo},\ and\ \citenamefont {Troncoso}}]{Oliva:2009ip}%
  \BibitemOpen
  \bibfield  {author} {\bibinfo {author} {\bibfnamefont {J.}~\bibnamefont
  {Oliva}}, \bibinfo {author} {\bibfnamefont {D.}~\bibnamefont {Tempo}}, \ and\
  \bibinfo {author} {\bibfnamefont {R.}~\bibnamefont {Troncoso}},\ }\href
  {\doibase 10.1088/1126-6708/2009/07/011} {\bibfield  {journal} {\bibinfo
  {journal} {JHEP}\ }\textbf {\bibinfo {volume} {07}},\ \bibinfo {pages} {011}
  (\bibinfo {year} {2009})},\ \Eprint {http://arxiv.org/abs/0905.1545}
  {arXiv:0905.1545 [hep-th]} \BibitemShut {NoStop}%
\bibitem [{\citenamefont {Barnich}\ and\ \citenamefont
  {Lambert}(2013)}]{Barnich:2013sxa}%
  \BibitemOpen
  \bibfield  {author} {\bibinfo {author} {\bibfnamefont {G.}~\bibnamefont
  {Barnich}}\ and\ \bibinfo {author} {\bibfnamefont {P.-H.}\ \bibnamefont
  {Lambert}},\ }\href {\doibase 10.1103/PhysRevD.88.103006} {\bibfield
  {journal} {\bibinfo  {journal} {Phys. Rev.}\ }\textbf {\bibinfo {volume}
  {D88}},\ \bibinfo {pages} {103006} (\bibinfo {year} {2013})},\ \Eprint
  {http://arxiv.org/abs/1310.2698} {arXiv:1310.2698 [hep-th]} \BibitemShut
  {NoStop}%
\bibitem [{\citenamefont {Bunster}\ and\ \citenamefont
  {Perez}(2015)}]{Bunster:2014cna}%
  \BibitemOpen
  \bibfield  {author} {\bibinfo {author} {\bibfnamefont {C.}~\bibnamefont
  {Bunster}}\ and\ \bibinfo {author} {\bibfnamefont {A.}~\bibnamefont
  {Perez}},\ }\href {\doibase 10.1103/PhysRevD.91.024029} {\bibfield  {journal}
  {\bibinfo  {journal} {Phys. Rev.}\ }\textbf {\bibinfo {volume} {D91}},\
  \bibinfo {pages} {024029} (\bibinfo {year} {2015})},\ \Eprint
  {http://arxiv.org/abs/1412.1492} {arXiv:1412.1492 [hep-th]} \BibitemShut
  {NoStop}%
\bibitem [{\citenamefont {Perez}\ \emph {et~al.}(2016)\citenamefont {Perez},
  \citenamefont {Riquelme}, \citenamefont {Tempo},\ and\ \citenamefont
  {Troncoso}}]{Perez:2015jxn}%
  \BibitemOpen
  \bibfield  {author} {\bibinfo {author} {\bibfnamefont {A.}~\bibnamefont
  {Perez}}, \bibinfo {author} {\bibfnamefont {M.}~\bibnamefont {Riquelme}},
  \bibinfo {author} {\bibfnamefont {D.}~\bibnamefont {Tempo}}, \ and\ \bibinfo
  {author} {\bibfnamefont {R.}~\bibnamefont {Troncoso}},\ }\href {\doibase
  10.1007/JHEP02(2016)015} {\bibfield  {journal} {\bibinfo  {journal} {JHEP}\
  }\textbf {\bibinfo {volume} {02}},\ \bibinfo {pages} {015} (\bibinfo {year}
  {2016})},\ \Eprint {http://arxiv.org/abs/1512.01576} {arXiv:1512.01576
  [hep-th]} \BibitemShut {NoStop}%
\bibitem [{\citenamefont {Henneaux}\ \emph {et~al.}(2013)\citenamefont
  {Henneaux}, \citenamefont {Perez}, \citenamefont {Tempo},\ and\ \citenamefont
  {Troncoso}}]{Henneaux:2013dra}%
  \BibitemOpen
  \bibfield  {author} {\bibinfo {author} {\bibfnamefont {M.}~\bibnamefont
  {Henneaux}}, \bibinfo {author} {\bibfnamefont {A.}~\bibnamefont {Perez}},
  \bibinfo {author} {\bibfnamefont {D.}~\bibnamefont {Tempo}}, \ and\ \bibinfo
  {author} {\bibfnamefont {R.}~\bibnamefont {Troncoso}},\ }\href {\doibase
  10.1007/JHEP12(2013)048} {\bibfield  {journal} {\bibinfo  {journal} {JHEP}\
  }\textbf {\bibinfo {volume} {1312}},\ \bibinfo {pages} {048} (\bibinfo {year}
  {2013})},\ \Eprint {http://arxiv.org/abs/1309.4362} {arXiv:1309.4362
  [hep-th]} \BibitemShut {NoStop}%
\bibitem [{\citenamefont {Bunster}\ \emph {et~al.}(2014)\citenamefont
  {Bunster}, \citenamefont {Henneaux}, \citenamefont {Perez}, \citenamefont
  {Tempo},\ and\ \citenamefont {Troncoso}}]{Bunster:2014mua}%
  \BibitemOpen
  \bibfield  {author} {\bibinfo {author} {\bibfnamefont {C.}~\bibnamefont
  {Bunster}}, \bibinfo {author} {\bibfnamefont {M.}~\bibnamefont {Henneaux}},
  \bibinfo {author} {\bibfnamefont {A.}~\bibnamefont {Perez}}, \bibinfo
  {author} {\bibfnamefont {D.}~\bibnamefont {Tempo}}, \ and\ \bibinfo {author}
  {\bibfnamefont {R.}~\bibnamefont {Troncoso}},\ }\href {\doibase
  10.1007/JHEP05(2014)031} {\bibfield  {journal} {\bibinfo  {journal} {JHEP}\
  }\textbf {\bibinfo {volume} {1405}},\ \bibinfo {pages} {031} (\bibinfo {year}
  {2014})},\ \Eprint {http://arxiv.org/abs/1404.3305} {arXiv:1404.3305
  [hep-th]} \BibitemShut {NoStop}%
\bibitem [{\citenamefont {Regge}\ and\ \citenamefont
  {Teitelboim}(1974)}]{Regge:1974zd}%
  \BibitemOpen
  \bibfield  {author} {\bibinfo {author} {\bibfnamefont {T.}~\bibnamefont
  {Regge}}\ and\ \bibinfo {author} {\bibfnamefont {C.}~\bibnamefont
  {Teitelboim}},\ }\href@noop {} {\bibfield  {journal} {\bibinfo  {journal}
  {Ann. Phys.}\ }\textbf {\bibinfo {volume} {88}},\ \bibinfo {pages} {286}
  (\bibinfo {year} {1974})}\BibitemShut {NoStop}%
\bibitem [{\citenamefont {Cornalba}\ and\ \citenamefont
  {Costa}(2002)}]{Cornalba:2002fi}%
  \BibitemOpen
  \bibfield  {author} {\bibinfo {author} {\bibfnamefont {L.}~\bibnamefont
  {Cornalba}}\ and\ \bibinfo {author} {\bibfnamefont {M.~S.}\ \bibnamefont
  {Costa}},\ }\href {\doibase 10.1103/PhysRevD.66.066001} {\bibfield  {journal}
  {\bibinfo  {journal} {Phys.Rev.}\ }\textbf {\bibinfo {volume} {D66}},\
  \bibinfo {pages} {066001} (\bibinfo {year} {2002})},\ \Eprint
  {http://arxiv.org/abs/hep-th/0203031} {arXiv:hep-th/0203031 [hep-th]}
  \BibitemShut {NoStop}%
\bibitem [{\citenamefont {Cornalba}\ and\ \citenamefont
  {Costa}(2004)}]{Cornalba:2003kd}%
  \BibitemOpen
  \bibfield  {author} {\bibinfo {author} {\bibfnamefont {L.}~\bibnamefont
  {Cornalba}}\ and\ \bibinfo {author} {\bibfnamefont {M.~S.}\ \bibnamefont
  {Costa}},\ }\href {\doibase 10.1002/prop.200310123} {\bibfield  {journal}
  {\bibinfo  {journal} {Fortsch.Phys.}\ }\textbf {\bibinfo {volume} {52}},\
  \bibinfo {pages} {145} (\bibinfo {year} {2004})},\ \Eprint
  {http://arxiv.org/abs/hep-th/0310099} {arXiv:hep-th/0310099 [hep-th]}
  \BibitemShut {NoStop}%
\bibitem [{\citenamefont {Barnich}\ and\ \citenamefont
  {Gonzalez}(2013)}]{Barnich:2013yka}%
  \BibitemOpen
  \bibfield  {author} {\bibinfo {author} {\bibfnamefont {G.}~\bibnamefont
  {Barnich}}\ and\ \bibinfo {author} {\bibfnamefont {H.~A.}\ \bibnamefont
  {Gonzalez}},\ }\href {\doibase 10.1007/JHEP05(2013)016} {\bibfield  {journal}
  {\bibinfo  {journal} {JHEP}\ }\textbf {\bibinfo {volume} {1305}},\ \bibinfo
  {pages} {016} (\bibinfo {year} {2013})},\ \Eprint
  {http://arxiv.org/abs/1303.1075} {arXiv:1303.1075 [hep-th]} \BibitemShut
  {NoStop}%
\bibitem [{\citenamefont {Afshar}(2013)}]{Afshar:2013bla}%
  \BibitemOpen
  \bibfield  {author} {\bibinfo {author} {\bibfnamefont {H.~R.}\ \bibnamefont
  {Afshar}},\ }\href {\doibase 10.1007/JHEP10(2013)027} {\bibfield  {journal}
  {\bibinfo  {journal} {JHEP}\ }\textbf {\bibinfo {volume} {1310}},\ \bibinfo
  {pages} {027} (\bibinfo {year} {2013})},\ \Eprint
  {http://arxiv.org/abs/1307.4855} {arXiv:1307.4855 [hep-th]} \BibitemShut
  {NoStop}%
\bibitem [{\citenamefont {Afshar}\ \emph
  {et~al.}(2016{\natexlab{b}})\citenamefont {Afshar}, \citenamefont
  {Detournay}, \citenamefont {Grumiller},\ and\ \citenamefont
  {Oblak}}]{Afshar:2015wjm}%
  \BibitemOpen
  \bibfield  {author} {\bibinfo {author} {\bibfnamefont {H.}~\bibnamefont
  {Afshar}}, \bibinfo {author} {\bibfnamefont {S.}~\bibnamefont {Detournay}},
  \bibinfo {author} {\bibfnamefont {D.}~\bibnamefont {Grumiller}}, \ and\
  \bibinfo {author} {\bibfnamefont {B.}~\bibnamefont {Oblak}},\ }\href
  {\doibase 10.1007/JHEP03(2016)187} {\bibfield  {journal} {\bibinfo  {journal}
  {JHEP}\ }\textbf {\bibinfo {volume} {03}},\ \bibinfo {pages} {187} (\bibinfo
  {year} {2016}{\natexlab{b}})},\ \Eprint {http://arxiv.org/abs/1512.08233}
  {arXiv:1512.08233 [hep-th]} \BibitemShut {NoStop}%
\bibitem [{\citenamefont {Barnich}\ and\ \citenamefont
  {Oblak}(2014)}]{Barnich:2014kra}%
  \BibitemOpen
  \bibfield  {author} {\bibinfo {author} {\bibfnamefont {G.}~\bibnamefont
  {Barnich}}\ and\ \bibinfo {author} {\bibfnamefont {B.}~\bibnamefont
  {Oblak}},\ }\href {\doibase 10.1007/JHEP06(2014)129} {\bibfield  {journal}
  {\bibinfo  {journal} {JHEP}\ }\textbf {\bibinfo {volume} {1406}},\ \bibinfo
  {pages} {129} (\bibinfo {year} {2014})},\ \Eprint
  {http://arxiv.org/abs/1403.5803} {arXiv:1403.5803 [hep-th]} \BibitemShut
  {NoStop}%
\bibitem [{\citenamefont {Campoleoni}\ \emph
  {et~al.}(2016{\natexlab{a}})\citenamefont {Campoleoni}, \citenamefont
  {Gonzalez}, \citenamefont {Oblak},\ and\ \citenamefont
  {Riegler}}]{Campoleoni:2015qrh}%
  \BibitemOpen
  \bibfield  {author} {\bibinfo {author} {\bibfnamefont {A.}~\bibnamefont
  {Campoleoni}}, \bibinfo {author} {\bibfnamefont {H.~A.}\ \bibnamefont
  {Gonzalez}}, \bibinfo {author} {\bibfnamefont {B.}~\bibnamefont {Oblak}}, \
  and\ \bibinfo {author} {\bibfnamefont {M.}~\bibnamefont {Riegler}},\ }\href
  {\doibase 10.1007/JHEP04(2016)034} {\bibfield  {journal} {\bibinfo  {journal}
  {JHEP}\ }\textbf {\bibinfo {volume} {04}},\ \bibinfo {pages} {034} (\bibinfo
  {year} {2016}{\natexlab{a}})},\ \Eprint {http://arxiv.org/abs/1512.03353}
  {arXiv:1512.03353 [hep-th]} \BibitemShut {NoStop}%
\bibitem [{\citenamefont {Campoleoni}\ \emph
  {et~al.}(2016{\natexlab{b}})\citenamefont {Campoleoni}, \citenamefont
  {Gonzalez}, \citenamefont {Oblak},\ and\ \citenamefont
  {Riegler}}]{Campoleoni:2016vsh}%
  \BibitemOpen
  \bibfield  {author} {\bibinfo {author} {\bibfnamefont {A.}~\bibnamefont
  {Campoleoni}}, \bibinfo {author} {\bibfnamefont {H.~A.}\ \bibnamefont
  {Gonzalez}}, \bibinfo {author} {\bibfnamefont {B.}~\bibnamefont {Oblak}}, \
  and\ \bibinfo {author} {\bibfnamefont {M.}~\bibnamefont {Riegler}},\
  }\bibfield  {booktitle} {\emph {\bibinfo {booktitle} {{Proceedings,
  International Workshop on Higher Spin Gauge Theories: Singapore, Singapore,
  November 4-6, 2015}}},\ }\href {\doibase 10.1142/S0217751X16500688,
  10.1142/9789813144101_0011} {\bibfield  {journal} {\bibinfo  {journal} {Int.
  J. Mod. Phys.}\ }\textbf {\bibinfo {volume} {A31}},\ \bibinfo {pages}
  {1650068} (\bibinfo {year} {2016}{\natexlab{b}})},\ \Eprint
  {http://arxiv.org/abs/1603.03812} {arXiv:1603.03812 [hep-th]} \BibitemShut
  {NoStop}%
\bibitem [{\citenamefont {Oblak}(2016)}]{Oblak:2016eij}%
  \BibitemOpen
  \bibfield  {author} {\bibinfo {author} {\bibfnamefont {B.}~\bibnamefont
  {Oblak}},\ }\href@noop {} {\  (\bibinfo {year} {2016})},\ \Eprint
  {http://arxiv.org/abs/1610.08526} {arXiv:1610.08526 [hep-th]} \BibitemShut
  {NoStop}%
\bibitem [{\citenamefont {Barnich}\ and\ \citenamefont
  {Comp{\`e}re}(2007)}]{Barnich:2006av}%
  \BibitemOpen
  \bibfield  {author} {\bibinfo {author} {\bibfnamefont {G.}~\bibnamefont
  {Barnich}}\ and\ \bibinfo {author} {\bibfnamefont {G.}~\bibnamefont
  {Comp{\`e}re}},\ }\href {\doibase 10.1088/0264-9381/24/5/F01,
  10.1088/0264-9381/24/11/C01} {\bibfield  {journal} {\bibinfo  {journal}
  {Class.Quant.Grav.}\ }\textbf {\bibinfo {volume} {24}},\ \bibinfo {pages}
  {F15} (\bibinfo {year} {2007})},\ \Eprint
  {http://arxiv.org/abs/gr-qc/0610130} {arXiv:gr-qc/0610130 [gr-qc]}
  \BibitemShut {NoStop}%
\bibitem [{\citenamefont {Gary}\ \emph {et~al.}(2015)\citenamefont {Gary},
  \citenamefont {Grumiller}, \citenamefont {Riegler},\ and\ \citenamefont
  {Rosseel}}]{Gary:2014ppa}%
  \BibitemOpen
  \bibfield  {author} {\bibinfo {author} {\bibfnamefont {M.}~\bibnamefont
  {Gary}}, \bibinfo {author} {\bibfnamefont {D.}~\bibnamefont {Grumiller}},
  \bibinfo {author} {\bibfnamefont {M.}~\bibnamefont {Riegler}}, \ and\
  \bibinfo {author} {\bibfnamefont {J.}~\bibnamefont {Rosseel}},\ }\href
  {\doibase 10.1007/JHEP01(2015)152} {\bibfield  {journal} {\bibinfo  {journal}
  {JHEP}\ }\textbf {\bibinfo {volume} {1501}},\ \bibinfo {pages} {152}
  (\bibinfo {year} {2015})},\ \Eprint {http://arxiv.org/abs/1411.3728}
  {arXiv:1411.3728 [hep-th]} \BibitemShut {NoStop}%
\bibitem [{\citenamefont {Matulich}\ \emph {et~al.}(2015)\citenamefont
  {Matulich}, \citenamefont {Perez}, \citenamefont {Tempo},\ and\ \citenamefont
  {Troncoso}}]{Matulich:2014hea}%
  \BibitemOpen
  \bibfield  {author} {\bibinfo {author} {\bibfnamefont {J.}~\bibnamefont
  {Matulich}}, \bibinfo {author} {\bibfnamefont {A.}~\bibnamefont {Perez}},
  \bibinfo {author} {\bibfnamefont {D.}~\bibnamefont {Tempo}}, \ and\ \bibinfo
  {author} {\bibfnamefont {R.}~\bibnamefont {Troncoso}},\ }\href {\doibase
  10.1007/JHEP05(2015)025} {\bibfield  {journal} {\bibinfo  {journal} {JHEP}\
  }\textbf {\bibinfo {volume} {05}},\ \bibinfo {pages} {025} (\bibinfo {year}
  {2015})},\ \Eprint {http://arxiv.org/abs/1412.1464} {arXiv:1412.1464
  [hep-th]} \BibitemShut {NoStop}%
\bibitem [{\citenamefont {Afshar}\ \emph
  {et~al.}(2016{\natexlab{c}})\citenamefont {Afshar}, \citenamefont
  {Grumiller},\ and\ \citenamefont {Sheikh-Jabbari}}]{Afshar:2016uax}%
  \BibitemOpen
  \bibfield  {author} {\bibinfo {author} {\bibfnamefont {H.}~\bibnamefont
  {Afshar}}, \bibinfo {author} {\bibfnamefont {D.}~\bibnamefont {Grumiller}}, \
  and\ \bibinfo {author} {\bibfnamefont {M.~M.}\ \bibnamefont
  {Sheikh-Jabbari}},\ }\href@noop {} {\  (\bibinfo {year}
  {2016}{\natexlab{c}})},\ \Eprint {http://arxiv.org/abs/1607.00009}
  {arXiv:1607.00009 [hep-th]} \BibitemShut {NoStop}%
\bibitem [{\citenamefont {Ashtekar}\ \emph {et~al.}(1997)\citenamefont
  {Ashtekar}, \citenamefont {Bicak},\ and\ \citenamefont
  {Schmidt}}]{Ashtekar:1996cd}%
  \BibitemOpen
  \bibfield  {author} {\bibinfo {author} {\bibfnamefont {A.}~\bibnamefont
  {Ashtekar}}, \bibinfo {author} {\bibfnamefont {J.}~\bibnamefont {Bicak}}, \
  and\ \bibinfo {author} {\bibfnamefont {B.~G.}\ \bibnamefont {Schmidt}},\
  }\href {\doibase 10.1103/PhysRevD.55.669} {\bibfield  {journal} {\bibinfo
  {journal} {Phys.Rev.}\ }\textbf {\bibinfo {volume} {D55}},\ \bibinfo {pages}
  {669} (\bibinfo {year} {1997})},\ \Eprint
  {http://arxiv.org/abs/gr-qc/9608042} {arXiv:gr-qc/9608042 [gr-qc]}
  \BibitemShut {NoStop}%
\bibitem [{\citenamefont {Wald}(1993)}]{Wald:1993nt}%
  \BibitemOpen
  \bibfield  {author} {\bibinfo {author} {\bibfnamefont {R.~M.}\ \bibnamefont
  {Wald}},\ }\href@noop {} {\bibfield  {journal} {\bibinfo  {journal} {Phys.
  Rev.}\ }\textbf {\bibinfo {volume} {D48}},\ \bibinfo {pages} {3427} (\bibinfo
  {year} {1993})},\ \Eprint {http://arXiv.org/abs/gr-qc/9307038}
  {gr-qc/9307038} \BibitemShut {NoStop}%
\bibitem [{\citenamefont {Iyer}\ and\ \citenamefont
  {Wald}(1994)}]{Iyer:1994ys}%
  \BibitemOpen
  \bibfield  {author} {\bibinfo {author} {\bibfnamefont {V.}~\bibnamefont
  {Iyer}}\ and\ \bibinfo {author} {\bibfnamefont {R.~M.}\ \bibnamefont
  {Wald}},\ }\href@noop {} {\bibfield  {journal} {\bibinfo  {journal} {Phys.
  Rev.}\ }\textbf {\bibinfo {volume} {D50}},\ \bibinfo {pages} {846} (\bibinfo
  {year} {1994})},\ \Eprint {http://arXiv.org/abs/gr-qc/9403028}
  {gr-qc/9403028} \BibitemShut {NoStop}%
\bibitem [{\citenamefont {Solodukhin}(1995)}]{Solodukhin:1994yz}%
  \BibitemOpen
  \bibfield  {author} {\bibinfo {author} {\bibfnamefont {S.~N.}\ \bibnamefont
  {Solodukhin}},\ }\href {\doibase 10.1103/PhysRevD.51.609} {\bibfield
  {journal} {\bibinfo  {journal} {Phys. Rev.}\ }\textbf {\bibinfo {volume}
  {D51}},\ \bibinfo {pages} {609} (\bibinfo {year} {1995})},\ \Eprint
  {http://arxiv.org/abs/hep-th/9407001} {arXiv:hep-th/9407001 [hep-th]}
  \BibitemShut {NoStop}%
\bibitem [{\citenamefont {Perez}\ \emph
  {et~al.}(2013{\natexlab{a}})\citenamefont {Perez}, \citenamefont {Tempo},\
  and\ \citenamefont {Troncoso}}]{Perez:2012cf}%
  \BibitemOpen
  \bibfield  {author} {\bibinfo {author} {\bibfnamefont {A.}~\bibnamefont
  {Perez}}, \bibinfo {author} {\bibfnamefont {D.}~\bibnamefont {Tempo}}, \ and\
  \bibinfo {author} {\bibfnamefont {R.}~\bibnamefont {Troncoso}},\ }\href
  {\doibase 10.1016/j.physletb.2013.08.038} {\bibfield  {journal} {\bibinfo
  {journal} {Phys. Lett.}\ }\textbf {\bibinfo {volume} {B726}},\ \bibinfo
  {pages} {444} (\bibinfo {year} {2013}{\natexlab{a}})},\ \Eprint
  {http://arxiv.org/abs/1207.2844} {arXiv:1207.2844 [hep-th]} \BibitemShut
  {NoStop}%
\bibitem [{\citenamefont {Perez}\ \emph
  {et~al.}(2013{\natexlab{b}})\citenamefont {Perez}, \citenamefont {Tempo},\
  and\ \citenamefont {Troncoso}}]{Perez:2013xi}%
  \BibitemOpen
  \bibfield  {author} {\bibinfo {author} {\bibfnamefont {A.}~\bibnamefont
  {Perez}}, \bibinfo {author} {\bibfnamefont {D.}~\bibnamefont {Tempo}}, \ and\
  \bibinfo {author} {\bibfnamefont {R.}~\bibnamefont {Troncoso}},\ }\href
  {\doibase 10.1007/JHEP04(2013)143} {\bibfield  {journal} {\bibinfo  {journal}
  {JHEP}\ }\textbf {\bibinfo {volume} {04}},\ \bibinfo {pages} {143} (\bibinfo
  {year} {2013}{\natexlab{b}})},\ \Eprint {http://arxiv.org/abs/1301.0847}
  {arXiv:1301.0847 [hep-th]} \BibitemShut {NoStop}%
\bibitem [{\citenamefont {de~Boer}\ and\ \citenamefont
  {Jottar}(2014)}]{deBoer:2013gz}%
  \BibitemOpen
  \bibfield  {author} {\bibinfo {author} {\bibfnamefont {J.}~\bibnamefont
  {de~Boer}}\ and\ \bibinfo {author} {\bibfnamefont {J.~I.}\ \bibnamefont
  {Jottar}},\ }\href {\doibase 10.1007/JHEP01(2014)023} {\bibfield  {journal}
  {\bibinfo  {journal} {JHEP}\ }\textbf {\bibinfo {volume} {1401}},\ \bibinfo
  {pages} {023} (\bibinfo {year} {2014})},\ \Eprint
  {http://arxiv.org/abs/1302.0816} {arXiv:1302.0816 [hep-th]} \BibitemShut
  {NoStop}%
\bibitem [{\citenamefont {Bloete}\ \emph {et~al.}(1986)\citenamefont {Bloete},
  \citenamefont {Cardy},\ and\ \citenamefont {Nightingale}}]{Bloete:1986qm}%
  \BibitemOpen
  \bibfield  {author} {\bibinfo {author} {\bibfnamefont {H.~W.~J.}\
  \bibnamefont {Bloete}}, \bibinfo {author} {\bibfnamefont {J.~L.}\
  \bibnamefont {Cardy}}, \ and\ \bibinfo {author} {\bibfnamefont {M.~P.}\
  \bibnamefont {Nightingale}},\ }\href@noop {} {\bibfield  {journal} {\bibinfo
  {journal} {Phys. Rev. Lett.}\ }\textbf {\bibinfo {volume} {56}},\ \bibinfo
  {pages} {742} (\bibinfo {year} {1986})}\BibitemShut {NoStop}%
\bibitem [{\citenamefont {Cardy}(1986)}]{Cardy:1986ie}%
  \BibitemOpen
  \bibfield  {author} {\bibinfo {author} {\bibfnamefont {J.~L.}\ \bibnamefont
  {Cardy}},\ }\href@noop {} {\bibfield  {journal} {\bibinfo  {journal} {Nucl.
  Phys.}\ }\textbf {\bibinfo {volume} {B270}},\ \bibinfo {pages} {186}
  (\bibinfo {year} {1986})}\BibitemShut {NoStop}%
\bibitem [{\citenamefont {Strominger}(1998)}]{Strominger:1997eq}%
  \BibitemOpen
  \bibfield  {author} {\bibinfo {author} {\bibfnamefont {A.}~\bibnamefont
  {Strominger}},\ }\href@noop {} {\bibfield  {journal} {\bibinfo  {journal}
  {JHEP}\ }\textbf {\bibinfo {volume} {02}},\ \bibinfo {pages} {009} (\bibinfo
  {year} {1998})},\ \Eprint {http://arxiv.org/abs/hep-th/9712251}
  {hep-th/9712251} \BibitemShut {NoStop}%
\bibitem [{\citenamefont {Sheikh-Jabbari}\ and\ \citenamefont
  {Yavartanoo}(2016{\natexlab{a}})}]{Sheikh-Jabbari:2016npa}%
  \BibitemOpen
  \bibfield  {author} {\bibinfo {author} {\bibfnamefont {M.~M.}\ \bibnamefont
  {Sheikh-Jabbari}}\ and\ \bibinfo {author} {\bibfnamefont {H.}~\bibnamefont
  {Yavartanoo}},\ }\href@noop {} {\  (\bibinfo {year} {2016}{\natexlab{a}})},\
  \Eprint {http://arxiv.org/abs/1608.01293} {arXiv:1608.01293 [hep-th]}
  \BibitemShut {NoStop}%
\bibitem [{\citenamefont {Gonzalez}\ \emph {et~al.}(2011)\citenamefont
  {Gonzalez}, \citenamefont {Tempo},\ and\ \citenamefont
  {Troncoso}}]{Gonzalez:2011nz}%
  \BibitemOpen
  \bibfield  {author} {\bibinfo {author} {\bibfnamefont {H.~A.}\ \bibnamefont
  {Gonzalez}}, \bibinfo {author} {\bibfnamefont {D.}~\bibnamefont {Tempo}}, \
  and\ \bibinfo {author} {\bibfnamefont {R.}~\bibnamefont {Troncoso}},\ }\href
  {\doibase 10.1007/JHEP11(2011)066} {\bibfield  {journal} {\bibinfo  {journal}
  {JHEP}\ }\textbf {\bibinfo {volume} {11}},\ \bibinfo {pages} {066} (\bibinfo
  {year} {2011})},\ \Eprint {http://arxiv.org/abs/1107.3647} {arXiv:1107.3647
  [hep-th]} \BibitemShut {NoStop}%
\bibitem [{\citenamefont {Barnich}(2012)}]{Barnich:2012xq}%
  \BibitemOpen
  \bibfield  {author} {\bibinfo {author} {\bibfnamefont {G.}~\bibnamefont
  {Barnich}},\ }\href {\doibase 10.1007/JHEP10(2012)095} {\bibfield  {journal}
  {\bibinfo  {journal} {JHEP}\ }\textbf {\bibinfo {volume} {1210}},\ \bibinfo
  {pages} {095} (\bibinfo {year} {2012})},\ \Eprint
  {http://arxiv.org/abs/1208.4371} {arXiv:1208.4371 [hep-th]} \BibitemShut
  {NoStop}%
\bibitem [{\citenamefont {Bagchi}\ \emph {et~al.}(2013)\citenamefont {Bagchi},
  \citenamefont {Detournay}, \citenamefont {Fareghbal},\ and\ \citenamefont
  {Simon}}]{Bagchi:2012xr}%
  \BibitemOpen
  \bibfield  {author} {\bibinfo {author} {\bibfnamefont {A.}~\bibnamefont
  {Bagchi}}, \bibinfo {author} {\bibfnamefont {S.}~\bibnamefont {Detournay}},
  \bibinfo {author} {\bibfnamefont {R.}~\bibnamefont {Fareghbal}}, \ and\
  \bibinfo {author} {\bibfnamefont {J.}~\bibnamefont {Simon}},\ }\href
  {\doibase 10.1103/PhysRevLett.110.141302} {\bibfield  {journal} {\bibinfo
  {journal} {Phys. Rev. Lett.}\ }\textbf {\bibinfo {volume} {110}},\ \bibinfo
  {pages} {141302} (\bibinfo {year} {2013})},\ \Eprint
  {http://arxiv.org/abs/1208.4372} {arXiv:1208.4372 [hep-th]} \BibitemShut
  {NoStop}%
\bibitem [{\citenamefont {Riegler}(2015)}]{Riegler:2014bia}%
  \BibitemOpen
  \bibfield  {author} {\bibinfo {author} {\bibfnamefont {M.}~\bibnamefont
  {Riegler}},\ }\href {\doibase 10.1103/PhysRevD.91.024044} {\bibfield
  {journal} {\bibinfo  {journal} {Phys.Rev.}\ }\textbf {\bibinfo {volume}
  {D91}},\ \bibinfo {pages} {024044} (\bibinfo {year} {2015})},\ \Eprint
  {http://arxiv.org/abs/1408.6931} {arXiv:1408.6931 [hep-th]} \BibitemShut
  {NoStop}%
\bibitem [{\citenamefont {Fareghbal}\ and\ \citenamefont
  {Naseh}(2015)}]{Fareghbal:2014qga}%
  \BibitemOpen
  \bibfield  {author} {\bibinfo {author} {\bibfnamefont {R.}~\bibnamefont
  {Fareghbal}}\ and\ \bibinfo {author} {\bibfnamefont {A.}~\bibnamefont
  {Naseh}},\ }\href {\doibase 10.1088/0264-9381/32/13/135013} {\bibfield
  {journal} {\bibinfo  {journal} {Class.Quant.Grav.}\ }\textbf {\bibinfo
  {volume} {32}},\ \bibinfo {pages} {135013} (\bibinfo {year} {2015})},\
  \Eprint {http://arxiv.org/abs/1408.6932} {arXiv:1408.6932 [hep-th]}
  \BibitemShut {NoStop}%
\bibitem [{\citenamefont {Castro}\ and\ \citenamefont
  {Rodriguez}(2012)}]{Castro:2012av}%
  \BibitemOpen
  \bibfield  {author} {\bibinfo {author} {\bibfnamefont {A.}~\bibnamefont
  {Castro}}\ and\ \bibinfo {author} {\bibfnamefont {M.~J.}\ \bibnamefont
  {Rodriguez}},\ }\href {\doibase 10.1103/PhysRevD.86.024008} {\bibfield
  {journal} {\bibinfo  {journal} {Phys.Rev.}\ }\textbf {\bibinfo {volume}
  {D86}},\ \bibinfo {pages} {024008} (\bibinfo {year} {2012})},\ \Eprint
  {http://arxiv.org/abs/1204.1284} {arXiv:1204.1284 [hep-th]} \BibitemShut
  {NoStop}%
\bibitem [{\citenamefont {Detournay}(2012)}]{Detournay:2012ug}%
  \BibitemOpen
  \bibfield  {author} {\bibinfo {author} {\bibfnamefont {S.}~\bibnamefont
  {Detournay}},\ }\href {\doibase 10.1103/PhysRevLett.109.031101} {\bibfield
  {journal} {\bibinfo  {journal} {Phys.Rev.Lett.}\ }\textbf {\bibinfo {volume}
  {109}},\ \bibinfo {pages} {031101} (\bibinfo {year} {2012})},\ \Eprint
  {http://arxiv.org/abs/1204.6088} {arXiv:1204.6088 [hep-th]} \BibitemShut
  {NoStop}%
\bibitem [{\citenamefont {Strominger}\ and\ \citenamefont
  {Zhiboedov}(2016)}]{Strominger:2014pwa}%
  \BibitemOpen
  \bibfield  {author} {\bibinfo {author} {\bibfnamefont {A.}~\bibnamefont
  {Strominger}}\ and\ \bibinfo {author} {\bibfnamefont {A.}~\bibnamefont
  {Zhiboedov}},\ }\href {\doibase 10.1007/JHEP01(2016)086} {\bibfield
  {journal} {\bibinfo  {journal} {JHEP}\ }\textbf {\bibinfo {volume} {01}},\
  \bibinfo {pages} {086} (\bibinfo {year} {2016})},\ \Eprint
  {http://arxiv.org/abs/1411.5745} {arXiv:1411.5745 [hep-th]} \BibitemShut
  {NoStop}%
\bibitem [{\citenamefont {Flanagan}\ and\ \citenamefont
  {Nichols}(2015)}]{Flanagan:2015pxa}%
  \BibitemOpen
  \bibfield  {author} {\bibinfo {author} {\bibfnamefont {E.~E.}\ \bibnamefont
  {Flanagan}}\ and\ \bibinfo {author} {\bibfnamefont {D.~A.}\ \bibnamefont
  {Nichols}},\ }\href@noop {} {\  (\bibinfo {year} {2015})},\ \Eprint
  {http://arxiv.org/abs/1510.03386} {arXiv:1510.03386 [hep-th]} \BibitemShut
  {NoStop}%
\bibitem [{\citenamefont {Comp{\`e}re}\ and\ \citenamefont
  {Long}(2016{\natexlab{a}})}]{Compere:2016hzt}%
  \BibitemOpen
  \bibfield  {author} {\bibinfo {author} {\bibfnamefont {G.}~\bibnamefont
  {Comp{\`e}re}}\ and\ \bibinfo {author} {\bibfnamefont {J.}~\bibnamefont
  {Long}},\ }\href {\doibase 10.1088/0264-9381/33/19/195001} {\bibfield
  {journal} {\bibinfo  {journal} {Class. Quant. Grav.}\ }\textbf {\bibinfo
  {volume} {33}},\ \bibinfo {pages} {195001} (\bibinfo {year}
  {2016}{\natexlab{a}})},\ \Eprint {http://arxiv.org/abs/1602.05197}
  {arXiv:1602.05197 [gr-qc]} \BibitemShut {NoStop}%
\bibitem [{\citenamefont {Penna}(2016)}]{Penna:2015gza}%
  \BibitemOpen
  \bibfield  {author} {\bibinfo {author} {\bibfnamefont {R.~F.}\ \bibnamefont
  {Penna}},\ }\href {\doibase 10.1007/JHEP03(2016)023} {\bibfield  {journal}
  {\bibinfo  {journal} {JHEP}\ }\textbf {\bibinfo {volume} {03}},\ \bibinfo
  {pages} {023} (\bibinfo {year} {2016})},\ \Eprint
  {http://arxiv.org/abs/1508.06577} {arXiv:1508.06577 [hep-th]} \BibitemShut
  {NoStop}%
\bibitem [{\citenamefont {'t~Hooft}(2016)}]{Hooft:2016itl}%
  \BibitemOpen
  \bibfield  {author} {\bibinfo {author} {\bibfnamefont {G.}~\bibnamefont
  {'t~Hooft}},\ }\href {\doibase 10.1007/s10701-016-0014-y} {\bibfield
  {journal} {\bibinfo  {journal} {Found. Phys.}\ }\textbf {\bibinfo {volume}
  {46}},\ \bibinfo {pages} {1185} (\bibinfo {year} {2016})},\ \Eprint
  {http://arxiv.org/abs/1601.03447} {arXiv:1601.03447 [gr-qc]} \BibitemShut
  {NoStop}%
\bibitem [{\citenamefont {Averin}\ \emph
  {et~al.}(2016{\natexlab{a}})\citenamefont {Averin}, \citenamefont {Dvali},
  \citenamefont {Gomez},\ and\ \citenamefont {Lust}}]{Averin:2016ybl}%
  \BibitemOpen
  \bibfield  {author} {\bibinfo {author} {\bibfnamefont {A.}~\bibnamefont
  {Averin}}, \bibinfo {author} {\bibfnamefont {G.}~\bibnamefont {Dvali}},
  \bibinfo {author} {\bibfnamefont {C.}~\bibnamefont {Gomez}}, \ and\ \bibinfo
  {author} {\bibfnamefont {D.}~\bibnamefont {Lust}},\ }\href {\doibase
  10.1007/JHEP06(2016)088} {\bibfield  {journal} {\bibinfo  {journal} {JHEP}\
  }\textbf {\bibinfo {volume} {06}},\ \bibinfo {pages} {088} (\bibinfo {year}
  {2016}{\natexlab{a}})},\ \Eprint {http://arxiv.org/abs/1601.03725}
  {arXiv:1601.03725 [hep-th]} \BibitemShut {NoStop}%
\bibitem [{\citenamefont {Comp{\`e}re}\ and\ \citenamefont
  {Long}(2016{\natexlab{b}})}]{Compere:2016jwb}%
  \BibitemOpen
  \bibfield  {author} {\bibinfo {author} {\bibfnamefont {G.}~\bibnamefont
  {Comp{\`e}re}}\ and\ \bibinfo {author} {\bibfnamefont {J.}~\bibnamefont
  {Long}},\ }\href {\doibase 10.1007/JHEP07(2016)137} {\bibfield  {journal}
  {\bibinfo  {journal} {JHEP}\ }\textbf {\bibinfo {volume} {07}},\ \bibinfo
  {pages} {137} (\bibinfo {year} {2016}{\natexlab{b}})},\ \Eprint
  {http://arxiv.org/abs/1601.04958} {arXiv:1601.04958 [hep-th]} \BibitemShut
  {NoStop}%
\bibitem [{\citenamefont {Sheikh-Jabbari}\ and\ \citenamefont
  {Yavartanoo}(2016{\natexlab{b}})}]{Sheikh-Jabbari:2016unm}%
  \BibitemOpen
  \bibfield  {author} {\bibinfo {author} {\bibfnamefont {M.~M.}\ \bibnamefont
  {Sheikh-Jabbari}}\ and\ \bibinfo {author} {\bibfnamefont {H.}~\bibnamefont
  {Yavartanoo}},\ }\href {\doibase 10.1140/epjc/s10052-016-4326-z} {\bibfield
  {journal} {\bibinfo  {journal} {Eur. Phys. J.}\ }\textbf {\bibinfo {volume}
  {C76}},\ \bibinfo {pages} {493} (\bibinfo {year} {2016}{\natexlab{b}})},\
  \Eprint {http://arxiv.org/abs/1603.05272} {arXiv:1603.05272 [hep-th]}
  \BibitemShut {NoStop}%
\bibitem [{\citenamefont {Sheikh-Jabbari}(2016)}]{Sheikh-Jabbari:2016lzm}%
  \BibitemOpen
  \bibfield  {author} {\bibinfo {author} {\bibfnamefont {M.~M.}\ \bibnamefont
  {Sheikh-Jabbari}},\ }\href {\doibase 10.1142/S0218271816440193} {\bibfield
  {journal} {\bibinfo  {journal} {Int. J. Mod. Phys.}\ }\textbf {\bibinfo
  {volume} {D25}},\ \bibinfo {pages} {1644019} (\bibinfo {year} {2016})},\
  \Eprint {http://arxiv.org/abs/1603.07862} {arXiv:1603.07862 [hep-th]}
  \BibitemShut {NoStop}%
\bibitem [{\citenamefont {Blau}\ and\ \citenamefont
  {O'Loughlin}(2016)}]{Blau:2016juv}%
  \BibitemOpen
  \bibfield  {author} {\bibinfo {author} {\bibfnamefont {M.}~\bibnamefont
  {Blau}}\ and\ \bibinfo {author} {\bibfnamefont {M.}~\bibnamefont
  {O'Loughlin}},\ }\href {\doibase 10.1142/S0218271816440107} {\bibfield
  {journal} {\bibinfo  {journal} {Int. J. Mod. Phys.}\ }\textbf {\bibinfo
  {volume} {D25}},\ \bibinfo {pages} {1644010} (\bibinfo {year} {2016})},\
  \Eprint {http://arxiv.org/abs/1604.01181} {arXiv:1604.01181 [hep-th]}
  \BibitemShut {NoStop}%
\bibitem [{\citenamefont {Eling}\ and\ \citenamefont
  {Oz}(2016)}]{Eling:2016xlx}%
  \BibitemOpen
  \bibfield  {author} {\bibinfo {author} {\bibfnamefont {C.}~\bibnamefont
  {Eling}}\ and\ \bibinfo {author} {\bibfnamefont {Y.}~\bibnamefont {Oz}},\
  }\href {\doibase 10.1007/JHEP07(2016)065} {\bibfield  {journal} {\bibinfo
  {journal} {JHEP}\ }\textbf {\bibinfo {volume} {07}},\ \bibinfo {pages} {065}
  (\bibinfo {year} {2016})},\ \Eprint {http://arxiv.org/abs/1605.00183}
  {arXiv:1605.00183 [hep-th]} \BibitemShut {NoStop}%
\bibitem [{\citenamefont {Ellis}\ \emph {et~al.}(2016)\citenamefont {Ellis},
  \citenamefont {Mavromatos},\ and\ \citenamefont
  {Nanopoulos}}]{Ellis:2016atb}%
  \BibitemOpen
  \bibfield  {author} {\bibinfo {author} {\bibfnamefont {J.}~\bibnamefont
  {Ellis}}, \bibinfo {author} {\bibfnamefont {N.~E.}\ \bibnamefont
  {Mavromatos}}, \ and\ \bibinfo {author} {\bibfnamefont {D.~V.}\ \bibnamefont
  {Nanopoulos}},\ }\href {\doibase 10.1103/PhysRevD.94.025007} {\bibfield
  {journal} {\bibinfo  {journal} {Phys. Rev.}\ }\textbf {\bibinfo {volume}
  {D94}},\ \bibinfo {pages} {025007} (\bibinfo {year} {2016})},\ \Eprint
  {http://arxiv.org/abs/1605.01653} {arXiv:1605.01653 [hep-th]} \BibitemShut
  {NoStop}%
\bibitem [{\citenamefont {Giddings}(2016)}]{Giddings:2016plq}%
  \BibitemOpen
  \bibfield  {author} {\bibinfo {author} {\bibfnamefont {S.~B.}\ \bibnamefont
  {Giddings}},\ }\href {\doibase 10.1142/S0218271816440144} {\bibfield
  {journal} {\bibinfo  {journal} {Int. J. Mod. Phys.}\ }\textbf {\bibinfo
  {volume} {D25}},\ \bibinfo {pages} {1644014} (\bibinfo {year} {2016})},\
  \Eprint {http://arxiv.org/abs/1605.05341} {arXiv:1605.05341 [gr-qc]}
  \BibitemShut {NoStop}%
\bibitem [{\citenamefont {Comp{\`e}re}(2016)}]{Compere:2016gwf}%
  \BibitemOpen
  \bibfield  {author} {\bibinfo {author} {\bibfnamefont {G.}~\bibnamefont
  {Comp{\`e}re}},\ }\href {\doibase 10.1142/S0218271816440065} {\bibfield
  {journal} {\bibinfo  {journal} {Int. J. Mod. Phys.}\ }\textbf {\bibinfo
  {volume} {D25}},\ \bibinfo {pages} {1644006} (\bibinfo {year} {2016})},\
  \Eprint {http://arxiv.org/abs/1606.00377} {arXiv:1606.00377 [hep-th]}
  \BibitemShut {NoStop}%
\bibitem [{\citenamefont {M{\"u}ck}(2016)}]{Muck:2016stv}%
  \BibitemOpen
  \bibfield  {author} {\bibinfo {author} {\bibfnamefont {W.}~\bibnamefont
  {M{\"u}ck}},\ }\href {\doibase 10.1140/epjc/s10052-016-4233-3} {\bibfield
  {journal} {\bibinfo  {journal} {Eur. Phys. J.}\ }\textbf {\bibinfo {volume}
  {C76}},\ \bibinfo {pages} {374} (\bibinfo {year} {2016})},\ \Eprint
  {http://arxiv.org/abs/1606.01790} {arXiv:1606.01790 [hep-th]} \BibitemShut
  {NoStop}%
\bibitem [{\citenamefont {Hotta}\ \emph {et~al.}(2016)\citenamefont {Hotta},
  \citenamefont {Trevison},\ and\ \citenamefont {Yamaguchi}}]{Hotta:2016qtv}%
  \BibitemOpen
  \bibfield  {author} {\bibinfo {author} {\bibfnamefont {M.}~\bibnamefont
  {Hotta}}, \bibinfo {author} {\bibfnamefont {J.}~\bibnamefont {Trevison}}, \
  and\ \bibinfo {author} {\bibfnamefont {K.}~\bibnamefont {Yamaguchi}},\ }\href
  {\doibase 10.1103/PhysRevD.94.083001} {\bibfield  {journal} {\bibinfo
  {journal} {Phys. Rev.}\ }\textbf {\bibinfo {volume} {D94}},\ \bibinfo {pages}
  {083001} (\bibinfo {year} {2016})},\ \Eprint
  {http://arxiv.org/abs/1606.02443} {arXiv:1606.02443 [gr-qc]} \BibitemShut
  {NoStop}%
\bibitem [{\citenamefont {Averin}\ \emph
  {et~al.}(2016{\natexlab{b}})\citenamefont {Averin}, \citenamefont {Dvali},
  \citenamefont {Gomez},\ and\ \citenamefont {Lust}}]{Averin:2016hhm}%
  \BibitemOpen
  \bibfield  {author} {\bibinfo {author} {\bibfnamefont {A.}~\bibnamefont
  {Averin}}, \bibinfo {author} {\bibfnamefont {G.}~\bibnamefont {Dvali}},
  \bibinfo {author} {\bibfnamefont {C.}~\bibnamefont {Gomez}}, \ and\ \bibinfo
  {author} {\bibfnamefont {D.}~\bibnamefont {Lust}},\ }\href@noop {} {\
  (\bibinfo {year} {2016}{\natexlab{b}})},\ \Eprint
  {http://arxiv.org/abs/1606.06260} {arXiv:1606.06260 [hep-th]} \BibitemShut
  {NoStop}%
\bibitem [{\citenamefont {Mirbabayi}\ and\ \citenamefont
  {Porrati}(2016)}]{Mirbabayi:2016axw}%
  \BibitemOpen
  \bibfield  {author} {\bibinfo {author} {\bibfnamefont {M.}~\bibnamefont
  {Mirbabayi}}\ and\ \bibinfo {author} {\bibfnamefont {M.}~\bibnamefont
  {Porrati}},\ }\href {\doibase 10.1103/PhysRevLett.117.211301} {\bibfield
  {journal} {\bibinfo  {journal} {Phys. Rev. Lett.}\ }\textbf {\bibinfo
  {volume} {117}},\ \bibinfo {pages} {211301} (\bibinfo {year} {2016})},\
  \Eprint {http://arxiv.org/abs/1607.03120} {arXiv:1607.03120 [hep-th]}
  \BibitemShut {NoStop}%
\bibitem [{\citenamefont {Donnay}\ \emph
  {et~al.}(2016{\natexlab{b}})\citenamefont {Donnay}, \citenamefont {Giribet},
  \citenamefont {Gonz\'alez},\ and\ \citenamefont {Pino}}]{Donnay:2016ejv}%
  \BibitemOpen
  \bibfield  {author} {\bibinfo {author} {\bibfnamefont {L.}~\bibnamefont
  {Donnay}}, \bibinfo {author} {\bibfnamefont {G.}~\bibnamefont {Giribet}},
  \bibinfo {author} {\bibfnamefont {H.~A.}\ \bibnamefont {Gonz\'alez}}, \ and\
  \bibinfo {author} {\bibfnamefont {M.}~\bibnamefont {Pino}},\ }\href {\doibase
  10.1007/JHEP09(2016)100} {\bibfield  {journal} {\bibinfo  {journal} {JHEP}\
  }\textbf {\bibinfo {volume} {09}},\ \bibinfo {pages} {100} (\bibinfo {year}
  {2016}{\natexlab{b}})},\ \Eprint {http://arxiv.org/abs/1607.05703}
  {arXiv:1607.05703 [hep-th]} \BibitemShut {NoStop}%
\bibitem [{\citenamefont {Cai}\ \emph {et~al.}(2016)\citenamefont {Cai},
  \citenamefont {Ruan},\ and\ \citenamefont {Zhang}}]{Cai:2016idg}%
  \BibitemOpen
  \bibfield  {author} {\bibinfo {author} {\bibfnamefont {R.-G.}\ \bibnamefont
  {Cai}}, \bibinfo {author} {\bibfnamefont {S.-M.}\ \bibnamefont {Ruan}}, \
  and\ \bibinfo {author} {\bibfnamefont {Y.-L.}\ \bibnamefont {Zhang}},\ }\href
  {\doibase 10.1007/JHEP09(2016)163} {\bibfield  {journal} {\bibinfo  {journal}
  {JHEP}\ }\textbf {\bibinfo {volume} {09}},\ \bibinfo {pages} {163} (\bibinfo
  {year} {2016})},\ \Eprint {http://arxiv.org/abs/1609.01056} {arXiv:1609.01056
  [gr-qc]} \BibitemShut {NoStop}%
\bibitem [{\citenamefont {Banerjee}\ \emph {et~al.}(2016)\citenamefont
  {Banerjee}, \citenamefont {Jatkar}, \citenamefont {Lodato}, \citenamefont
  {Mukhi},\ and\ \citenamefont {Neogi}}]{Banerjee:2016nio}%
  \BibitemOpen
  \bibfield  {author} {\bibinfo {author} {\bibfnamefont {N.}~\bibnamefont
  {Banerjee}}, \bibinfo {author} {\bibfnamefont {D.~P.}\ \bibnamefont
  {Jatkar}}, \bibinfo {author} {\bibfnamefont {I.}~\bibnamefont {Lodato}},
  \bibinfo {author} {\bibfnamefont {S.}~\bibnamefont {Mukhi}}, \ and\ \bibinfo
  {author} {\bibfnamefont {T.}~\bibnamefont {Neogi}},\ }\href {\doibase
  10.1007/JHEP11(2016)059} {\bibfield  {journal} {\bibinfo  {journal} {JHEP}\
  }\textbf {\bibinfo {volume} {11}},\ \bibinfo {pages} {059} (\bibinfo {year}
  {2016})},\ \Eprint {http://arxiv.org/abs/1609.09210} {arXiv:1609.09210
  [hep-th]} \BibitemShut {NoStop}%
\bibitem [{\citenamefont {Setare}\ and\ \citenamefont
  {Adami}(2016)}]{Setare:2016qob}%
  \BibitemOpen
  \bibfield  {author} {\bibinfo {author} {\bibfnamefont {M.~R.}\ \bibnamefont
  {Setare}}\ and\ \bibinfo {author} {\bibfnamefont {H.}~\bibnamefont {Adami}},\
  }\href@noop {} {\  (\bibinfo {year} {2016})},\ \Eprint
  {http://arxiv.org/abs/1611.04259} {arXiv:1611.04259 [hep-th]} \BibitemShut
  {NoStop}%
\bibitem [{\citenamefont {Setare}\ and\ \citenamefont
  {Adami}(2017)}]{Setare:2016vhy}%
  \BibitemOpen
  \bibfield  {author} {\bibinfo {author} {\bibfnamefont {M.~R.}\ \bibnamefont
  {Setare}}\ and\ \bibinfo {author} {\bibfnamefont {H.}~\bibnamefont {Adami}},\
  }\href@noop {} {\bibfield  {journal} {\bibinfo  {journal} {Nucl. Phys.}\
  }\textbf {\bibinfo {volume} {B914}},\ \bibinfo {pages} {220} (\bibinfo {year}
  {2017})},\ \Eprint {http://arxiv.org/abs/1606.05260} {arXiv:1606.05260
  [hep-th]} \BibitemShut {NoStop}%
\bibitem [{\citenamefont {Banados}(1999)}]{Banados:1998wy}%
  \BibitemOpen
  \bibfield  {author} {\bibinfo {author} {\bibfnamefont {M.}~\bibnamefont
  {Banados}},\ }\href {\doibase 10.1103/PhysRevLett.82.2030} {\bibfield
  {journal} {\bibinfo  {journal} {Phys. Rev. Lett.}\ }\textbf {\bibinfo
  {volume} {82}},\ \bibinfo {pages} {2030} (\bibinfo {year} {1999})},\ \Eprint
  {http://arxiv.org/abs/hep-th/9811162} {arXiv:hep-th/9811162 [hep-th]}
  \BibitemShut {NoStop}%
\bibitem [{\citenamefont {Bagchi}\ \emph {et~al.}(2012)\citenamefont {Bagchi},
  \citenamefont {Detournay},\ and\ \citenamefont {Grumiller}}]{Bagchi:2012yk}%
  \BibitemOpen
  \bibfield  {author} {\bibinfo {author} {\bibfnamefont {A.}~\bibnamefont
  {Bagchi}}, \bibinfo {author} {\bibfnamefont {S.}~\bibnamefont {Detournay}}, \
  and\ \bibinfo {author} {\bibfnamefont {D.}~\bibnamefont {Grumiller}},\ }\href
  {\doibase 10.1103/PhysRevLett.109.151301} {\bibfield  {journal} {\bibinfo
  {journal} {Phys.Rev.Lett.}\ }\textbf {\bibinfo {volume} {109}},\ \bibinfo
  {pages} {151301} (\bibinfo {year} {2012})},\ \Eprint
  {http://arxiv.org/abs/1208.1658} {arXiv:1208.1658 [hep-th]} \BibitemShut
  {NoStop}%
\bibitem [{\citenamefont {Afshar}\ \emph {et~al.}(2013)\citenamefont {Afshar},
  \citenamefont {Bagchi}, \citenamefont {Fareghbal}, \citenamefont
  {Grumiller},\ and\ \citenamefont {Rosseel}}]{Afshar:2013vka}%
  \BibitemOpen
  \bibfield  {author} {\bibinfo {author} {\bibfnamefont {H.}~\bibnamefont
  {Afshar}}, \bibinfo {author} {\bibfnamefont {A.}~\bibnamefont {Bagchi}},
  \bibinfo {author} {\bibfnamefont {R.}~\bibnamefont {Fareghbal}}, \bibinfo
  {author} {\bibfnamefont {D.}~\bibnamefont {Grumiller}}, \ and\ \bibinfo
  {author} {\bibfnamefont {J.}~\bibnamefont {Rosseel}},\ }\href {\doibase
  10.1103/PhysRevLett.111.121603} {\bibfield  {journal} {\bibinfo  {journal}
  {Phys.Rev.Lett.}\ }\textbf {\bibinfo {volume} {111}},\ \bibinfo {pages}
  {121603} (\bibinfo {year} {2013})},\ \Eprint {http://arxiv.org/abs/1307.4768}
  {arXiv:1307.4768 [hep-th]} \BibitemShut {NoStop}%
\bibitem [{\citenamefont {Gonzalez}\ \emph {et~al.}(2013)\citenamefont
  {Gonzalez}, \citenamefont {Matulich}, \citenamefont {Pino},\ and\
  \citenamefont {Troncoso}}]{Gonzalez:2013oaa}%
  \BibitemOpen
  \bibfield  {author} {\bibinfo {author} {\bibfnamefont {H.~A.}\ \bibnamefont
  {Gonzalez}}, \bibinfo {author} {\bibfnamefont {J.}~\bibnamefont {Matulich}},
  \bibinfo {author} {\bibfnamefont {M.}~\bibnamefont {Pino}}, \ and\ \bibinfo
  {author} {\bibfnamefont {R.}~\bibnamefont {Troncoso}},\ }\href {\doibase
  10.1007/JHEP09(2013)016} {\bibfield  {journal} {\bibinfo  {journal} {JHEP}\
  }\textbf {\bibinfo {volume} {1309}},\ \bibinfo {pages} {016} (\bibinfo {year}
  {2013})},\ \Eprint {http://arxiv.org/abs/1307.5651} {arXiv:1307.5651
  [hep-th]} \BibitemShut {NoStop}%
\end{thebibliography}

%

\end{document}